\PassOptionsToPackage{british}{babel}
\documentclass{aa}  
\let\orienddocument\enddocument
\let\enddocument\orienddocument
\AtBeginDocument{\selectlanguage{british}}
\usepackage{graphicx}
\usepackage[breaklinks=true]{hyperref}
\usepackage{rotating}
\usepackage{acronym}
\usepackage{longtable}
\usepackage{booktabs}

\usepackage{txfonts}
\usepackage{array}
\usepackage{capt-of}
\makeatletter\let\hyper@natlinkstart\@gobble\let\hyper@natlinkend\@empty\def\hyper@natlinkbreak#1#2{#1}\makeatother
\begin{document}

   \title{The NEMESIS general YSO catalogue}\subtitle{I. Supervised classification with deep learning methods}

   \author{G. Marton
          \inst{\ref{konkoly},\ref{mtaexcellence}}\thanks{Corresponding author: \email{marton.gabor@csfk.org}}
          \and
          M. Madarász\inst{\ref{konkoly},\ref{mtaexcellence},\ref{szte}}
          \and
          J. Roquette\inst{\ref{inst-ch-obs},\ref{inst-ch-neuro}}
          \and
          M. Audard\inst{\ref{inst-ch-obs}}
          \and
          I. Gezer\inst{\ref{konkoly},\ref{mtaexcellence}}
          \and          
          D. Hernandez\inst{\ref{univie}}
          \and
          O. Dionatos\inst{\ref{univie}}
          }

   \institute{Konkoly Observatory, Research Centre for Astronomy and Earth Sciences, Hungarian Research Network (HUN-REN), Konkoly Thege Miklós út 15--17, 1121 Budapest, Hungary\label{konkoly}        
         \and
CSFK, MTA Centre of Excellence, Konkoly Thege Miklós út 15--17, 1121 Budapest, Hungary\label{mtaexcellence}
\and
Department of Experimental Physics, Institute of Physics, University of Szeged, D{\'o}m t{\'e}r 9, 6720 Szeged, Hungary\label{szte}
\and
Université de Genève, Department of Astronomy, Chemin Pegasi 51, 1290 Versoix, Switzerland\label{inst-ch-obs}
\and
Université de Genève, Department of Basic Neuroscience, 1211 Geneva, Switzerland\label{inst-ch-neuro}
\and
Universität Wien, Institut für Astrophysik, Türkenschanzstrasse 17, 1180 Vienna, Austria\label{univie}
             }

\date{Received 9 April 2026 / Accepted 24 September 24}

  \abstract
   {In modern large-scale photometric surveys, young stellar objects (YSOs) are mostly identified by the infrared excess seen in their spectral energy distributions (SEDs). To this end, YSO identification and classification were mostly based on certain colours specific to a dataset, resulting in wavelength-specific methods and catalogues. Using all available data allows us to construct a uniformly classified YSO catalogue by applying the same identification method to heterogeneous literature samples, thereby reducing systematic differences between catalogues and enabling more consistent studies of YSO populations and star formation.}
   {Our primary goal is to develop a method that takes advantage of the heterogeneous data available in the VizieR database and to provide a tool that can be easily used to identify YSOs with high accuracy. We also aim to create the largest homogeneous catalogue of YSOs to date.}
   {We use the CDS VizieR and NASA/IPAC Infrared Science Archive (IRSA) databases to collect data for a large number of celestial objects of various types. The data collected included SEDs, photometric light curves, and direct imaging data and were used to build different image representations for our list of sources. These image representations were used to teach deep learning algorithms to differentiate between YSOs and all other types of sources. Established architectures from the \texttt{PyTorch} library and custom-built convolutional neural networks (CNNs) were applied to create an ensemble of binary classifiers specialised at identifying YSOs against other types of astrophysical sources.}
   {The resulting method is able to automatically identify YSO candidates with an accuracy higher than 90\% while contamination is identified with an accuracy higher than 99\%. We applied our classifier to numerous YSO catalogues from the literature to confirm the youth of the stars listed there and to create the largest catalogue of homogeneously identified potential YSOs ever -- the NEMESIS General YSO (NGYSO) catalogue, which contains 274\,408 unique, high reliability YSOs. Their positions on the colour-magnitude diagrams, and their distribution relative to the local Galactic structures, were also used to test their young nature.}
   {}


   \maketitle
%

\section{Introduction}

Studying young stellar objects (YSOs) allows us to better understand the physical processes and circumstances regulating star formation, which also has implications for planetary system formation \citep{1987ARA&A..25...23S,KennicuttEvans2012,Krumholz2014}. Various complex physical processes, such as turbulence, magnetic fields, and interactions with the interstellar medium, influence star formation \citep[e.g.,][]{2007ARA&A..45..565M,Krumholz2014,Padoan2014}. Also, YSOs have several distinctive features, including accretion disks, outflows, and jets, which offer important insights into the dynamics and interactions that take place during these formative phases \citep[e.g.,][]{1998ApJ...495..385H, 2016ARA&A..54..491B}.

Stars exhibit many characteristics in their spectral energy distribution (SED) throughout their early existence \citep{1993ApJ...406..122A, 1987IAUS..115....1L}. In the early stages protostars are first completely embedded in their parental clouds and are detectable only at submillimetre and far-infrared wavelengths due to the dense dust and gas envelopes that surround them. As their envelopes dissolve, young pre-main-sequence stars become visible at optical wavelengths as well. Their protoplanetary disks are still forming, they still have circumstellar matter, the stellar magnetosphere drives accretion columns, and accretion shocks are visible at these evolutionary phases, and can be observed at various wavelengths \citep[e.g.,][]{2014ApJ...783...29D,2014prpl.conf..387A,2014A&A...561A...2A,2015ApJS..220...11D}. Because of all the above processes, YSOs are intrinsically variable, and their photometric variability also helps us to reveal the key physics behind their formation and evolution \citep{2014prpl.conf..387A, 2014AJ....147...82C, 2014AJ....147...83S}.  To fully understand the underlying mechanisms and statistical trends, studies require statistically significant sample sizes.

Previous large-scale YSO catalogues \citep[e.g.,][]{2016MNRAS.458.3479M, 2019MNRAS.487.2522M, 2018A&A...620A.172Z, 2021A&A...647A.116C, 2021AJ....162..282M} relied on predefined colours and limited-band photometry, or a combination of photometry and astrometry \citep{2022A&A...664A.175P, 2021ApJ...917...23K, 2024A&A...686A..42H}. Although these selection methods have produced valuable samples, the resulting catalogues remain survey-specific and heterogeneous. Moreover, most machine learning studies have focused on measurement values organised in tables, rather than exploiting the information content of SEDs rendered as images, compatible with convolutional neural networks (CNNs; {\citealt{CNN726791}}). CNNs are nowadays widely used in many fields of astrophysics, for example, for star--galaxy classification \citep{2017MNRAS.464.4463K}, for analysis of gravitational lenses \citep{2017Natur.548..555H}, for cosmological application \citep{2019A&C....27..130P}, or for transient detection \citep{2017ApJ...836...97C}. Unlike classifiers using simple numerical arrays that treat each photometric point as an independent feature, CNNs applied to SED images can capture the full spectral morphology, including the shape, slope, and wavelength position of infrared excess, as spatially coherent patterns, without requiring explicit feature engineering. The combination of \textit{Gaia} \citep{2016A&A...595A...1G}, all-sky infrared surveys, and time-domain facilities now makes it possible to identify YSOs across the entire Galaxy, but their heterogeneous data coverage demands new, data-driven approaches to classification.

The NEMESIS (Novel Evolutionary Model for the Early Stages of Star Formation) project aimed to build a truly multi-wavelength, machine-learning-driven framework for revisiting YSOs, their evolutionary classification, and timescales of early star-system development. Across several companion papers, NEMESIS has assembled panchromatic YSO catalogues in the Orion Star Formation Complex \citep[OSFC,][]{2025A&A...702A..63R}, applied specialised self-organising maps \citep{Kohonen1989, Kohonen2001} to image data for morphology-based classification methods \citep{2026A&A...707A..23H}, refined the mapping between observational class (e.g., spectral index) and physical stage \citep{2025A&A...696A.196G}, explored YSO variability \citep{2026A&A...705A.172M}, and provided invaluable information on the far-infrared properties of early stellar evolution \citep[][]{2024A&A...688A.203M, 2025A&A...696A..37M}.

As part of the NEMESIS project, in this study we aim at providing the most accurate all-sky YSO catalogue to date, and \texttt{DLYSO}, the deep learning tool that can be used for identifying YSOs. To do so, we gathered all photometric data from VizieR to construct SEDs, complemented by available time-series data from the Zwicky Transient Facility \citep[ZTF,][]{2019PASP..131a8002B}, and infrared direct imaging data. These data are used by independent classifiers trained on a large number of sources. The present work is one of the most important outcomes of NEMESIS: a homogeneous, all-sky re-classification of literature YSO candidates using image-based deep-learning methods, producing the NEMESIS General YSO (NGYSO) catalogue containing 274\,408 objects, helping to underpin the future studies of star- and planet formation, and early stellar evolution. The \texttt{DLYSO} software is released as open-source and is publicly available at GitHub\footnote{\url{https://github.com/martongabor/DLYSO}}. The repository contains the complete pipeline together with documentation and usage examples, allowing users to classify their own source catalogues using the same methodology presented in this work.

The paper is organised as follows. In Sect.~\ref{sect:data} we describe our dataset, the way the training and validation samples were created, how the photometric and imaging data were collected, and the methods we used to construct the input representations for our machine--learning models. Section~\ref{sect:methods} outlines the supervised classification methods, including both standard off-the-shelf and custom-built CNN architectures. In Sect.~\ref{sect:results} we present the performance metrics of the different methods, the resulting NGYSO catalogue, and we give details about the contamination. In Sect.~\ref{sect:discussion} we discuss the astrophysical properties of the classified sources, such as their distribution on the Hertzsprung--Russell diagram, their inferred evolutionary classes, and their spatial relation to the Local Bubble and the Radcliffe wave \citep{2020Natur.578..237A}. The summary and outlook are given in Sect.~\ref{sect:summary}.

This paper begins the NEMESIS general YSO catalogue series, which extends the earlier regional NEMESIS studies to an all-sky catalogue. Paper~II will present SED modelling of the NGYSO sources and discuss the results of the fits. Paper~III will address the unsupervised classification of NGYSO sources.

\section{Data}\label{sect:data}

The reliability of any machine--learning classification critically depends on the quality and representativeness of the training data. Our goal was to assemble a comprehensive, representative dataset that captures the diversity of both YSOs and their most common astrophysical contaminants. To achieve this, we combined two high-fidelity YSO catalogues --- the Konkoly Optical YSO catalogue \citep[KYSO,][]{2023A&A...674A..21M} and the NEMESIS OSFC catalogue --- with a broad compilation of non--YSO sources covering main--sequence stars, evolved objects, and extragalactic contaminants. For all training and inference targets, we gathered photometric measurements from the VizieR database to construct SEDs, as well as contextual information such as interstellar reddening, variability features, and direct imaging data.

\subsection{Source selection}\label{sect:training}

\subsubsection{YSO selection}

Our YSO training sample contains sources from two well-curated lists of young stars. The \href{https://vizier.u-strasbg.fr/viz-bin/VizieR?-source=J/A+A/674/A21&-to=3}{KYSO catalogue} was created to help identify young stars among the \textit{Gaia} DR3 variable stars (see Appendix~A of {\citealt{2023A&A...674A..21M}}). It contains nearly 12,000 objects carefully selected as young stars based on optical data, with their young nature mostly confirmed using spectroscopic data.

The second catalogue used to build our YSO training sample is the \href{https://vizier.u-strasbg.fr/viz-bin/VizieR?-source=J/A+A/702/A63&-to=2}{NEMESIS Catalogue of YSOs for the OSFC} \citep{2025A&A...702A..63R}. It reviewed the literature focused on the OSFC to compile a thorough catalogue of previously identified YSO candidates in the region, including the curation of observables relevant to probe their youth. The NEMESIS OSFC catalogue includes data collated for 27\,879 sources located in the region. It also reports flags and probabilities for sources indicating if they are probable extragalactic sources, or Galactic giants. Of these sources, 534 have one of the flags. The remaining objects cover the whole mass spectrum and the various stages of pre-main sequence evolution from protostars to diskless young stars. The catalogue contains additional ancillary information, such as infrared spectral slope ($\alpha_{\rm IR}$), disk properties, spectral types, accretion properties, and properties of several spectral lines indicating the young nature of the stars. The catalogue also uses flags which help users to identify probable contaminants, such as extragalactic objects, giants and main-sequence stars. After removing these sources from our sample, we restricted our training sample to the truly young objects by applying a cut on the $\alpha_{\rm IR}$ value. Only YSOs with $\alpha_{\rm IR}>-2.5$ were considered for further usage -- 15\,109 true YSOs. According to the classification boundaries accepted in the literature \citep[e.g.,][]{Grossschedl2019}, YSOs with $\alpha_{\rm IR}>-2.5$ are those that are still surrounded by a protoplanetary disk. The top panel of Fig.~\ref{fig:trainingyso} shows the sky distribution of the KYSO and NEMESIS sources in Galactic coordinates.

\subsubsection{Non-YSO selection}

The purpose of collecting non-YSO objects is to create a reliable training sample for the supervised classification approach, in order for the classifiers to learn to distinguish between YSOs and other object types. Therefore, we also collected a large number of objects from the literature that were classified as any type of object but not as YSO. These include extragalactic objects, variable stars, M dwarfs, main-sequence stars and finally a sample of evolved stars containing asymptotic giant branch (AGB) and post-AGB stars. For more details, see Table~\ref{tvt}.

During the compilation of the non-YSO training sample our goal was to cover different areas of the sky, but with more emphasis on regions where YSOs are most likely to be found, i.e., close to the Galactic mid-plane and low galactic latitude regions where the nearby star forming regions are located. The sky-distribution of the collected sources is shown in the middle panel of Fig.~\ref{fig:trainingyso}. Having a large number of non-YSOs in regions where YSOs are mostly present forced the methods to learn the difference between YSOs and non-YSOs, and to account for the reddening affecting background sources, as well as foreground objects seen in the direction of extended dust emission.

\subsubsection{Literature YSO catalogues}\label{literatureysos}

To construct a comprehensive and heterogeneous sample of YSO candidates, we compiled a large set of literature catalogues spanning a wide range of selection methods, wavelengths, and environments. These catalogues include objects identified through infrared excess, variability, spectroscopic signatures, and machine-learning-based classifications, and therefore represent a diverse mixture of YSO populations as well as varying levels of contamination. The catalogues include targeted areas, and all-sky surveys, as well. They also include all stages of YSO evolution from the embedded protostars to all members of young stellar clusters.

\begin{figure}[h!tb]
    \centering
    \includegraphics[width=\linewidth]{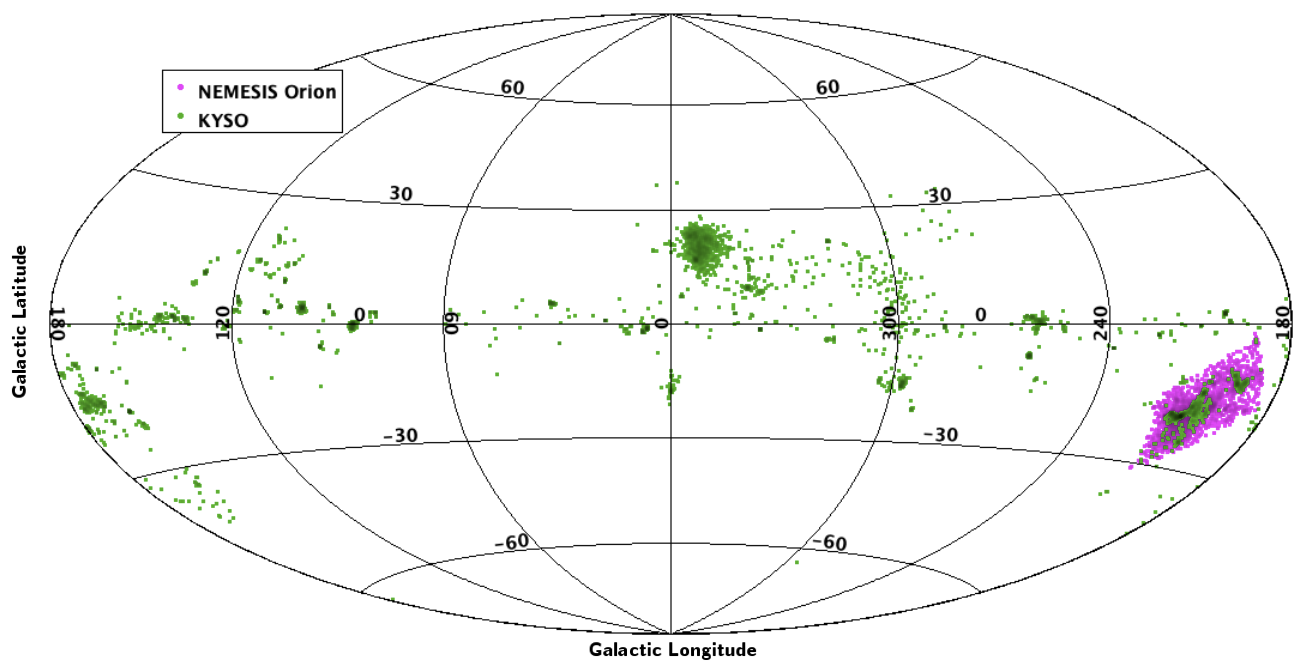} \includegraphics[width=\linewidth]{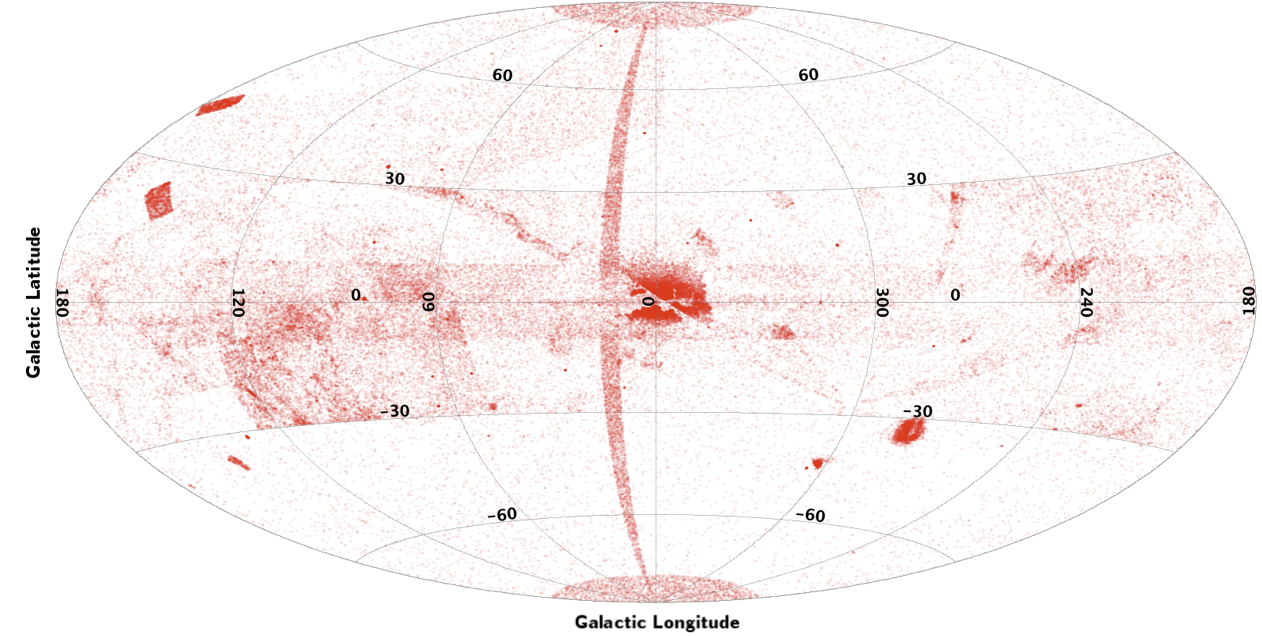} \includegraphics[width=\linewidth]{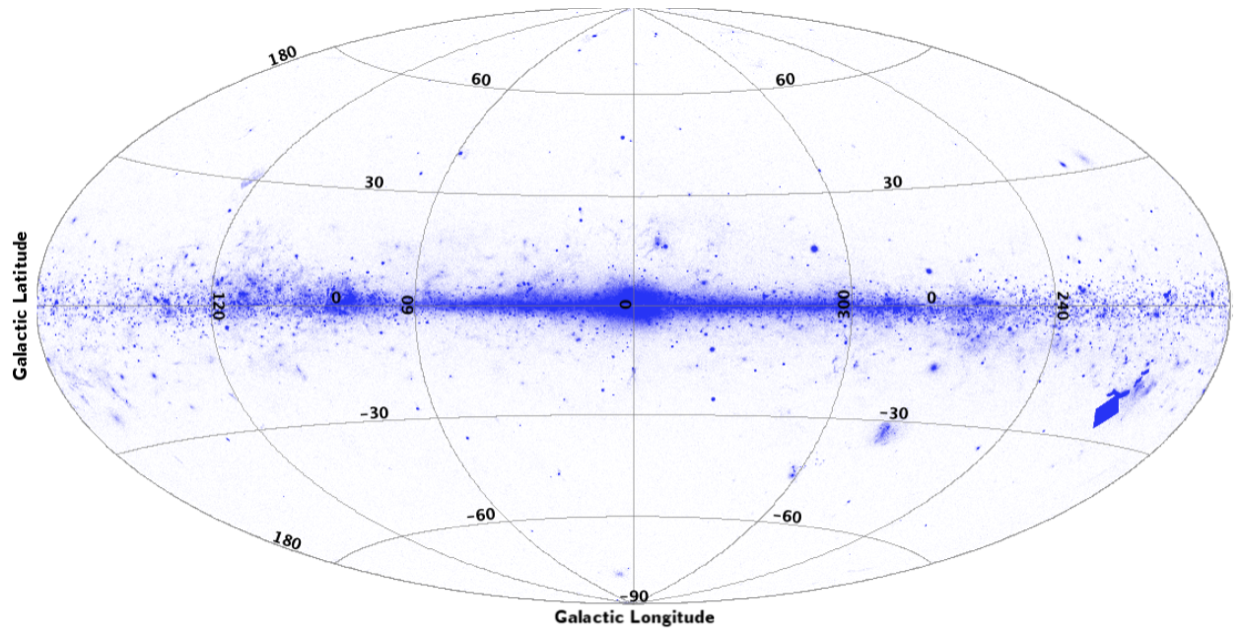}\\
    \caption{Top: The sky distribution of the YSO training sample in Aitoff projection. Magenta dots represent the sources from the NEMESIS Orion compilation, and green dots are the YSOs from the KYSO catalogue. Middle: Sources of the non-YSO training sample in Aitoff projection, in Galactic coordinates. The train--validation--test split and catalogue references are summarised in Table~\ref{tvt}. Bottom: 3.9 million YSOs, YSO candidates, and cluster members collected from the literature.}
    \label{fig:trainingyso}
\end{figure}

In total, the combined sample contains $\sim3.9\times10^6$ sources drawn from nearly 90 independent studies (Table~\ref{tab:sed_yso}). The final NGYSO catalogue (see Sect.~\ref{sect:results}) was created from data gathered for all these literature YSOs and YSO candidates. The sky distribution of the sources is shown on the bottom panel of Fig.~\ref{fig:trainingyso}.

\subsection{Data types}\label{sect:datacollection}

For each source in the dataset, we constructed and used several different image-based representations with the aim of training independent CNNs to discriminate between YSOs and non-YSOs: SED plots, ZTF light curve variability maps, and direct imaging cutouts. Below we describe how the data were collected and how each data representation was constructed and what information it encodes. An example of each type is shown in Fig.~\ref{sed}.

\subsubsection{SED data}

To collect the most complete data possible for our SEDs, we relied on the VizieR database. The \href{https://vizier.cds.unistra.fr/vizier/sed/sedapi.gml}{VizieR SED API} provides a Python snippet that can query the whole database in a region defined around input RA and Dec coordinates. For our case, we used a search radius of 2$^{\prime\prime}$. This value is sufficiently large to account for astrometric uncertainties, proper motion offsets, and resolution differences between the contributing surveys, while remaining small enough to minimise spurious associations in crowded star-forming regions. The query returns a table with the name of the filters, the central frequency of the passbands, the measured flux density, and their uncertainty where available. As a first step, we identified and excluded catalogues that appeared to contain data which cannot be processed directly, likely because of different conversion, or do not fulfil the quality requirements needed for our approach. These catalogues are related to the following publications: \citet{2007MNRAS.375.1220M}, \citet{2007MNRAS.379.1599L}, \citet{2008MNRAS.391..136L}, \citet{2008AJ....136..735L}, \citet{2011AJ....142...15G}, \citet{2013ApJS..207...10R}, \citet{2013Msngr.154...35M}, \citet{2017ApJ...836...34M}, \citet{2017ApJ...848...97R}, \citet{2017AJ....153..166Z}, \citet{2019AJ....158..138S}, \citet{2020ApJS..249....3A}, \citet{2021ApJS..255...20A}, \citet{2021AJ....161..234M}, \citet{2021yCat.2368....0S}, \citet{2021arXiv210804778P}.

In the second step, if multiple measurements with the same filter were found, we kept the one having the smallest angular separation to our input coordinates. While this may in some cases favour a closer but lower-quality measurement, it ensures a consistent and reproducible selection criterion across the full dataset. However, the adopted matching radius of 2$^{\prime\prime}$ mitigates this effect in most cases, and the large-scale statistical nature of our analysis reduces the impact of individual mismatches.

Finally, we also removed photometric entries from outdated catalogue versions where updated measurements are available, such as \textit{Gaia} DR2, which was replaced by \textit{Gaia} DR3. When unWISE data were available, they were used instead of the older AllWISE $W1$ and $W2$ data. We were also aware of the inaccuracies of the AllWISE flux density values in the $W3$ and $W4$ bands \citep{2014ApJ...791..131K, 2025A&A...702A..63R}, therefore, when possible, we replaced those values with the \textit{Spitzer} IRAC 8.0 and MIPS 24.0 band measurements. Additionally, WISE $W3$ and $W4$ measurements with missing photometric uncertainties were discarded. The SED data were then used to build different representations. We note that the constructed SEDs are limited by the availability and sensitivity of the contributing surveys across different wavelength regimes. The mid- and far-infrared domains suffer from incompleteness due to limited sensitivity, spatial resolution, and heterogeneous sky coverage. On the other hand, in the optical and near-infrared, deeply embedded young sources may remain undetected due to strong extinction, resulting in SEDs that are only partially sampled at shorter wavelengths. The bands most frequently represented were the 2MASS, the Pan-STARRS, the \textit{Gaia}, the \textit{Spitzer}/IRAC and the WISE bands. Additional optical, near-infrared, mid-infrared, far-infrared, and submillimetre measurements were included whenever available. The median number of represented filters was 16 in the collected SED data.

As a first representation, each source was plotted on a fixed logarithmic grid, with wavelength on the x-axis and $\nu F_\nu$ on the y-axis, covering 0.01--2000~$\mathrm{\mu m}$ and $10^{-15}$--$10^{-6}$~erg~cm$^{-2}$~s$^{-1}$ ranges, respectively. The resulting images were used as single-channel greyscale image data. Using fixed axis ranges ensures a consistent representation across all sources: the flux scale is comparable between images, and the overall shape and extent of the SED are always rendered in the same dynamic range, making the training more stable and the learnt features more transferable. The photometric measurements were binned into 300 log-spaced wavelength bins, and only sources with at least 10 populated bins were kept. If multiple measurements fell in the same bin, their mean was taken. Consecutive bins were then connected by lines rather than plotted as individual symbols. Although this interpolation is not physically motivated, preliminary experiments showed that it results in better classification performance, as the continuous spectral shape provides more information for the CNN to extract than isolated photometric points. An example is shown in the leftmost panel of Fig.~\ref{sed}.

In the next representation we used the same SED plotting method with the addition of interstellar extinction information encoded in the background colour of the plot. The purpose of adding the extinction information as the image background colour was to supply the CNN with additional information about the Galactic environment in which the source is located. While two sources may exhibit very similar observed SEDs, their astrophysical interpretation can differ depending on whether they are projected onto a region of high interstellar extinction, where star formation is expected, or onto a relatively unobscured field, where contaminants such as reddened giants or background galaxies are more likely. The line-of-sight reddening was estimated using the \texttt{dustmaps} Python package \citep{2018JOSS....3..695M} with the corrected Schlegel-Finkbeiner-Davis maps \citep{2023ApJ...958..118C}. With the choice of a 2D rather than a 3D extinction map, our primary motivation was to maintain a homogeneous input representation for all sources. While \textit{Gaia} parallaxes are available for a large fraction of the sample, they are absent or highly uncertain for many faint and embedded YSOs. Employing a 3D extinction map only for a subset of the sources would have introduced heterogeneous inputs into the training set. The map returns $E(B-V)$ values, which were then mapped to a red colour scale using a logarithmic normalisation between values of 0.01 and 400, where light red represents low reddening and dark red represents high reddening. While the value of 400 is unrealistic, we wanted to cover areas in the $E(B-V)$ map where the calculated values also seem to be unphysical. This background colouring approach avoids modifying the standard CNN architectures to accept an additional scalar input, while still providing the reddening information as a visually encoded channel. An example is shown in the second panel of Fig.~\ref{sed}.

\subsubsection{Light curves}

We downloaded ZTF light curves for each target using their RA and Dec coordinates from IPAC’s ZTF database with the help of the \texttt{IRSA ZTF-LC-API}\footnote{\url{https://irsa.ipac.caltech.edu/docs/program_interface/ztf_lightcurve_api.html}}. For each source, the data were filtered using ZTF photometric flags, keeping only those measurements with \texttt{catflag=0}, which selects the safest measurements, free from potential issues like blending, bad pixels, or any artefacts. We then split by the three ZTF filters ($g$, $r$, and $i$) before constructing the delta time--delta magnitude (DTDM) variability representations. To ensure that the light curves are not heavily contaminated, we used a 2$^{\prime\prime}$ matching radius with the \href{https://irsa.ipac.caltech.edu/data/ZTF/docs/releases/dr24/ztf_release_notes_dr24.pdf}{ZTF DR24}. In the next step, we downloaded all the available observations, and selected only those ZTF object ids (per band) which had the most unflagged measurements. Data from a band were used only if at least 5 data points were left after filtering.

DTDM images were created following the method described in \citet{2017SSCI....1M}. Briefly, the method calculates pairwise differences in both observation time (DT) and magnitude (DM) for all pairs of measurements within a light curve. These differences were then organised into two-dimensional histograms (magnitude difference versus time difference) separately for each band, which are $g$, $r$ and $i$ filters for ZTF \citep[$\lambda_{\rm eff}$=4746.48\AA, 6366.38\AA \, and 7829.03\AA, respectively,][]{2019PASP..131a8002B}. Finally, the three histograms were combined into a single RGB image, providing a compact representation of the variability properties of each source across the three filters. We used 70 bins along the DT axis covering time differences between 1 second and 3000 days, and 86 bins along the DM axis covering magnitude differences between $-10$ and $10$. On each axis an arbitrary scale was used to cover differences that are meaningful for the human brain, like seconds, minutes, hours, and days on the DT axis and to equally cover very small and large variations on the DM axis. Because ZTF is ground-based, it does not provide data for sources below a declination of $-30^\circ$ due to its location, therefore, we could not collect ZTF time series data for all of our sources. Another limiting factor was the wavelength domain of ZTF. Since it is an optical survey, cold and deeply embedded objects such as Class~0/I protostars, whose emission peaks in the IR regime are frequently missed, meaning that the very early stages of stellar evolution cannot be observed with ZTF. An example DTDM image is shown in the third panel of Fig.~\ref{sed}.

\subsubsection{Direct imaging data}

Direct imaging data were also downloaded for all sources possible. We collected the stamp images of the sources through the Hierarchical Progressive Surveys (HiPS) protocol using the \texttt{hips2fits} interface in \texttt{Astroquery}\footnote{\url{https://astroquery.readthedocs.io/en/latest/hips2fits/hips2fits.html}}. HiPS is a data format developed by CDS that allows seamless browsing of astronomical surveys at different resolutions, and it converts images stored in the HiPS tile structure into standard FITS files or RGB png images. While preprocessed and compressed images are not suitable for conserving the flux density of the sources, the goal with the images was to gather information on the extension and the environment of the sources. YSOs tend to be embedded in environments with background structures, while this is less prominent for extragalactic sources or main-sequence stars.

While data for many surveys are available through the \texttt{hips2fits} service, we queried the AllWISE RGB image data \citep{2013wise.rept....1C} for cutouts centred around our sources, because WISE offers all-sky coverage, and it includes near- and mid-infrared observations, crucial for identifying YSOs. In preliminary tests ZTF images were also used, and they provided good results, but the lack of all-sky data led us to the decision to rely on AllWISE. While the AllWISE survey carried out observations in 4 photometric bands ($W1$, $W2$, $W3$ and $W4$, centred around $\lambda$= 3.4, 4.6, 11.6, and 22.1 $\mu$m, respectively), the RGB stamp images contain only the $W1$, $W2$ and $W4$ band data.

\begin{figure*}[tbp]
   \sidecaption
   \includegraphics[width=12cm]{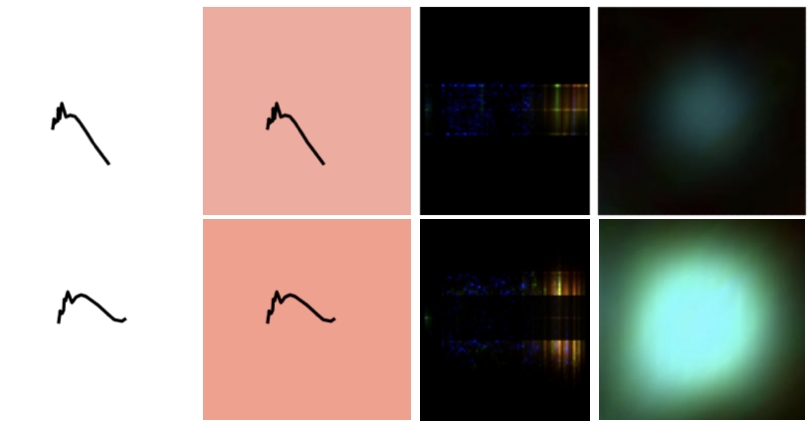}
      \caption{The different image-based representations used for classification. The upper row shows data for the same source located at RA=98.804, Dec=12.166 degrees, and the lower row shows data for the source at RA=0.55529, Dec=64.90703 degrees. From left to right: the simple SED plot, the SED plot with background colour encoding the interstellar extinction seen in the direction of the source, the DTDM variability image, where the RGB layers represent the ZTF $i$, $r$ and $g$ band light curve data, and the AllWISE cutout stamp image of the source, respectively.
              }
         \label{sed}
\end{figure*}

If a source is too faint, and no emission in the background is detected, the images are black, while sources that saturate the detectors appear fully white. To avoid using these images, which contain no information, we included cutouts only if in all 3 image layers the median value was above zero, and below 250 (in this image encoding the maximum value is 255). The cutout size was chosen to be 30$^{\prime\prime} \times$30$^{\prime\prime}$ with 200$\times$200 pixel resolution. While brighter sources appear as larger objects on the images, we did not change the cutout size, because we wanted the CNNs to capture context at a fixed angular scale. Example stamp images are shown in the fourth column of Fig.~\ref{sed}.

\section{Analysis}\label{sect:methods}

\subsection{Standard CNN architectures}
\label{sect:2dcnns}

We applied several established CNN architectures to classify the image-based representations described in Sect.~\ref{sect:datacollection}. CNNs are well suited for this task because they can learn characteristic patterns directly from pixel data, without requiring explicit feature extraction or parametric modelling. In the case of SED images, these patterns correspond to the overall spectral shape, including slopes, curvature, and infrared excess, as well as structural differences between YSOs and contaminants in the DTDM and AllWISE images.

To assess the robustness of the classification with respect to network architecture and model complexity, we trained a set of CNNs spanning a wide range of parameter counts and design principles, available in the {\tt PyTorch} Python framework. The tested architectures include computationally efficient, low-parameter networks such as \texttt{SqueezeNet~1.1} \citep{iandola2016squeezenet} (1.2M parameters), \texttt{ShuffleNet V2} $\times$ 0.5 \citep{ma2018shufflenet} (1.4M), \texttt{MNASNet 0.5} \citep{tan2019mnasnet} (2.3M), \texttt{MobileNet V3 Small} \citep{howard2019searching} (2.5M), as well as medium-complexity models such as \texttt{RegNet-Y 400MF} \citep{radosavovic2020regnet} (4.3M), \texttt{EfficientNet-B0} \citep{tan2019efficientnet} (5.3M), and deeper residual architectures including \texttt{ResNet-18} and \texttt{ResNet-50} \citep{he2016resnet} (11.7M and 25.6M, respectively), \texttt{EfficientNet-V2-S} \citep{tan2021efficientnetv2} (21.5M), and \texttt{ResNeXt-50} 32$\times$4d \citep{xie2017resnext} (25M).

All these standard CNNs were trained from scratch with all parameters optimised, and for each architecture, the original classification layer was replaced by a binary output layer. Input images were resized to 224$\times$224 pixels and converted to RGB format. The resizing ensures a uniform input representation across the different data modalities used in this work (AllWISE images, DTDM maps, and SED-based images), which have intrinsically different native resolutions and aspect ratios. The chosen size is also compatible with standard CNN architectures, which employ successive downsampling by factors of two, and benefit from input dimensions that preserve integer feature map sizes. Training was performed using the cross-entropy loss function with label smoothing ($\epsilon = 0.05$) and the \texttt{AdamW} optimiser \citep{loshchilov2019adamw}. A One-Cycle learning rate schedule was applied over 30 epochs. Model selection was based on the macro-averaged $F_1$ score measured on the validation set.

\subsection{Custom-built CNN architectures}
\label{sect:customcnns}

In addition to the off-the-shelf architectures described in Sect.~\ref{sect:2dcnns}, we developed two custom CNN models, referred to as \texttt{SmallResNet} and \texttt{ResConvAttnClassifier} (\texttt{RCA}). Both were designed to remain lightweight, with approximately 1.3 and 1.67 million trainable parameters, respectively, while incorporating architectural choices designed for the relatively small image sizes. They can handle both single-channel greyscale inputs (SED and SED$_r$) and three-channel RGB inputs (AllWISE and DTDM) within a unified framework.

\subsubsection{\texttt{SmallResNet}}

\texttt{SmallResNet} is a compact, configurable residual network. It consists of a stem with a 7$\times$7 convolution at stride 2, followed by a sequence of four residual stages each containing two residual blocks \citep{he2016resnet}. Each residual block follows a standard two-layer design, a $3\times3$ convolution followed by batch normalisation \citep{ioffe2015batchnorm} and \texttt{ReLU} activation \citep{glorot2011relu}, repeated twice, with the input added back via a skip connection. The number of channels at each stage $i$ is set to $\min(\text{base\_width} \times 2^i, \text{max\_width})$, with $\text{base\_width}=32$ and $\text{max\_width}=128$, resulting in channel widths of 32, 64, 128, and 128 across the four stages. Downsampling between stages is performed by the first block of each stage with stride 2. The final feature map is reduced to a fixed-size vector via global average pooling \citep{lin2014nin}, followed by a linear classification layer. All convolutional weights are initialised using He initialisation \citep{he2015init}.

\subsubsection{\texttt{ResConvAttnClassifier}}

The \texttt{RCA} model is a custom classification architecture combining residual convolutional blocks with multi-head self-attention modules at multiple spatial scales \citep{vaswani2017attention}. Self-attention allows the network to capture relationships between different spatial regions of the image, complementing the local patterns learnt by the convolutional blocks. It consists of a stride-2 stem convolution, followed by five stages with channel multipliers $(1, 2, 4, 8, 8)$ applied to a base channel width of 16, resulting in channel widths of 16, 32, 64, 128, and 128. Each stage contains two residual blocks using group normalisation \citep{wu2018groupnorm} and \texttt{SiLU} activations \citep{hendrycks2016silu, ramachandran2017swish}, with \texttt{LayerScale} \citep{touvron2021layerscale} applied to each residual branch to improve training stability. Downsampling between stages is performed by strided convolutions. Multi-head self-attention \citep{vaswani2017attention} with two heads is applied at selected spatial downsampling factors, where the spatial extent is small enough for full attention to be tractable. The attention blocks include a learnable per-channel gating parameter initialised to a small value to ensure stable early training. Following the feature extractor, global average pooling produces a fixed-size representation, which is passed to a multilayer perceptron classification head consisting of a \texttt{LayerNorm} layer, a linear projection with \texttt{GELU} activation \citep{hendrycks2016silu}, dropout, and a final linear output layer.

\subsubsection{Input configuration and training details}

Both architectures were configured independently for each data representation, with differences in input size, number of channels, and attention placement. The SED and SED$_r$ images were treated as single-channel greyscale inputs at $190 \times 190$ pixels, while the AllWISE and DTDM images were three-channel RGB inputs. AllWISE images were resized to $128 \times 128$ pixels, while DTDM images to $128 \times 128$ pixels. For the \texttt{RCA} model, self-attention was applied at spatial downsampling factors of 8, 16, and 32 for all data products, except for the DTDM configuration, where an additional attention level at downsampling factor 4 was included and a smaller Multilayer Perceptron head (hidden size 128 instead of 256) was used, reflecting the different spatial and temporal structure of the variability maps.

All pixel values were normalised to the range $[0,~1]$ before being passed to the networks. Training was performed using binary cross-entropy loss and the \texttt{AdamW} optimiser with a learning rate of $10^{-3}$, weight decay of $5\times10^{-4}$, and $(\beta_1, \beta_2) = (0.9, 0.999)$. A \texttt{ReduceLROnPlateau} learning rate schedule monitored the validation loss, reducing the learning rate by a factor of 0.5 after 2 epochs without improvement, down to a minimum of $10^{-6}$. Training ran for a maximum of 50 epochs, with early stopping triggered after 5 epochs without improvement in validation $F_1$ score.

\subsubsection{Train/validation/test split}

The success of supervised machine learning methods heavily relies on the training, validation and test samples. Our final training/test/validation dataset, comprising only sources for which a usable SED could be constructed (see Sect.~\ref{sect:datacollection}), and for which AllWISE stamp images were available according to the criteria described in Sect.~\ref{sect:datacollection}, contains a YSO/non-YSO ratio of approximately 1:6, drawn from a broad mixture of contaminants, as detailed in Table~\ref{tvt}. The train/validation/test splits were then created at the source level, such that each object appears in exactly one subset, preventing any information leakage between training and evaluation. This imbalance is qualitatively consistent with the fact that genuine YSOs are rare compared to other Galactic and extragalactic sources seen at wavelengths similar to those where YSOs mostly appear. We deliberately preserve it during training so that the network is exposed to a wide variety of non-YSOs and learns to minimise false positives. Final performance metrics (see Sect.~\ref{evaluation}) are computed on the holdout test set that has class ratios similar to the training set.

\subsection{Evaluation methods}\label{evaluation}

We evaluated each classifier using a standard set of binary classification metrics computed on the test set, designed to assess both the completeness (recovery of true YSOs) and purity (rejection of contaminants) of the predicted samples in a consistent way. For all models, we report the true positive rate (TP), true negative rate (TN), false positive rate (FP), and false negative rate (FN), where TP and FN are fractions of labelled YSOs correctly and incorrectly classified, respectively, TN is the fraction of labelled non-YSOs classified also as non-YSOs, and FP is the fraction of non-YSOs misclassified as YSOs. We also report the $F_1$ score, defined as the harmonic mean of precision and recall:

\begin{equation}
    F_1 = \frac{2\,\mathrm{TP}}{2\,\mathrm{TP} + \mathrm{FP} + \mathrm{FN}},
\end{equation}

\noindent where $\mathrm{Precision} = \mathrm{TP}/(\mathrm{TP}+\mathrm{FP})$ and $\mathrm{Recall} = \mathrm{TP}/(\mathrm{TP}+\mathrm{FN})$.

The $F_1$ score is the primary metric used in this work, as it rewards models that maintain both high completeness and low contamination, and is robust to class imbalance. In the context of YSO identification, where genuine objects are relatively rare, false positives are particularly problematic: contaminating sources (e.g. galaxies or evolved stars) can bias population-level analyses, and distort inferred evolutionary class fractions. For these reasons, $F_1$ is more informative than simple accuracy. For example, a classifier predicting non-YSO for each source in a $90\%-10\%$ non-YSO/YSO split achieves 90\% accuracy but an $F_1$ of zero.

However, the $F_1$ score depends on the classification threshold. We optimised it by performing a threshold sweep on the validation set. For each threshold, precision and recall were computed and combined into $F_1$, and the threshold maximising $F_1$ was adopted for each model.

\section{Results}\label{sect:results}

\subsection{Classification performance}\label{sec:methodperf}

We evaluated the performance of all CNN architectures introduced in Sect.~\ref{sect:methods} across the four image representations described in Sect.~\ref{sect:datacollection}. Table~\ref{sedclassificationcnnresults} summarises the TP, TN, FP, FN, thresholds where the maximum $F_1$ scores were obtained (TH), and $F_1$ scores for all architectures and data products. We note that thresholds deviating substantially from 0.5 indicate miscalibrated posterior probabilities, where the model's raw output does not reliably reflect the true class probability. Such miscalibration is common when training on imbalanced datasets and does not necessarily affect classification quality, but it does mean that the default threshold of 0.5 would yield suboptimal results. Overall, all classifiers achieved consistently high performance, with typical $F_1$ scores above 90\%, except for the AllWISE stamp images where scores fell slightly below this level.

To have an overall picture about the performances, we calculated average rates and errors, which are the standard deviations of the achieved values. Tests on the standard SED images yielded an average TP percentage of $95.37\pm0.53$\% and a FP percentage of $0.54\pm0.080$\%, with an average $F_1$ score of $96.06\pm0.39$. The best-performing standard model was \texttt{EfficientNet-V2-S} ($F_1 = 96.80$, TH$=0.508$, 21.5M parameters), indicating relatively well-calibrated probabilities. Our custom \texttt{SmallResNet} reached $F_1 = 96.44$ (TH$=0.470$) and \texttt{RCA} $F_1 = 96.13$ (TH$=0.650$), both within 0.7\% of the best standard model while using approximately 16 and 13 times fewer parameters, respectively. Including line-of-sight reddening as a background colour layer (SED$_r$) shifted classifiers towards higher purity at the expense of a modest loss in completeness, reflecting its ability to better distinguish true YSOs from reddened non-YSOs. The average TP percentage slightly decreased to $91.5\pm1.1$\%, while the FP percentage improved to $0.22\pm0.076$\%, with an average $F_1$-score of $94.92\pm0.57$. This data product shows the most extreme threshold behaviour of all four representations. \texttt{MNASNet~0.5} required a threshold of 0.938 and \texttt{MobileNet-V3~Small} 0.857. By contrast, our custom models show more moderate thresholds (\texttt{SmallResNet}: 0.65, \texttt{RCA}: 0.71), suggesting better calibrated outputs. The best-performing model was \texttt{ResNet-18} ($F_1 = 96.25$, TH$=0.653$, 11.7M parameters), while our \texttt{SmallResNet} achieved $F_1 = 95.64$ at one ninth of the parameter count.

The RGB stamp images from the AllWISE survey resulted in somewhat lower performance overall, as expected for direct imaging data affected by background structure and crowding, with limited information on their physical nature. Still, among the models, the average TP rate was $87.3\pm1.3$\%, the averaged FP percentage was $1.68\pm0.22$\%, and the average $F_1$-score was $88.51\pm0.85$. Thresholds for this product are generally close to 0.5, indicating better-calibrated models compared to SED$_r$. Our \texttt{RCA} model was the best-performing architecture overall, achieving $F_1 = 90.35$ (TH$=0.580$), outperforming all standard models. We attribute \texttt{RCA}'s advantage on AllWISE images to its multi-scale attention mechanism, which captures local morphological features of sources embedded in complex backgrounds.

Finally, the ZTF DTDM representations, which encode variability rather than SED morphology, achieved comparably high performance, demonstrating that time-domain information alone can serve as an effective discriminant of youth when sufficient photometric sampling is available. The average $F_1$ across the architectures was $92.93\pm0.60$, with an average TP of $92.28\pm0.87$\% and a FP of $2.37\pm0.43$\%. Thresholds were generally moderate. The \texttt{RCA} model achieved the best overall performance ($F_1 = 93.84$, TH$=0.500$), with the highest TP percentage of 93.95\%, surpassing the best standard model, \texttt{ResNeXt-50} ($F_1 = 93.73$, 25M parameters), while using approximately 15 times fewer parameters. The perfectly calibrated threshold of 0.500 for \texttt{RCA} on DTDM suggests that the model's probability outputs are well aligned with the true class posterior for this data product.

All image-based classifiers achieved $F_1$ scores close to, or above 90\%, with contamination levels below 2.5\%. However, instead of selecting a single best classifier, we combined all 12 models trained on each data product into an ensemble voting system in which each model casts a binary vote, and a source is accepted as a YSO candidate if at least half of the models vote positively. Ensemble voting outperforms any individual classifier because different architectures make independent errors, shaped by their different inductive biases, parameter counts, and regularisation strategies. Requiring majority agreement suppresses false positives that no single model can consistently avoid, while the diversity of thresholds across models means the ensemble is less sensitive to any individual calibration issue. The results of the voting system are listed in Table~\ref{voteresults}, calculated on the test dataset. In all cases, ensemble voting improved both the $F_1$ score and the FP percentage relative to the best individual classifier. The best $F_1$-score was achieved based on the SED images (without the encoded interstellar reddening), therefore this product was used for the final classification.

\begin{table}[htb]
\centering
\caption{Classification performance of the ensemble voting system for each data product.}\label{voteresults}
\begin{tabular}{lccccc}
\hline\hline
  \multicolumn{1}{c}{Data product} &
  \multicolumn{1}{c}{TP} &
  \multicolumn{1}{c}{TN} &
  \multicolumn{1}{c}{FP} &
  \multicolumn{1}{c}{FN} &
  \multicolumn{1}{c}{$F_1$} \\

\hline
    SEDplot & 96.36	& 99.57	& 0.43 & 3.63 & 96.90  \\
    SED$_r$plot & 93.43 & 99.91 & 0.09 & 6.57 & 96.33 \\
    ZTF DTDM & 94.59 & 98.28 & 1.72 & 5.40 & 94.98 \\
    AllWISE & 89.20 & 98.75 & 1.24 & 10.79 & 90.74  \\
\hline
\end{tabular}
       \tablefoot{Reported values are true-positive (TP), true-negative (TN), false-positive (FP), and false-negative (FN) rates, in percent, together with the $F_1$ score. TP and FN are computed with respect to the true YSO sample, while TN and FP are computed with respect to the true non-YSO sample. In the ensemble voting scheme, a source is classified as a YSO candidate if at least 6 of the 12 models vote for the YSO class.}
\end{table}

\subsection{The NEMESIS General YSO catalogue}
To create the largest and most reliable YSO catalogue to date, the NEMESIS General YSO (NGYSO) catalogue \footnote{The NGYSO catalogue and its supporting tables are only available in electronic form at the CDS via anonymous ftp to cdsarc.u-strasbg.fr (130.79.128.5) or via http://cdsweb.u-strasbg.fr/cgi-bin/qcat?J/A+A/}, we applied our method to a compilation of literature catalogues containing confirmed YSOs, YSO candidates, and other samples expected to contain significant young stellar populations, described in Sect.~\ref{literatureysos} and shown in Table~\ref{tab:sed_yso}. We want to note that our goal was not to revisit and evaluate literature YSO catalogues, but to identify the most reliable YSO candidates in well-known star forming regions, and in such large-scale catalogues, that potentially contain a large number of previously unidentified young stars.

For each source we attempted a uniform classification based on the SED plots, retaining only sources with SEDs containing at least ten photometric points after cleaning (see Sect.~\ref{sect:datacollection}). The column ``Confirmed [\%]'' in Table~\ref{tab:sed_yso} gives the reconfirmation fraction relative to the subset of objects for which a usable SED could be constructed, rather than relative to the total number of objects in the original literature catalogue. Since some literature sources lack sufficient photometric measurements to build an SED, they cannot be evaluated by our classifier and are therefore excluded from this percentage. The wide range of values---from $\sim$5--10\% for early all-sky colour selections up to $\gtrsim$90\% for recent spectroscopically curated samples---quantifies, with a single metric, the heterogeneous reliability of legacy catalogues. High reconfirmation fractions are found among programmes with spectroscopy, vetted membership, and/or multi-band infrared coverage \citep[e.g.,][]{2014A&A...561A...2A,2011ApJS..193...25R}, while lower fractions are typical of shallow colour cuts or regions with heavy crowding and diffuse emission where mid-IR fluxes are uncertain (e.g.\ some WISE-only selections). After merging all sources classified as YSOs by at least six of the twelve SED classifiers (see Sect.~\ref{sec:methodperf}) across the input catalogues, and removing duplicate entries corresponding to the same sky position appearing in multiple source lists, we obtain 274\,408 unique candidates, forming the NEMESIS General YSO catalogue (NGYSO). To find and merge duplicate entries we were using a 2$^{\prime\prime}$ radius, and we simply kept the coordinates of the first appearance of the object while sweeping through our list of catalogues. To identify sources which can be unresolved multiple sources we matched the final list of coordinates to the \textit{Gaia} DR3 catalogue using the same 2$^{\prime\prime}$ radius and flagged the sources with multiple matches.

Table~\ref{voteresults} shows that the ensemble voting system achieves the highest $F_1$ score for the all-sky SED data while keeping the false-positive percentage $\lesssim 0.5\%$. The posterior probabilities of individual classifiers are provided in the released catalogue, enabling users to apply custom selection criteria for their science case.

A qualitative trend emerges when the reconfirmation fractions in Table~\ref{tab:sed_yso} are considered in the context of the selection methods of the input catalogues. Rather than providing a direct measure of the reliability of the literature samples, the confirmation fraction primarily reflects the degree to which their selection functions overlap with that of our SED-based classifier.

Catalogues constructed explicitly from infrared excess generally show the highest (re)confirmation fractions. Region-focused \textit{Spitzer} studies, which identify YSOs through their characteristic near- and mid-infrared colours and SED shapes, are well recovered. Examples include \citet{2008ApJ...674..336G} and \citet{2009ApJS..184...18G}, for which 99.1\% and 96.1\%, respectively, of the sources with usable SEDs are recovered by our classifier. Similarly high fractions are found for several other infrared-selected samples. This strong agreement is expected, because the presence and wavelength dependence of infrared excess constitute prominent features in the SED images used by our CNNs.

In contrast, catalogues designed to identify the broader young stellar population using astrometry, colour--magnitude position, spatial clustering, or optical and near-infrared photometry generally exhibit lower reconfirmation fractions. Examples include the young-cluster members of \citet{2024A&A...686A..42H} (4.8\%), \citet{2021ApJ...917...23K} (14.3\%), \citet{2022A&A...664A.175P} (20.3\%), \citet{2018A&A...620A.172Z} (21.1\%), and \citet{2021AJ....162..282M} (25.9\%). These catalogues can contain substantial populations of diskless or weak-excess pre-main-sequence stars whose SEDs approach that of normal stellar photospheres. Their lower recovery fractions therefore do not imply that these catalogues are strongly contaminated, but rather demonstrate the different selection functions of the methods.

Overall, these systematic trends reflect the methodology of the input catalogues rather than any regional bias in our model.  The NGYSO framework therefore provides a uniform re-assessment of classical and early machine-learning YSO searches under a common, reproducible standard. It is important to emphasise here that the catalogues used in building the NGYSO catalogue are a mix of region-based surveys, focusing on the completeness of the given sky area, but we also used large all-sky catalogues, as well. This means that the NGYSO is not the most complete, but combines depth in well-studied areas with a wider sampling of the Galactic YSO population.

Although the ZTF DTDM ensemble achieved a competitive performance ($F_1$=94.98, compared to 96.90 for the SED ensemble), we did not use it as the primary basis of the final NGYSO catalogue because its applicability is more limited. ZTF light curves are unavailable below $\delta\sim -30^\circ$, and, being optical, they are frequently missing or incomplete for the most deeply embedded Class 0/I sources. Thus, DTDM variability maps probe a complementary aspect of YSO physics and provide an effective youth diagnostic where sufficient sampling exists, but they cannot deliver the same homogeneous all-sky coverage as the SED-based representation. For this reason, we adopted the SED ensemble as the backbone of the catalogue and provide the posterior probabilities of the individual classifiers in the released table for users who wish to impose modality-specific selections.

\begin{figure*}[h!tb]
   \centering

   \includegraphics[width=0.95\hsize]{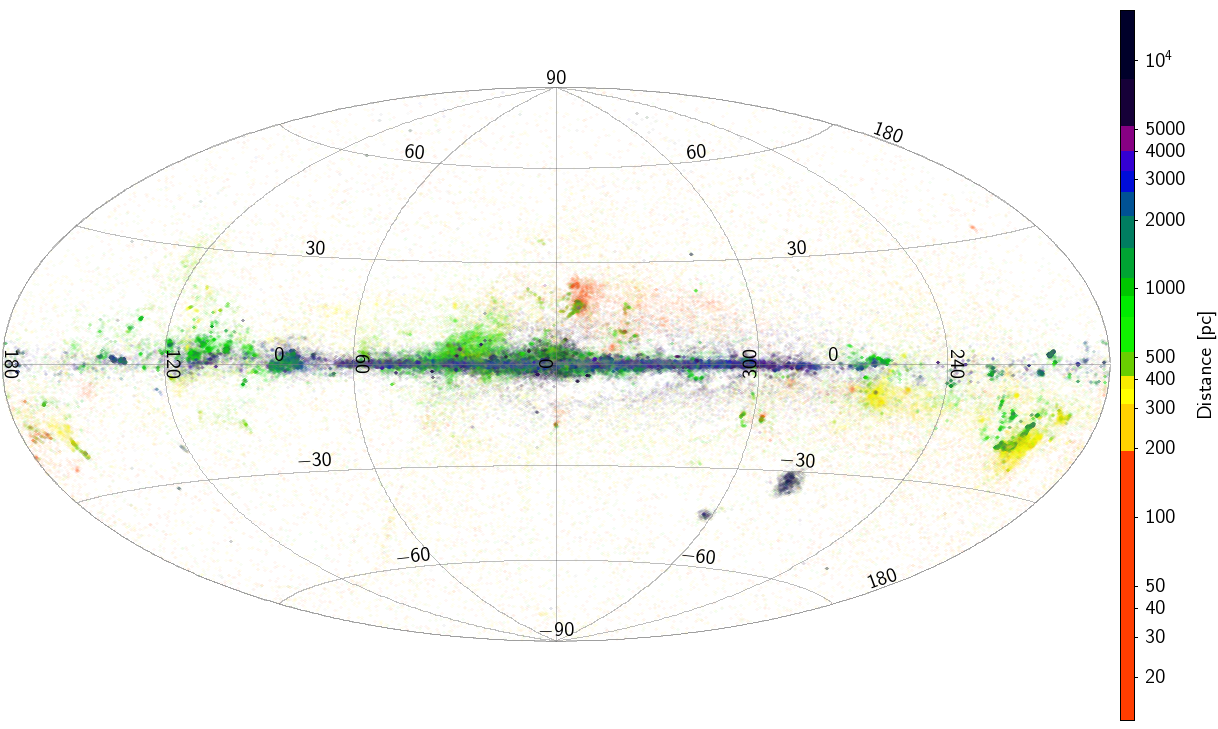}
      \caption{All-sky distribution of reliable YSO candidates with distance estimates from \citet{2021AJ....161..147B}, shown in Galactic coordinates using an Aitoff projection. Distance values were available for 223\,587 out of the 274\,408 sources of our final NGYSO catalogue. Colour scale is shown on the right.}
         \label{allysoaitoff}
\end{figure*}

Figure~\ref{allysoaitoff} shows the distribution of NGYSO in Galactic coordinates, coloured according to their heliocentric distance values from the \citet{2021AJ....161..147B} catalogue. The most salient feature is the narrow, high-contrast ridge along the mid-plane, punctuated by concentrations at the longitudes of the major star-forming complexes in the solar neighbourhood and inner disk (e.g.\ Orion $l\simeq 210^\circ$--$220^\circ$, Perseus/Cassiopeia $l\simeq 140^\circ$--$160^\circ$, Cygnus~X $l\simeq 80^\circ$, Aquila Rift $l\simeq 30^\circ$--$40^\circ$, Carina $l\simeq 285^\circ$, Vela $l\simeq 265^\circ$). Outside the plane, prominent high-latitude groupings correspond to nearby clouds (Taurus, Lupus, Chamaeleon, $\rho$~Oph, Corona Australis), in line with expectations from resolved Gould Belt studies. The large-scale morphology---a thin disk component plus clustered nearby clouds---provides an independent, purely morphological sanity check that the classifier is indeed mapping canonical star-forming regions rather than generic reddened backgrounds. Fainter, patchy overdensities at $|b|\gtrsim 30^\circ$ are consistent with the distributed, older pre-main-sequence population surrounding Sco--Cen and other OB associations. The Large and Small Magellanic Clouds (LMC and SMC) are also visible, as many all-sky catalogues and the Simbad database contained sources from these two galaxies. The distribution of the NGYSO sources presented in Galactic Cartesian coordinates is shown in Fig.~\ref{xyz}. A more detailed discussion of the 3D distribution of the nearby star forming regions is presented in Sects.~\ref{localbubble} and~\ref{radcliffewave}.

Because we require a minimum SED length and remove problematic mid-IR points (e.g.\ $W3$/$W4$ with missing uncertainties), NGYSO is naturally biased towards sources with multi-band coverage and reliable photometry. This disfavours the most crowded, high-background inner-plane fields and very deeply embedded protostars lacking near-IR detections. Conversely, optically visible, more evolved pre-main-sequence (PMS) stars are more likely to meet the SED-length criterion. Users studying intrinsic population ratios should therefore model (or marginalise over) these effects; we provide the number of photometric points per source to facilitate such selection-function work.

\begin{figure}[tbp]
   \centering
   \includegraphics[width=0.95\hsize]{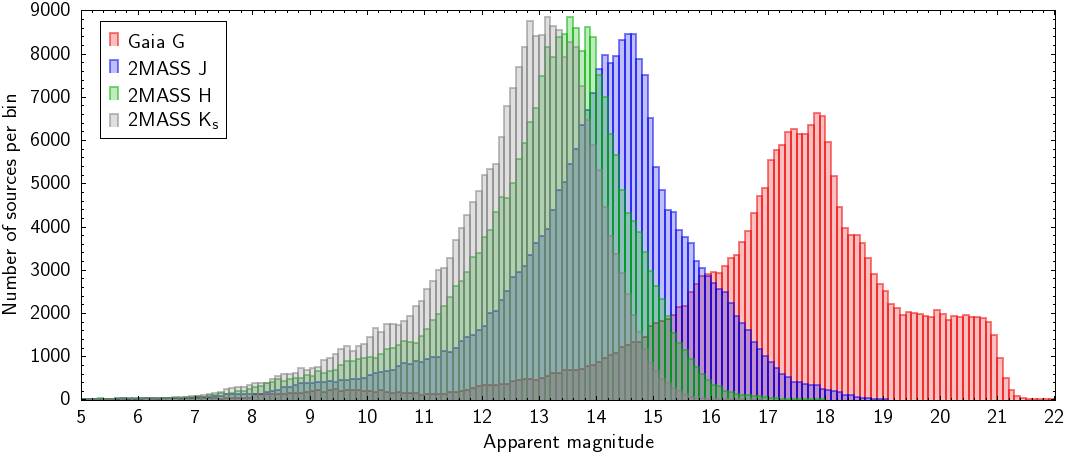}
      \caption{Apparent-magnitude distributions of the NGYSO sources in \textit{Gaia} $G$ and the 2MASS $J$, $H$, and $K_s$ bands. The vertical axis shows the number of sources per 0.1-mag bins. The near-infrared distributions peak at progressively brighter magnitudes from $J$ to $K_s$, whereas the broader \textit{Gaia} $G$-band distribution peaks around $G\approx17.5$--18 mag and extends beyond $G$=21 mag.}
         \label{ngysogmaghist}
\end{figure}

As a result of the above, the completeness of the NGYSO catalogue is hard to estimate. The observed magnitude distributions provide an empirical characterisation of the photometric coverage. Figure ~\ref{ngysogmaghist} shows the apparent-magnitude distributions in \textit{Gaia} $G$ and the 2MASS $J$, $H$, and $K_s$ bands. The near-infrared distributions peak at approximately $K_s\approx13.0$, $H\approx13.4$, and $J\approx14.5$ mag, while the broader \textit{Gaia} distribution reaches its maximum around $G\approx17.5$--18 mag. The substantially fainter $G$-band distribution reflects the greater optical depth of \textit{Gaia}, although \textit{Gaia} detections preferentially represent optically visible and less embedded YSOs. Conversely, the near-infrared measurements provide better coverage of extincted sources, but their distributions turn over at brighter magnitudes. The distribution shapes result from a combination of sensitivity and sky coverage of the contributing surveys, extinction, source crowding, photometric-quality requirements, the heterogeneous selection functions of the literature catalogues, and the requirement that a sufficiently sampled SED could be constructed. The structure visible in the \textit{Gaia} $G$-band distribution probably reflects the mixture of source populations and input catalogues probing different distances and Galactic environments. Nevertheless, the distributions indicate the effective magnitude ranges most strongly represented in the final catalogue.

In Table~\ref{tab:sed_yso} we list the number of sources originally listed in each catalogue and also the number of sources for which we were able to construct the SEDs for further analysis. In total 59.17\% of the listed objects had enough data points to have a proper SED, but this percentage was highly variable depending on the wavelength coverage of the original study. In those based on infrared data, focusing on actual YSOs instead of young clusters, we managed to build an SED for more than 80\% of the sources \citep[e.g.][]{2011ApJ...736..133A,2016ApJ...832...87B,2009ApJS..181..321E}, but in cases where the input catalogue was more focused on young clusters and based on optical data \citep[e.g.][]{2024A&A...686A..42H,2021ApJ...917...23K,2021AJ....162..282M} the fraction of the SEDs meeting our minimum requirements was in some cases well below 50\%. The completeness is also affected by the TP percentage of our method, which was found to be 96.36\% based on the SED plots. A more detailed completeness estimate can be based on the infrared slope ($\alpha_{\rm IR}$) value. We found that our method classifies as contaminants 96.7\% of those sources for which the $\alpha_{\rm IR}$ cannot be calculated. For the sources being in the very first stages of their evolution the recovery rate is $\sim$35\%, which drops down to $\sim$20\% for evolutionary stages that are closer to the main-sequence. However, statistics done on selected, reliable catalogues \citep{2014ApJ...783...29D, 2009ApJS..181..321E, 2008ApJ...674..336G} showed that our method is able to recover $\sim$80\% of the earliest evolutionary stages and $\sim$20\% of the later stages.

\subsection{Contamination}

Based on internal train--validation--test splits, only $0.43$\% of the labelled non--YSO sources in the test set were misclassified as YSOs, indicating a very low contamination rate.  To obtain an independent and more realistic estimate, we applied the same voting system to a large sample of non-YSO sources from completely external catalogues (independent from our train--validation--test samples), covering all major contaminant classes (see Table~\ref{tab:sedr_nonyso}). These include evolved, late-type stars (AGB, post--AGB, and RGB stars), a variety of variable stars (long--period variables and Miras, RR Lyrae stars, Cepheids), extragalactic sources (galaxies, quasars, AGNs), M~dwarfs, and main--sequence stars. The total sample comprises 92\,543 objects with sufficient photometric coverage for classification.

Only 914 of these sources ($0.99$\,\%) were falsely assigned a YSO label using the voting system described in Sect.~\ref{sec:methodperf}. This number is fully consistent with the internal validation and implies that fewer than 1\,\% of NGYSO entries are likely contaminants, even under conservative assumptions. By comparison, early all--sky YSO catalogues based on colour cuts or simple ML classifiers typically reported contamination fractions of 10--30\,\% \citep[e.g.][]{2011ApJ...726...18K,2016MNRAS.458.3479M,2019MNRAS.487.2522M}. The more than one--order--of--magnitude reduction achieved here primarily stems from the richer input representation (full SED images rather than discrete colours) and the inclusion of reddened main--sequence and M--dwarf populations in the negative training set. This allows the network to learn the subtle spectral shape differences that distinguish genuine circumstellar excess from interstellar reddening.

A closer inspection of the few misclassified sources offers further insight into the limits of the method.  More than half of the false positives originate from evolved stars (Cepheids, Miras, semi--regulars, and AGBs) whose dense circumstellar envelopes mimic the mid--infrared excess of embedded YSOs. These objects show SEDs rising beyond 10--20\,$\mu$m, similar to protostars, and can be virtually indistinguishable without variability, parallax, or spectroscopic information.  A smaller subset consists of compact galaxies or AGNs with steep near-- to mid--IR continua, and a handful of Wolf--Rayet and B[e] stars with hot--dust emission. The contamination rate among \textit{Gaia} main--sequence and RR~Lyrae stars is negligible ($<1$\,\%), confirming that the classifier does not spuriously select normal photospheric SEDs.  The small fraction ($\sim6$\,\%) of FPs among classical M~dwarfs likely reflects genuine young, low--mass stars misclassified in their original catalogues rather than actual contamination.

Because the architectures used heterogeneous photometry from the VizieR database, some residual contamination may also arise from mismatched or blended sources in crowded regions.  A positional crossmatch tolerance of 2$^{\prime\prime}$ inevitably produces occasional photometric confusion, especially for bright evolved stars with extended haloes or for faint sources projected on high--background Galactic fields.  Nonetheless, the contamination rate of $\approx$1\,\% represents an upper limit, as some of the apparently spurious classifications may correspond to real YSOs absent from the external training lists.

In summary, both internal and external tests confirm that NGYSO reaches a global purity of $\gtrsim99$\%. This level of reliability marks a substantial improvement over earlier all--sky YSO catalogues and is comparable to that of region--specific \textit{Spitzer} surveys. The small residual contamination is dominated by evolved red giants and AGB stars with genuine circumstellar dust emission---a physical degeneracy rather than a methodological artefact---and therefore defines the practical limit of purely photometric classification.  Further refinement can be achieved by incorporating parallax, proper motion, and variability features directly into the training process.

\section{Discussion}\label{sect:discussion}

\subsection{Colour-magnitude diagram}
\begin{figure}

\includegraphics[width=\hsize]{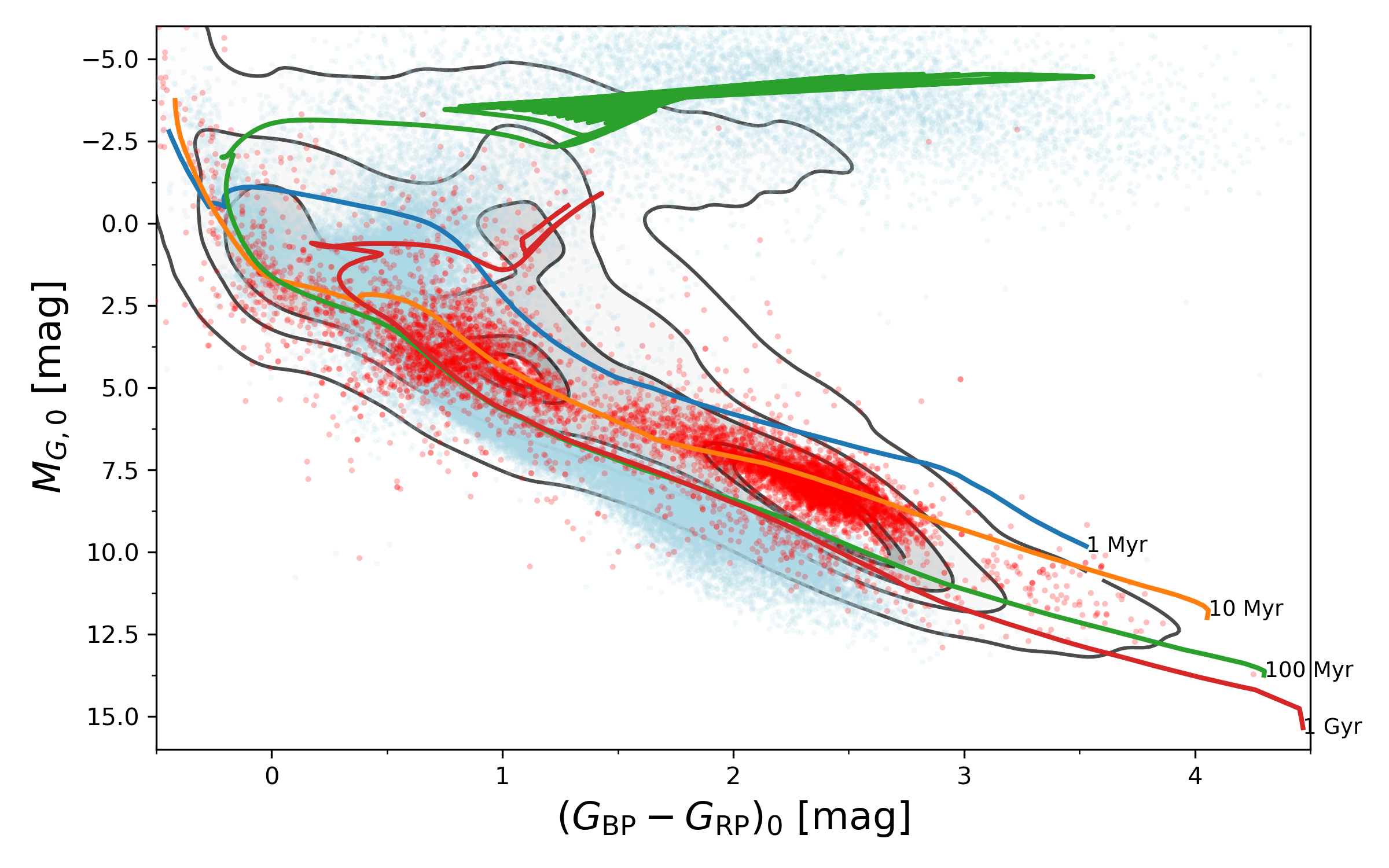}
\caption{Distribution of the YSO training sample (red dots), the non-YSO training sample (light blue dots), and the sources classified as YSOs (contour lines) on the de-reddened \textit{Gaia} colour--absolute magnitude diagram. Contour levels are at the 0.8, 0.9, 0.95, 0.99, and 0.995 density quantiles. PARSEC evolutionary isochrones of 1, 10, 100\,Myr, and 1\,Gyr are shown with blue, orange, green, and red lines, respectively.}
\label{hrd}
\end{figure}

As part of the validation of our YSO candidate catalogue, we checked the distribution of the training sample stars and the NGYSO candidates on the \textit{Gaia} colour-absolute magnitude diagram, shown in Fig.~\ref{hrd}. Distance values used for calculating the absolute $G$ magnitude are the geometric distance values from the \textit{Gaia} EDR3 distance catalogue of \citet{2021AJ....161..147B}. The absolute magnitude was also corrected with the \textit{Gaia} $G$ band extinction value ($A_G$, extinction in $G$ band from GSP-Phot Aeneas best library using BP/RP spectra). As a result Fig.~\ref{hrd} shows 82\,844 non-YSO training sources, 14\,744 YSO training sources and the density contours of 135\,078 sources classified as YSOs.

We used the PARSEC v2.0 evolutionary tracks \citep{2022A&A...665A.126N, 2025A&A...701A.258N} to identify the 1, 10, and 100\,Myr isochrones on the \textit{Gaia} HRD, also shown on Fig.~\ref{hrd}. We found that the lower contour of the NGYSO distribution is very close to the 100\,Myr track, while the densest part of the distribution is close to the 10\,Myr track, and below the 1\,Myr isochrone. While numerous sources are found above the 1\,Myr track, and in regions of the HRD where YSOs are not directly expected, we found that this is most probably caused by the incorrect \textit{Gaia} extinction estimation, as the $A_G$ value shows a monotonic increase farther away from the main sequence, reflecting known uncertainties of the DR3 $A_G$ value estimation \citep{2023A&A...674A..28F}. The HRD morphology confirms that the majority of NGYSO sources occupy the expected PMS region for ages 1--10\,Myr, providing independent astrophysical validation of the NGYSO catalogue.

\subsection{Evolutionary classes}\label{sect:evolutionaryclasses}

\begin{figure}
   \centering
\includegraphics[width=\hsize]{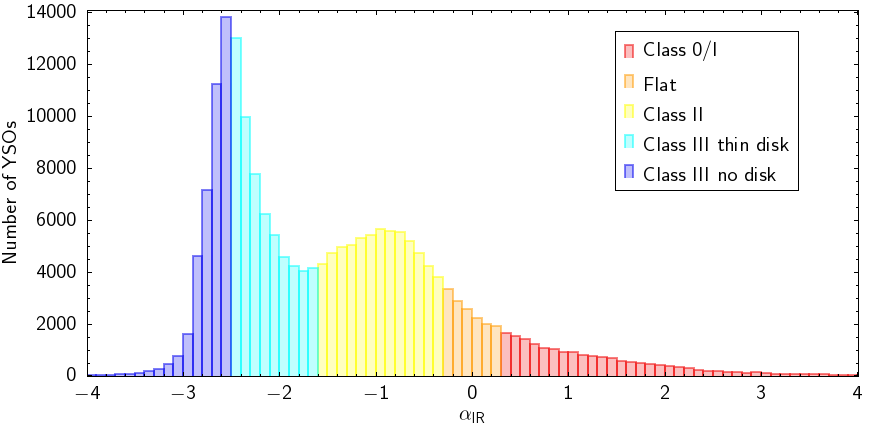}
\caption{Distribution of the $\alpha_\mathrm{IR}$ values for 195\,927 NGYSO sources. Different observational classes are presented with the different colours. Class boundaries are adopted from \citet{Grossschedl2019}.}
\label{classhist}
\end{figure}

To investigate the NGYSO candidates in more detail, we used the slope of the infrared spectral energy distribution between 2 and 24~$\mu$m ($\alpha_{\rm IR}$) to assign each source to an observational class following the scheme of \citet{Grossschedl2019}, which refines the original definition by \citet{1987IAUS..115....1L}. Only sources with at least four distinct photometric points within this wavelength range were considered. Table~\ref{tab:classes} lists the adopted $\alpha_\mathrm{IR}$ boundaries, and the number and fraction of NGYSO objects in each class. Figure~\ref{classhist} shows the distribution of the $\alpha_\mathrm{IR}$ values with the boundaries of the different observational classes for the 195\,927 NGYSO sources that had a sufficient number of data points in their SED to calculate the slope, and which have $-4<\alpha_{\rm IR}<4$ values.  Here, thin disk refers to the anaemic or thin circumstellar disks described by \citet{Grossschedl2019}, rather than the Galactic thin-disk population.

The overall distribution shows that $8.9$\,\% of the candidates are classified as Class~0/I and $7.6$\,\% as flat--spectrum sources, giving a combined protostellar fraction of $\approx17$\,\%. About $32.8$\,\% belong to Class~II, while $50.7$\,\% fall into the two Class~III categories. Compared to nearby--cloud statistics based on \textit{Spitzer} \citep[e.g.][]{2009ApJS..181..321E, 2015ApJS..220...11D}, which typically yield 10--15\,\% protostars and $\sim$50--60\,\% Class~II stars, NGYSO contains a slightly lower proportion of disk sources and a correspondingly higher fraction of Class~III stars. This trend is expected for an all--sky catalogue that includes many optically visible, more evolved pre--main--sequence stars as well as the dispersed populations surrounding the large molecular complexes. A comparison with additional studies of nearby star forming regions is shown in Table~\ref{tab:classes}. The NGYSO class mix differs systematically from region-focused \textit{Spitzer} surveys because of both selection bandpass and class definition. Orion-focused catalogues \citep[e.g.][]{2012AJ....144..192M, Grossschedl2019} target IR-excess YSOs and thus exclude diskless Class III stars; nearby-cloud compilations \citep[e.g.][]{2015ApJS..220...11D} include Class III but caution that this bin is susceptible to contamination. Mid-IR, inner-plane catalogues built from GLIMPSE/$W4$ slopes \citep[][]{2021ApJS..254...33K} naturally favour Class II sources. In contrast, NGYSO incorporates Class III objects and applies a classifier trained explicitly against reddened main-sequence and M-dwarf contaminants, yielding a higher, but plausibly cleaner Class III fraction in the all-sky context. The modest differences in Flat/II fractions across studies are consistent with (i) whether $\alpha$ is de-reddened, (ii) the exact wavelength range used for the slope, and (iii) regional age/environment. Overall, the class distribution indicates that the catalogue samples the full sequence of early stellar evolution from embedded protostars to diskless pre--main--sequence stars.

Figure~\ref{hrdclasses} illustrates the location of the different classes on the \textit{Gaia} colour--absolute magnitude diagram. Class~0/I and flat--spectrum objects populate the region above the 1\,Myr isochrone, as expected for the youngest, most embedded sources. Class~II members are concentrated just above and around the 1\,Myr line, while the two Class~III groups lie predominantly below it. The brighter, high--mass tail of the Class~III--no--disk distribution may include a small number of reddened main--sequence or giant contaminants, but the majority are likely genuine weak--lined T~Tauri stars or PMS stars with dispersed disks, consistent with the low false--positive rate observed with the voting system.

\begin{table*}[tbp]
    \caption{Numbers and fractions of observational evolutionary classes in YSO catalogues compared with NGYSO.}\label{tab:classes}
    \centering
    \begin{tabular}{@{}lccccc@{}}\hline\hline
Catalogue/reference & Class 0/I & Flat & Class II & Class III & Class III\\
    \hline
    NGYSO & 17\,500 (8.9\%) & 14\,846 (7.6\%) & 64\,195 (32.8\%) & 59\,180 (30.2\%) & 40\,206 (20.5\%) \\

    {\citealt{Grossschedl2019}} & 219 (12.2\%) & 223 (12.4\%) & 1\,318 (73.3\%) & 39 (2.2\%) & 0 (0\%) \\

    {\citealt{2015ApJS..220...11D}} & 375 (12.7\%) & 256 (8.6\%) & 1\,409 (47.6\%) & 913 (30.8\%) & 9 (0.3\%) \\

    {\citealt{2009ApJS..181..321E}} & 162 (15.9\%) & 122 (11.9\%) & 613 (60\%) & 122 (11.9\%) & 2 (0.2\%) \\

    {\citealt{2012AJ....144..192M}} & 288 (11\%) & 257 (9.8\%) & 1\,739 (66.6\%) & 298 (11.4\%) & 31 (1.2\%) \\

    {\citealt{2008ApJ...674..336G}} & 26 (20.5\%) & 21 (16.5\%) & 72 (56.7\%) & 8 (6.3\%) & 0 (0\%) \\

    {\citealt{2017ApJ...839..108S}} & 2\,148 (2.9\%) & 1\,554 (2.1\%) & 3\,372 (4.6\%) & 709 (1\%) & 65\,944 (89.4\%) \\

    {\citealt{2021AJ....162..236M}} & 67 (8.2\%) & 103 (12.6\%) & 573 (70.1\%) & 73 (8.9\%) & 1 (0\%) \\

\hline
    \end{tabular}
    \tablefoot{$\alpha_\mathrm{IR}$ ranges for the different classes are the following, adopted from \citet{Grossschedl2019}: Class 0/I: $\alpha_\mathrm{IR} > 0.3$; Flat: $0.3 \geq \alpha_\mathrm{IR} > -0.3$; Class II: $-0.3 \geq \alpha_\mathrm{IR} > -1.6$; Class III thin disk: $-1.6 \geq \alpha_\mathrm{IR} > -2.5$; Class III no disk: $-2.5 \geq \alpha_\mathrm{IR}$.}

\end{table*}

\subsection{3D distribution and relation to the Local Bubble}\label{localbubble}

The Local Bubble (LB) is a $\sim$200 pc-wide cavity of hot, low-density gas thought to have been carved by multiple supernovae over the past $\sim$10--15\,Myr. Recent dust-tomography studies \citep[e.g.][]{2024ApJ...973..136O, 2024A&A...685A..82E} revealed a thin, fragmented shell that hosts several nearby star-forming regions, suggesting that the expanding LB has compressed molecular gas and possibly triggered new star formation along its rim. Testing this scenario requires a uniform, all-sky census of YSOs with reliable distances---precisely what NGYSO provides.

To investigate whether the nearby population of YSOs traces the structure of the LB cavity, we reconstructed the three-dimensional shape of the LB shell following the methodology of \citet{2024ApJ...973..136O}, using the \citet{2024A&A...685A..82E} 3D dust map, providing local extinction. The 3D dust map was sampled along every \texttt{HEALPix} (Hierarchical Equal Area isoLatitude Pixelization; {\citealt{Gorski2005}}) line of sight with resolution parameter \texttt{NSIDE} = 256 ($\sim$13.7$^{\prime}$) and a radial step of 0.5 pc out to 850 pc. Each extinction profile was smoothed with a Gaussian kernel of $\sigma$=5 pc, and the first significant peak in the smoothed profile---identified by a prominence threshold of $2.0\times10^{-6}$ mag pc$^{-1}$---was taken as LB wall location in that direction.  The \texttt{NSIDE} parameter specifies the number of subdivisions along each side of a base pixel.

The NGYSO catalogue was then projected onto this 3D structure. A representation of the Local Bubble and the spatial density of the catalogue sources (along with the Radcliffe Wave) is shown in Fig.~\ref{fig:lb3d}. We analysed the line-of-sight distances of the NGYSO sources relative to the reconstructed shell geometry. For each direction on the sky, the wall model provides characteristic distances corresponding to the inner edge ($d_{\mathrm{inner}}$), the peak density ($d_{\mathrm{peak}}$), and the outer edge ($d_{\mathrm{outer}}$). Each YSO was assigned to a specific line-of-sight using a \texttt{HEALPix} tessellation, and the corresponding wall distances were extracted. We define a signed radial offset relative to the shell peak as

\begin{equation}
\Delta = d_{\mathrm{YSO}} - d_{\mathrm{peak}},
\end{equation}
\noindent
where $d_{\mathrm{YSO}}$ is the distance of the source. Negative values of $\Delta$ correspond to locations interior to the present-day shell, while positive values indicate positions exterior to it.

The distribution of $\Delta$ was computed for all YSOs within 600~pc. To assess the statistical significance of any apparent correlation, we constructed a Monte Carlo (MC) null model in which the YSO distances were kept fixed, but their angular association with the wall was randomised. In practice, for each MC realisation, each YSO was assigned a randomly selected line-of-sight from the wall map, and the corresponding offset $\Delta$ was recomputed. This procedure preserves the intrinsic distance distribution of the YSO sample while removing any physical spatial correlation with the shell.

\begin{figure}

\includegraphics[width=\hsize]{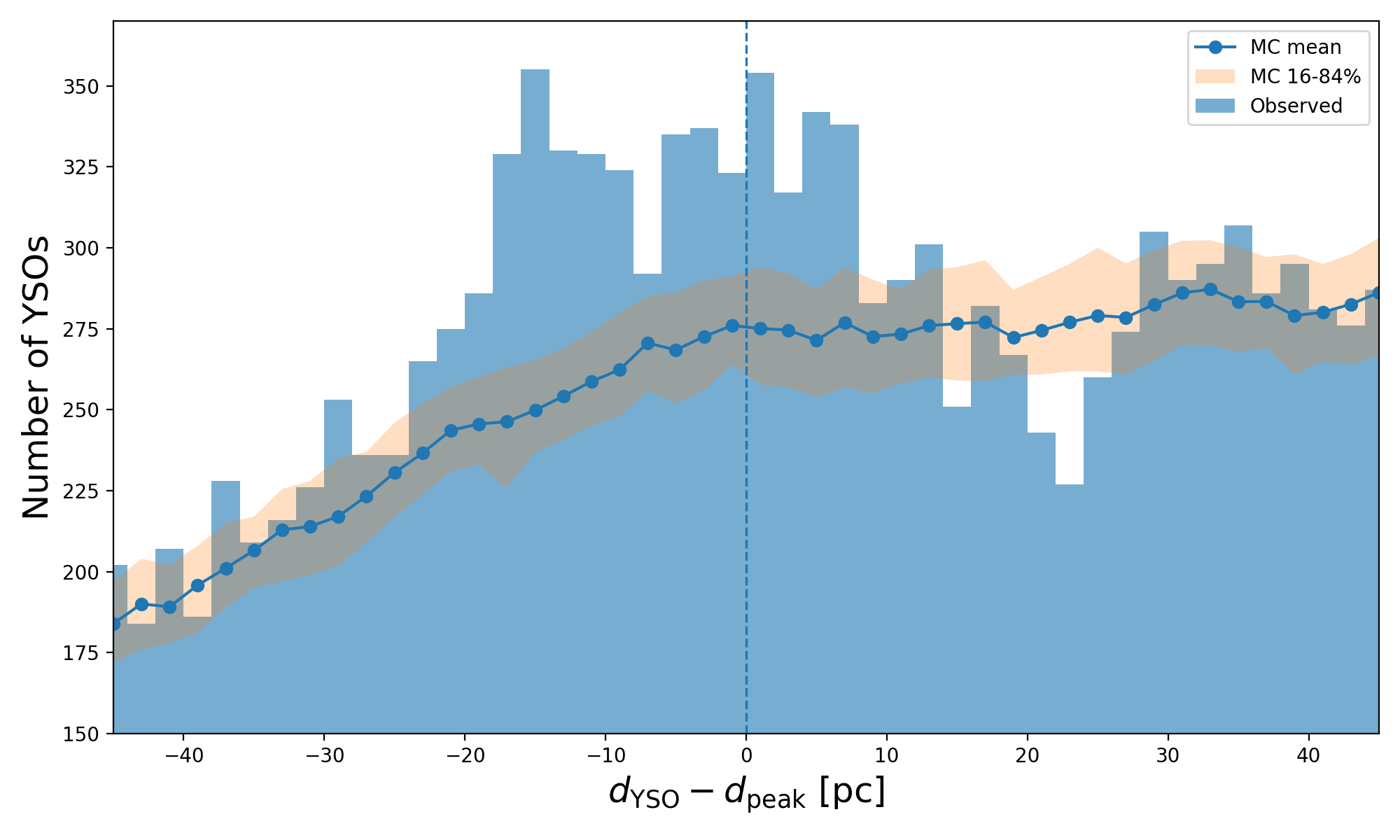}
\caption{Distribution of YSO distances relative to the Local Bubble wall. The horizontal axis shows the signed offset $\Delta = d_{\mathrm{YSO}} - d_{\mathrm{peak}}$, where $d_{\mathrm{peak}}$ is the distance of the shell density maximum along each line of sight. Negative values correspond to positions interior to the present-day shell, while positive values indicate locations exterior to it. The blue histogram shows the observed distribution of YSOs, while the solid line indicates the mean of the Monte Carlo (MC) realisations in which YSO distances are preserved but their angular association with the wall is randomised. The shaded region marks the 16th--84th percentile range of the MC distribution. The vertical dashed line denotes $\Delta = 0$, i.e. the location of the present-day shell peak.}

\label{fig:lbysohist}
\end{figure}

The observed offset distribution was then compared to the ensemble of MC realisations, see Fig.~\ref{fig:lbysohist}. We find a statistically significant excess of YSOs in the vicinity of the Local Bubble wall. The strongest overdensity occurs in the interval $-20~\mathrm{pc} < \Delta < -10~\mathrm{pc}$, where the observed number of YSOs exceeds the MC expectation by $\sim 30\%$. This corresponds to a deviation of approximately $11\sigma$ relative to the MC distribution. Significant excesses are also detected in the adjacent intervals $-10~\mathrm{pc} < \Delta < 0$ and $0 < \Delta < 10~\mathrm{pc}$, with amplitudes of $\sim 20\%$ above the null expectation and significances of $\sim 7\sigma$. In contrast, no statistically significant excess is found at larger positive offsets, and a mild deficit is observed beyond $\Delta \gtrsim 20$~pc.

The fact that the maximum excess is located slightly interior to the present-day shell ($\Delta < 0$), rather than exactly at $\Delta = 0$, suggests that the current position of the wall does not coincide with the location of peak YSO density. This offset can be explained if star formation was triggered during an earlier phase of shell expansion, and the shell has since moved outward relative to the newly formed population. The observed offset scale of $\sim 10$--20~pc is consistent with plausible expansion velocities of several to $\sim 10$~km~s$^{-1}$ over timescales of $\sim 1$\,Myr \citep{2022Natur.601..334Z}.

Overall, this analysis provides strong statistical evidence that the spatial distribution of YSOs is not independent of the Local Bubble structure, and supports a scenario in which the expanding shell has played a role in shaping recent star formation in the solar neighbourhood, in good agreement with the results of \citet{2022Natur.601..334Z}.

\subsection{Spatial association with the Radcliffe Wave}\label{radcliffewave}

\begin{figure*}

\includegraphics[width=\hsize]{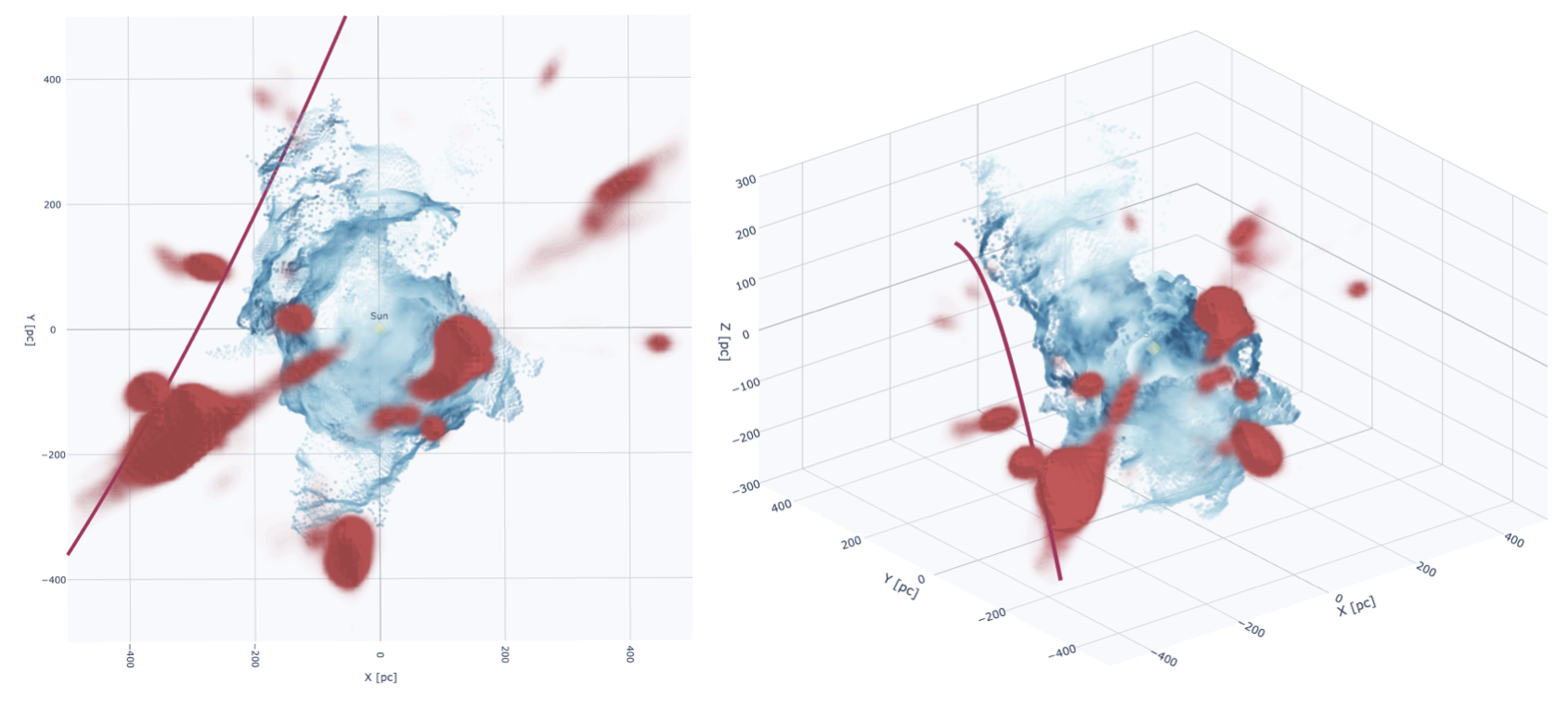}
\caption{Top-down (left) and oblique (right) views of the three-dimensional distribution of YSOs around the Local Bubble and the Radcliffe Wave. The Local Bubble shell is represented by extinction-weighted blue points placed at the peak shell distance along each line of sight, with darker blue indicating higher extinction. The Radcliffe Wave model is shown by the red curve. The red volume shows the observed spatial density of NGYSO sources, calculated on a 10 pc Cartesian grid and smoothed with a Gaussian kernel of $\sigma=13.5$~pc. The colour of the YSO distribution is fixed, while opacity increases with local density. Coordinates are heliocentric Galactic Cartesian coordinates, with $X$ directed towards the Galactic centre, $Y$ in the direction of Galactic rotation, and $Z$ towards the north Galactic pole. Most of the high density NGYSO clusters are seen close to the edges of the Local Bubble, or to the Radcliffe Wave.}
\label{fig:lb3d}
\end{figure*}

Building upon the 3D spatial framework established in Sect.~\ref{localbubble}, we extended our analysis to another major local star-forming structure, the Radcliffe Wave (see Fig.~\ref{fig:lb3d}). Discovered by \citet{2020Natur.578..237A}, this continuous filament of gas and young stars spans roughly 2.7 kpc across the solar neighbourhood, and was primarily identified from the 3D distribution of interstellar dust, with known star-forming regions serving as additional tracers. While some of these regions are also represented in our input catalogues, the NGYSO catalogue provides a more homogeneous and significantly expanded sample of candidate YSOs. Validating the spatial correlation between the NGYSO catalogue and the Radcliffe Wave serves as a robust test of both our classification reliability and the wave's ongoing star-forming nature.

We cross-matched the YSOs with the 3D backbone model of the Radcliffe Wave presented by \citet{2024Natur.628...62K}. To minimise dynamical noise caused by older pre-main-sequence stars that have had millions of years to drift away from their birth sites due to standard velocity dispersion, we applied an evolutionary stage cut. Based on the infrared spectral index ($\alpha$), we isolated the youngest, most embedded populations by requiring $\alpha \ge -1.6$, effectively filtering out the older Class III sources.

To quantitatively assess the significance of this spatial association, we employed a Monte Carlo permutation test. First, using a 3D nearest-neighbour algorithm, we measured the actual number of these filtered YSOs located within a 100 pc radius of the Radcliffe Wave's spline. We then constructed a null hypothesis to represent the background galactic field. We generated 500 mock distributions by rotating the YSO coordinates by a random angle $\theta \in [0, 2\pi]$ around the Galactic $Z$-axis. This rotation preserves the heliocentric distance and vertical ($Z$) distributions of the catalogue, thus maintaining inherent observational and survey biases, while completely decoupling the stars from the Radcliffe Wave's specific geometry.

\begin{figure}

\includegraphics[width=\hsize]{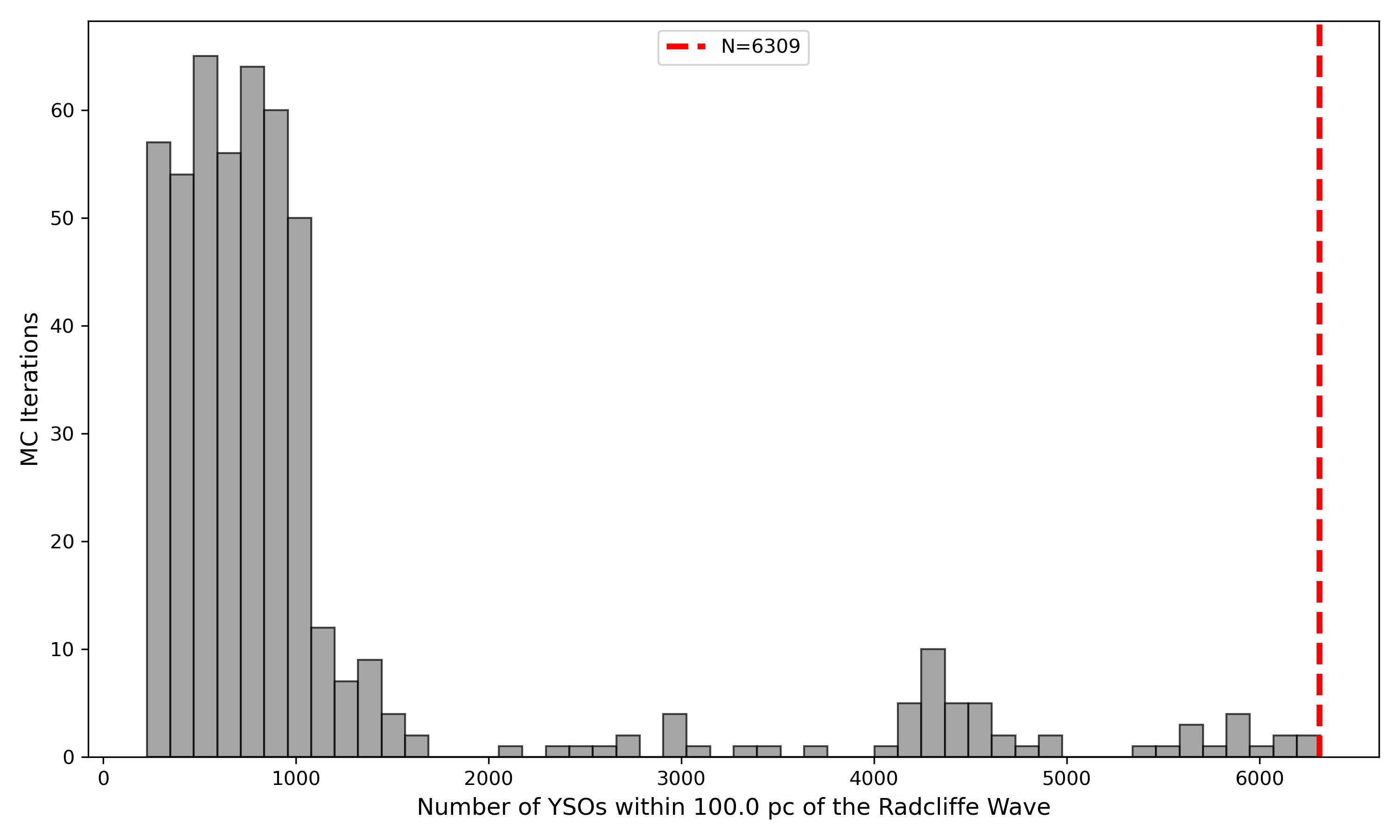}
\caption{Results of the Monte Carlo test between the NGYSO catalogue and the Radcliffe Wave. The grey histogram shows the expected background distribution of YSOs generated from 500 randomised rotations of the dataset around the Galactic $Z$-axis. The red vertical dashed line indicates the actual number of YSOs located within a 100 pc radius of the wave's 3D spatial spline.}
\label{fig:radcliffeysohist}
\end{figure}

The resulting statistical distribution is shown in Fig.~\ref{fig:radcliffeysohist}. The randomised samples form a normal background distribution, whereas the actual NEMESIS YSO sample exhibits a significant overdensity along the wave. The real spatial configuration yields a Z-score of 3.7 (a $p$-value of approximately 2.5$\times10^{-4}$). This $> 3\sigma$ confirmation provides strong evidence that the youngest stellar objects in the NGYSO catalogue are truly physically associated with the 3D topology of the Radcliffe Wave.

\section{Summary}\label{sect:summary}
In this work we present the NEMESIS General YSO (NGYSO) catalogue, which provides the largest homogeneous re-assessment of young stellar object candidates to date. Our aim was to unify the multitude of heterogeneous YSO catalogues by applying a single, consistent, and data-driven classification scheme that exploits the full information content of SEDs. To achieve this, we utilised a deep-learning approach that treats SEDs as images and applies an ensemble of both standard off-the-shelf and custom-built CNN architectures in a voting system to distinguish YSOs from non-YSOs with high fidelity.

Training relied on two well-curated reference samples: the KYSO, a large catalogue of optically detected, bona fide YSOs, and the NEMESIS Orion Star Formation Complex catalogue covering the full pre-main-sequence range from protostars to diskless sources. A broad non-YSO training set, including main-sequence stars, M dwarfs, evolved stars, quasars, and variable stars ensured robust discrimination against common contaminants. For each source, all available photometric data were gathered from VizieR to construct SEDs, complemented by contextual information such as interstellar reddening, variability, and direct imaging.

Across a wide suite of CNN architectures, the best-performing method was found to be a system based on voting from each CNN using the SEDplot images. Accepting decisions if at least half of the models voted a source as YSO achieved $F_1 \simeq 96.9$ with contamination below 0.5\%. The method was applied to more than 3.9 million literature candidates drawn from nearly one hundred catalogues in VizieR, producing SEDs of sufficient quality for 2.3 million sources. Of these, 501\,051 were reclassified as YSOs with at least 6 votes. After removing duplicates across catalogues, the final NGYSO catalogue contains 274\,408 unique YSO candidates, representing the largest homogeneous and reproducible all-sky sample to date.

The classifier performance was further validated using independent non-YSO catalogues comprising $>$92\,000 objects. Only 0.99\% of these were misclassified as YSOs, confirming the very low contamination rate. The \textit{Gaia} colour--absolute magnitude diagram of NGYSO objects shows a distribution consistent with pre-main-sequence evolutionary tracks for ages between 1 and 10\,Myr, demonstrating that the identified sources occupy the expected PMS regime.

Using the slope of the infrared SED between 2 and 24~$\mu$m, we assigned each NGYSO source with sufficient infrared photometry to an observational class following the \citet{Grossschedl2019} scheme. The overall class fractions are 8.9\% (Class~0/I), 7.6\% (Flat), 32.8\% (Class~II), and 50.7\% (Class~III, thin and no-disk combined). Compared with nearby-cloud surveys such as the c2d and Gould Belt projects \citep{2009ApJS..181..321E,2015ApJS..220...11D}, NGYSO shows a slightly lower disk fraction and a higher proportion of Class~III sources, as expected for an all-sky catalogue including dispersed and optically visible populations.

Projecting the NGYSO sources onto the three-dimensional dust structure of the Local Bubble reveals an enhanced YSO density near and slightly inside the bubble wall, suggesting that star formation in the solar vicinity was likely triggered along the compressed molecular shell bounding the cavity. This spatial correlation provides an additional astrophysical validation of the catalogue.

We also investigated the spatial association between the NGYSO catalogue and the Radcliffe Wave. Using the 3D backbone model of the wave, we cross-matched the youngest NGYSO populations ($\alpha \geq -1.6$). Based on 500 Monte Carlo simulations, we found a clear and statistically significant excess of YSOs along the wave. This $3.7\sigma$ detection demonstrates that the youngest NGYSO objects are not randomly distributed, but instead trace the three-dimensional geometry of the Radcliffe Wave, supporting its interpretation as an actively star-forming structure.

The NGYSO catalogue thus delivers (i) a reproducible, deep-learning--based re-classification of literature YSO lists; (ii) a clean, all-sky sample spanning the full pre-main-sequence evolutionary sequence; and (iii) a quantitative framework for comparing the reliability and completeness of existing YSO catalogues. It can serve as a reference dataset for future region-specific studies, for calibrating star-formation rate estimates, and for guiding spectroscopic or variability-based follow-up of specific subclasses such as Flat-spectrum and transitional-disk sources.

Finally, our \texttt{DLYSO} pipeline offers an easy-to-use, coordinate-driven solution for YSO classification. By integrating automated data acquisition, image generation, and inference across different data modalities, it enables fast, homogeneous identification of YSO candidates from any input catalogue without manual preprocessing.

\begin{acknowledgements}
We thank the anonymous referee for the careful reading of the manuscript and for the constructive comments and suggestions, which significantly improved the quality and clarity of this work.
This project is funded by the European Union’s Horizon 2020 research and innovation programme under grant agreement No. 101004141. This research was supported by the International Space Science Institute (ISSI) in Bern, through ISSI International Team project \#521 (Star Formation with Big Data). G.M. acknowledges support from the János Bolyai Research Scholarship of the Hungarian Academy of Sciences. M.M. acknowledges support from the ESA PRODEX contract nr. 4000132054.
This research used the software \texttt{TOPCAT/STILTS} \citep{2005ASPC..347...29T}, the Simbad database and the VizieR catalogue access tool operated at Centre de données astronomiques de Strasbourg (CDS), France (DOI: 10.26093/cds/vizier), and the NASA/IPAC Extragalactic Database (NED), which is operated by the Jet Propulsion Laboratory, California Institute of Technology, under contract with the National Aeronautics and Space Administration.
During manuscript revision, the authors used OpenAI's ChatGPT and Codex (GPT 6 Astra) to assist with language editing, LaTeX formatting, and checks of references and manuscript consistency. The authors reviewed the suggested changes and take full responsibility for the final content.

\end{acknowledgements}

\bibliographystyle{aa}
\bibliography{aanda}

\begin{thebibliography}{201}
\expandafter\ifx\csname natexlab\endcsname\relax\def\natexlab#1{#1}\fi

\bibitem[{{Abbott} {et~al.}(2021){Abbott}, {Adam{\'o}w}, {Aguena}, {Allam},
  {Amon}, {Annis}, {Avila}, {Bacon}, {Banerji}, {Bechtol}, {Becker},
  {Bernstein}, {Bertin}, {Bhargava}, {Bridle}, {Brooks}, {Burke}, {Carnero
  Rosell}, {Carrasco Kind}, {Carretero}, {Castander}, {Cawthon}, {Chang},
  {Choi}, {Conselice}, {Costanzi}, {Crocce}, {da Costa}, {Davis}, {De Vicente},
  {DeRose}, {Desai}, {Diehl}, {Dietrich}, {Drlica-Wagner}, {Eckert},
  {Elvin-Poole}, {Everett}, {Evrard}, {Ferrero}, {Fert{\'e}}, {Flaugher},
  {Fosalba}, {Friedel}, {Frieman}, {Garc{\'\i}a-Bellido}, {Gaztanaga},
  {Gelman}, {Gerdes}, {Giannantonio}, {Gill}, {Gruen}, {Gruendl}, {Gschwend},
  {Gutierrez}, {Hartley}, {Hinton}, {Hollowood}, {Honscheid}, {Huterer},
  {James}, {Jeltema}, {Johnson}, {Kent}, {Kron}, {Kuehn}, {Kuropatkin},
  {Lahav}, {Li}, {Lidman}, {Lin}, {MacCrann}, {Maia}, {Manning}, {Maloney},
  {March}, {Marshall}, {Martini}, {Melchior}, {Menanteau}, {Miquel}, {Morgan},
  {Myles}, {Neilsen}, {Ogando}, {Palmese}, {Paz-Chinch{\'o}n}, {Petravick},
  {Pieres}, {Plazas}, {Pond}, {Rodriguez-Monroy}, {Romer}, {Roodman}, {Rykoff},
  {Sako}, {Sanchez}, {Santiago}, {Scarpine}, {Serrano}, {Sevilla-Noarbe},
  {Smith}, {Smith}, {Soares-Santos}, {Suchyta}, {Swanson}, {Tarle}, {Thomas},
  {To}, {Tremblay}, {Troxel}, {Tucker}, {Turner}, {Varga}, {Walker},
  {Wechsler}, {Weller}, {Wester}, {Wilkinson}, {Yanny}, {Zhang}, {Nikutta},
  {Fitzpatrick}, {Jacques}, {Scott}, {Olsen}, {Huang}, {Herrera}, {Juneau},
  {Nidever}, {Weaver}, {Adean}, {Correia}, {de Freitas}, {Freitas},
  {Singulani}, {Vila-Verde}, \& {Linea Science Server}}]{2021ApJS..255...20A}
{Abbott}, T.~M.~C., {Adam{\'o}w}, M., {Aguena}, M., {et~al.} 2021, \apjs, 255,
  20

\bibitem[{{Ahumada} {et~al.}(2020){Ahumada}, {Allende Prieto}, {Almeida},
  {Anders}, {Anderson}, {Andrews}, {Anguiano}, {Arcodia}, {Armengaud},
  {Aubert}, {Avila}, {Avila-Reese}, {Badenes}, {Balland}, {Barger},
  {Barrera-Ballesteros}, {Basu}, {Bautista}, {Beaton}, {Beers}, {Benavides},
  {Bender}, {Bernardi}, {Bershady}, {Beutler}, {Bidin}, {Bird}, {Bizyaev},
  {Blanc}, {Blanton}, {Boquien}, {Borissova}, {Bovy}, {Brandt}, {Brinkmann},
  {Brownstein}, {Bundy}, {Bureau}, {Burgasser}, {Burtin}, {Cano-D{\'\i}az},
  {Capasso}, {Cappellari}, {Carrera}, {Chabanier}, {Chaplin}, {Chapman},
  {Cherinka}, {Chiappini}, {Doohyun Choi}, {Chojnowski}, {Chung}, {Clerc},
  {Coffey}, {Comerford}, {Comparat}, {da Costa}, {Cousinou}, {Covey}, {Crane},
  {Cunha}, {Ilha}, {Dai}, {Damsted}, {Darling}, {Davidson}, {Davies}, {Dawson},
  {De}, {de la Macorra}, {De Lee}, {Queiroz}, {Deconto Machado}, {de la Torre},
  {Dell'Agli}, {du Mas des Bourboux}, {Diamond-Stanic}, {Dillon}, {Donor},
  {Drory}, {Duckworth}, {Dwelly}, {Ebelke}, {Eftekharzadeh}, {Davis Eigenbrot},
  {Elsworth}, {Eracleous}, {Erfanianfar}, {Escoffier}, {Fan}, {Farr},
  {Fern{\'a}ndez-Trincado}, {Feuillet}, {Finoguenov}, {Fofie},
  {Fraser-McKelvie}, {Frinchaboy}, {Fromenteau}, {Fu}, {Galbany}, {Garcia},
  {Garc{\'\i}a-Hern{\'a}ndez}, {Garma Oehmichen}, {Ge}, {Geimba Maia},
  {Geisler}, {Gelfand}, {Goddy}, {Gonzalez-Perez}, {Grabowski}, {Green},
  {Grier}, {Guo}, {Guy}, {Harding}, {Hasselquist}, {Hawken}, {Hayes}, {Hearty},
  {Hekker}, {Hogg}, {Holtzman}, {Horta}, {Hou}, {Hsieh}, {Huber}, {Hunt}, {Ider
  Chitham}, {Imig}, {Jaber}, {Jimenez Angel}, {Johnson}, {Jones},
  {J{\"o}nsson}, {Jullo}, {Kim}, {Kinemuchi}, {Kirkpatrick}, {Kite}, {Klaene},
  {Kneib}, {Kollmeier}, {Kong}, {Kounkel}, {Krishnarao}, {Lacerna}, {Lan},
  {Lane}, {Law}, {Le Goff}, {Leung}, {Lewis}, {Li}, {Lian}, {Lin}, {Long},
  {Longa-Pe{\~n}a}, {Lundgren}, {Lyke}, {Mackereth}, {MacLeod}, {Majewski},
  {Manchado}, {Maraston}, {Martini}, {Masseron}, {Masters}, {Mathur},
  {McDermid}, {Merloni}, {Merrifield}, {M{\'e}sz{\'a}ros}, {Miglio}, {Minniti},
  {Minsley}, {Miyaji}, {Mohammad}, {Mosser}, {Mueller}, {Muna},
  {Mu{\~n}oz-Guti{\'e}rrez}, {Myers}, {Nadathur}, {Nair}, {Nandra}, {Correa do
  Nascimento}, {Nevin}, {Newman}, {Nidever}, {Nitschelm}, {Noterdaeme},
  {O'Connell}, {Olmstead}, {Oravetz}, {Oravetz}, {Osorio}, {Pace}, {Padilla},
  {Palanque-Delabrouille}, \& {Palicio}}]{2020ApJS..249....3A}
{Ahumada}, R., {Allende Prieto}, C., {Almeida}, A., {et~al.} 2020, \apjs, 249,
  3

\bibitem[{{Alcal{\'a}} {et~al.}(2014){Alcal{\'a}}, {Natta}, {Manara}, {Spezzi},
  {Stelzer}, {Frasca}, {Biazzo}, {Covino}, {Randich}, {Rigliaco}, {Testi},
  {Comer{\'o}n}, {Cupani}, \& {D'Elia}}]{2014A&A...561A...2A}
{Alcal{\'a}}, J.~M., {Natta}, A., {Manara}, C.~F., {et~al.} 2014, \aap, 561, A2

\bibitem[{{Allen} {et~al.}(2012){Allen}, {Gutermuth}, {Kryukova}, {Megeath},
  {Pipher}, {Naylor}, {Jeffries}, {Wolk}, {Spitzbart}, \&
  {Muzerolle}}]{2012ApJ...750..125A}
{Allen}, T.~S., {Gutermuth}, R.~A., {Kryukova}, E., {et~al.} 2012, \apj, 750,
  125

\bibitem[{{Alonso-Floriano} {et~al.}(2015){Alonso-Floriano}, {Morales},
  {Caballero}, {Montes}, {Klutsch}, {Mundt}, {Cort{\'e}s-Contreras}, {Ribas},
  {Reiners}, {Amado}, {Quirrenbach}, \& {Jeffers}}]{2015A&A...577A.128A}
{Alonso-Floriano}, F.~J., {Morales}, J.~C., {Caballero}, J.~A., {et~al.} 2015,
  \aap, 577, A128

\bibitem[{{Alves} {et~al.}(2020){Alves}, {Zucker}, {Goodman}, {Speagle},
  {Meingast}, {Robitaille}, {Finkbeiner}, {Schlafly}, \&
  {Green}}]{2020Natur.578..237A}
{Alves}, J., {Zucker}, C., {Goodman}, A.~A., {et~al.} 2020, \nat, 578, 237

\bibitem[{{An} {et~al.}(2011){An}, {Ram{\'\i}rez}, {Sellgren}, {Arendt}, {Adwin
  Boogert}, {Robitaille}, {Schultheis}, {Cotera}, {Smith}, \&
  {Stolovy}}]{2011ApJ...736..133A}
{An}, D., {Ram{\'\i}rez}, S.~V., {Sellgren}, K., {et~al.} 2011, \apj, 736, 133

\bibitem[{{Andre} {et~al.}(1993){Andre}, {Ward-Thompson}, \&
  {Barsony}}]{1993ApJ...406..122A}
{Andre}, P., {Ward-Thompson}, D., \& {Barsony}, M. 1993, \apj, 406, 122

\bibitem[{{Asmus} {et~al.}(2020){Asmus}, {Greenwell}, {Gandhi}, {Boorman},
  {Aird}, {Alexander}, {Assef}, {Baldi}, {Davies}, {H{\"o}nig}, {Ricci},
  {Rosario}, {Salvato}, {Shankar}, \& {Stern}}]{2020MNRAS.494.1784A}
{Asmus}, D., {Greenwell}, C.~L., {Gandhi}, P., {et~al.} 2020, \mnras, 494, 1784

\bibitem[{{Audard} {et~al.}(2014){Audard}, {{\'A}brah{\'a}m}, {Dunham},
  {Green}, {Grosso}, {Hamaguchi}, {Kastner}, {K{\'o}sp{\'a}l}, {Lodato},
  {Romanova}, {Skinner}, {Vorobyov}, \& {Zhu}}]{2014prpl.conf..387A}
{Audard}, M., {{\'A}brah{\'a}m}, P., {Dunham}, M.~M., {et~al.} 2014, in
  Protostars and Planets VI, ed. H.~{Beuther}, R.~S. {Klessen}, C.~P.
  {Dullemond}, \& T.~{Henning}, 387

\bibitem[{{Bailer-Jones} {et~al.}(2021){Bailer-Jones}, {Rybizki}, {Fouesneau},
  {Demleitner}, \& {Andrae}}]{2021AJ....161..147B}
{Bailer-Jones}, C.~A.~L., {Rybizki}, J., {Fouesneau}, M., {Demleitner}, M., \&
  {Andrae}, R. 2021, \aj, 161, 147

\bibitem[{{Bally}(2016)}]{2016ARA&A..54..491B}
{Bally}, J. 2016, \araa, 54, 491

\bibitem[{{Balog} {et~al.}(2016){Balog}, {Siegler}, {Rieke}, {Kiss},
  {Muzerolle}, {Gutermuth}, {Bell}, {Vink{\'o}}, {Su}, {Young}, \&
  {G{\'a}sp{\'a}r}}]{2016ApJ...832...87B}
{Balog}, Z., {Siegler}, N., {Rieke}, G.~H., {et~al.} 2016, \apj, 832, 87

\bibitem[{{Bellm} {et~al.}(2019){Bellm}, {Kulkarni}, {Graham}, {Dekany},
  {Smith}, {Riddle}, {Masci}, {Helou}, {Prince}, {Adams}, {Barbarino},
  {Barlow}, {Bauer}, {Beck}, {Belicki}, {Biswas}, {Blagorodnova}, {Bodewits},
  {Bolin}, {Brinnel}, {Brooke}, {Bue}, {Bulla}, {Burruss}, {Cenko}, {Chang},
  {Connolly}, {Coughlin}, {Cromer}, {Cunningham}, {De}, {Delacroix}, {Desai},
  {Duev}, {Eadie}, {Farnham}, {Feeney}, {Feindt}, {Flynn}, {Franckowiak},
  {Frederick}, {Fremling}, {Gal-Yam}, {Gezari}, {Giomi}, {Goldstein},
  {Golkhou}, {Goobar}, {Groom}, {Hacopians}, {Hale}, {Henning}, {Ho}, {Hover},
  {Howell}, {Hung}, {Huppenkothen}, {Imel}, {Ip}, {Ivezi{\'c}}, {Jackson},
  {Jones}, {Juric}, {Kasliwal}, {Kaspi}, {Kaye}, {Kelley}, {Kowalski},
  {Kramer}, {Kupfer}, {Landry}, {Laher}, {Lee}, {Lin}, {Lin}, {Lunnan},
  {Giomi}, {Mahabal}, {Mao}, {Miller}, {Monkewitz}, {Murphy}, {Ngeow},
  {Nordin}, {Nugent}, {Ofek}, {Patterson}, {Penprase}, {Porter}, {Rauch},
  {Rebbapragada}, {Reiley}, {Rigault}, {Rodriguez}, {van Roestel}, {Rusholme},
  {van Santen}, {Schulze}, {Shupe}, {Singer}, {Soumagnac}, {Stein}, {Surace},
  {Sollerman}, {Szkody}, {Taddia}, {Terek}, {Van Sistine}, {van Velzen},
  {Vestrand}, {Walters}, {Ward}, {Ye}, {Yu}, {Yan}, \&
  {Zolkower}}]{2019PASP..131a8002B}
{Bellm}, E.~C., {Kulkarni}, S.~R., {Graham}, M.~J., {et~al.} 2019, \pasp, 131,
  018002

\bibitem[{{Billot} {et~al.}(2010){Billot}, {Noriega-Crespo}, {Carey}, {Guieu},
  {Shenoy}, {Paladini}, \& {Latter}}]{2010ApJ...712..797B}
{Billot}, N., {Noriega-Crespo}, A., {Carey}, S., {et~al.} 2010, \apj, 712, 797

\bibitem[{{Cabrera-Vives} {et~al.}(2017){Cabrera-Vives}, {Reyes},
  {F{\"o}rster}, {Est{\'e}vez}, \& {Maureira}}]{2017ApJ...836...97C}
{Cabrera-Vives}, G., {Reyes}, I., {F{\"o}rster}, F., {Est{\'e}vez}, P.~A., \&
  {Maureira}, J.-C. 2017, \apj, 836, 97

\bibitem[{{Cambr{\'e}sy} {et~al.}(2013){Cambr{\'e}sy}, {Marton}, {Feher},
  {T{\'o}th}, \& {Schneider}}]{2013A&A...557A..29C}
{Cambr{\'e}sy}, L., {Marton}, G., {Feher}, O., {T{\'o}th}, L.~V., \&
  {Schneider}, N. 2013, \aap, 557, A29

\bibitem[{{Chavarr{\'\i}a} {et~al.}(2014){Chavarr{\'\i}a}, {Allen}, {Brunt},
  {Hora}, {Muench}, \& {Fazio}}]{2014MNRAS.439.3719C}
{Chavarr{\'\i}a}, L., {Allen}, L., {Brunt}, C., {et~al.} 2014, \mnras, 439,
  3719

\bibitem[{{Chavarr{\'\i}a} {et~al.}(2008){Chavarr{\'\i}a}, {Allen}, {Hora},
  {Brunt}, \& {Fazio}}]{2008ApJ...682..445C}
{Chavarr{\'\i}a}, L.~A., {Allen}, L.~E., {Hora}, J.~L., {Brunt}, C.~M., \&
  {Fazio}, G.~G. 2008, \apj, 682, 445

\bibitem[{{Chiang}(2023)}]{2023ApJ...958..118C}
{Chiang}, Y.-K. 2023, \apj, 958, 118

\bibitem[{{Christy} {et~al.}(2023){Christy}, {Jayasinghe}, {Stanek},
  {Kochanek}, {Thompson}, {Shappee}, {Holoien}, {Prieto}, {Dong}, \&
  {Giles}}]{2023MNRAS.519.5271C}
{Christy}, C.~T., {Jayasinghe}, T., {Stanek}, K.~Z., {et~al.} 2023, \mnras,
  519, 5271

\bibitem[{{Cody} \& {Hillenbrand}(2018)}]{2018AJ....156...71C}
{Cody}, A.~M. \& {Hillenbrand}, L.~A. 2018, \aj, 156, 71

\bibitem[{{Cody} {et~al.}(2014){Cody}, {Stauffer}, {Baglin}, {Micela},
  {Rebull}, {Flaccomio}, {Morales-Calder{\'o}n}, {Aigrain}, {Bouvier},
  {Hillenbrand}, {Gutermuth}, {Song}, {Turner}, {Alencar}, {Zwintz},
  {Plavchan}, {Carpenter}, {Findeisen}, {Carey}, {Terebey}, {Hartmann},
  {Calvet}, {Teixeira}, {Vrba}, {Wolk}, {Covey}, {Poppenhaeger}, {G{\"u}nther},
  {Forbrich}, {Whitney}, {Affer}, {Herbst}, {Hora}, {Barrado}, {Holtzman},
  {Marchis}, {Wood}, {Medeiros Guimar{\~a}es}, {Lillo Box}, {Gillen},
  {McQuillan}, {Espaillat}, {Allen}, {D'Alessio}, \&
  {Favata}}]{2014AJ....147...82C}
{Cody}, A.~M., {Stauffer}, J., {Baglin}, A., {et~al.} 2014, \aj, 147, 82

\bibitem[{{Comer{\'o}n} {et~al.}(2022){Comer{\'o}n}, {Djupvik}, \&
  {Schneider}}]{2022A&A...665A..76C}
{Comer{\'o}n}, F., {Djupvik}, A.~A., \& {Schneider}, N. 2022, \aap, 665, A76

\bibitem[{{Connelley} \& {Greene}(2010)}]{2010AJ....140.1214C}
{Connelley}, M.~S. \& {Greene}, T.~P. 2010, \aj, 140, 1214

\bibitem[{{Contreras Pe{\~n}a} {et~al.}(2020){Contreras Pe{\~n}a}, {Johnstone},
  {Baek}, {Herczeg}, {Mairs}, {Scholz}, {Lee}, \& {JCMT Transient
  Team}}]{2020MNRAS.495.3614C}
{Contreras Pe{\~n}a}, C., {Johnstone}, D., {Baek}, G., {et~al.} 2020, \mnras,
  495, 3614

\bibitem[{{Contreras Pe{\~n}a} {et~al.}(2025){Contreras Pe{\~n}a}, {Lee},
  {Herczeg}, {Johnstone}, {{\'A}brah{\'a}m}, {Antoniucci}, {Audard}, {Ashraf},
  {Baek}, {Garatti}, {Carvalho}, {Cieza}, {Cruz-Sa{\'e}nz de Miera},
  {Eisl{\"o}ffel}, {Froebrich}, {Giannini}, {Green}, {Ghosh}, {Guo},
  {Hillenbrand}, {Hodapp}, {Jheonn}, {Jose}, {Kim}, {K{\'o}sp{\'a}l}, {Lee},
  {Lucas}, {Magakian}, {Nagy}, {Naylor}, {Ninan}, {Peneva}, {Reipurth},
  {Scholz}, {Semkov}, {Sicilia-Aguilar}, {Singh}, {Siwak}, {Stecklum},
  {Szab{\'o}}, {Wolf}, \& {Yoon}}]{2025JKAS...58..209C}
{Contreras Pe{\~n}a}, C., {Lee}, J.-E., {Herczeg}, G., {et~al.} 2025, Journal
  of Korean Astronomical Society, 58, 209

\bibitem[{{Cornu} \& {Montillaud}(2021)}]{2021A&A...647A.116C}
{Cornu}, D. \& {Montillaud}, J. 2021, \aap, 647, A116

\bibitem[{{Cutri} {et~al.}(2013){Cutri}, {Wright}, {Conrow}, {Fowler},
  {Eisenhardt}, {Grillmair}, {Kirkpatrick}, {Masci}, {McCallon}, {Wheelock},
  {Fajardo-Acosta}, {Yan}, {Benford}, {Harbut}, {Jarrett}, {Lake}, {Leisawitz},
  {Ressler}, {Stanford}, {Tsai}, {Liu}, {Helou}, {Mainzer}, {Gettings},
  {Gonzalez}, {Hoffman}, {Marsh}, {Padgett}, {Skrutskie}, {Beck}, {Papin}, \&
  {Wittman}}]{2013wise.rept....1C}
{Cutri}, R.~M., {Wright}, E.~L., {Conrow}, T., {et~al.} 2013, {Explanatory
  Supplement to the AllWISE Data Release Products}, Explanatory Supplement to
  the AllWISE Data Release Products, by R. M. Cutri et al.

\bibitem[{{Delfini} {et~al.}(2025){Delfini}, {Vioque}, {Ribas}, \&
  {Hodgkin}}]{2025A&A...699A.145D}
{Delfini}, L., {Vioque}, M., {Ribas}, {\'A}., \& {Hodgkin}, S. 2025, \aap, 699,
  A145

\bibitem[{{Drake} {et~al.}(2013){Drake}, {Catelan}, {Djorgovski}, {Torrealba},
  {Graham}, {Mahabal}, {Prieto}, {Donalek}, {Williams}, {Larson},
  {Christensen}, \& {Beshore}}]{2013ApJ...765..154D}
{Drake}, A.~J., {Catelan}, M., {Djorgovski}, S.~G., {et~al.} 2013, \apj, 765,
  154

\bibitem[{{Dunham} {et~al.}(2015){Dunham}, {Allen}, {Evans},
  {Broekhoven-Fiene}, {Cieza}, {Di Francesco}, {Gutermuth}, {Harvey},
  {Hatchell}, {Heiderman}, {Huard}, {Johnstone}, {Kirk}, {Matthews}, {Miller},
  {Peterson}, \& {Young}}]{2015ApJS..220...11D}
{Dunham}, M.~M., {Allen}, L.~E., {Evans}, Neal~J., I., {et~al.} 2015, \apjs,
  220, 11

\bibitem[{{Dunham} {et~al.}(2014){Dunham}, {Arce}, {Mardones}, {Lee},
  {Matthews}, {Stutz}, \& {Williams}}]{2014ApJ...783...29D}
{Dunham}, M.~M., {Arce}, H.~G., {Mardones}, D., {et~al.} 2014, \apj, 783, 29

\bibitem[{{Edenhofer} {et~al.}(2024){Edenhofer}, {Zucker}, {Frank}, {Saydjari},
  {Speagle}, {Finkbeiner}, \& {En{\ss}lin}}]{2024A&A...685A..82E}
{Edenhofer}, G., {Zucker}, C., {Frank}, P., {et~al.} 2024, \aap, 685, A82

\bibitem[{{Evans} {et~al.}(2009){Evans}, {Dunham}, {J{\o}rgensen}, {Enoch},
  {Mer{\'\i}n}, {van Dishoeck}, {Alcal{\'a}}, {Myers}, {Stapelfeldt}, {Huard},
  {Allen}, {Harvey}, {van Kempen}, {Blake}, {Koerner}, {Mundy}, {Padgett}, \&
  {Sargent}}]{2009ApJS..181..321E}
{Evans}, Neal~J., I., {Dunham}, M.~M., {J{\o}rgensen}, J.~K., {et~al.} 2009,
  \apjs, 181, 321

\bibitem[{{Eyer} {et~al.}(2023){Eyer}, {Audard}, {Holl}, {Rimoldini},
  {Carnerero}, {Clementini}, {De Ridder}, {Distefano}, {Evans}, {Gavras},
  {Gomel}, {Lebzelter}, {Marton}, {Mowlavi}, {Panahi}, {Ripepi}, {Wyrzykowski},
  {Nienartowicz}, {Jevardat de Fombelle}, {Lecoeur-Taibi}, {Rohrbasser},
  {Riello}, {Garc{\'\i}a-Lario}, {Lanzafame}, {Mazeh}, {Raiteri}, {Zucker},
  {{\'A}brah{\'a}m}, {Aerts}, {Aguado}, {Anderson}, {Bashi}, {Binnenfeld},
  {Faigler}, {Garofalo}, {Karbevska}, {K{\'o}sp{\'a}l}, {Kruszy{\'n}ska},
  {Kun}, {Lanza}, {Leccia}, {Marconi}, {Messina}, {Molinaro}, {Moln{\'a}r},
  {Muraveva}, {Musella}, {Nagy}, {Pagano}, {Palaversa}, {Plachy}, {Pr{\v{s}}a},
  {Rybicki}, {Shahaf}, {Szabados}, {Szegedi-Elek}, {Trabucchi}, {Barblan},
  {Grenon}, {Roelens}, \& {S{\"u}veges}}]{2023A&A...674A..13E}
{Eyer}, L., {Audard}, M., {Holl}, B., {et~al.} 2023, \aap, 674, A13

\bibitem[{{Fang} {et~al.}(2020){Fang}, {Hillenbrand}, {Kim}, {Findeisen},
  {Herczeg}, {Carpenter}, {Rebull}, \& {Wang}}]{2020ApJ...904..146F}
{Fang}, M., {Hillenbrand}, L.~A., {Kim}, J.~S., {et~al.} 2020, \apj, 904, 146

\bibitem[{{Fischer} {et~al.}(2016){Fischer}, {Padgett}, {Stapelfeldt}, \&
  {Sewi{\l}o}}]{2016ApJ...827...96F}
{Fischer}, W.~J., {Padgett}, D.~L., {Stapelfeldt}, K.~L., \& {Sewi{\l}o}, M.
  2016, \apj, 827, 96

\bibitem[{{Flaherty} {et~al.}(2013){Flaherty}, {Muzerolle}, {Rieke},
  {Gutermuth}, {Balog}, {Herbst}, \& {Megeath}}]{2013AJ....145...66F}
{Flaherty}, K.~M., {Muzerolle}, J., {Rieke}, G., {et~al.} 2013, \aj, 145, 66

\bibitem[{{Flesch}(2015)}]{2015PASA...32...10F}
{Flesch}, E.~W. 2015, \pasa, 32, e010

\bibitem[{{Fouesneau} {et~al.}(2023){Fouesneau}, {Fr{\'e}mat}, {Andrae},
  {Korn}, {Soubiran}, {Kordopatis}, {Vallenari}, {Heiter}, {Creevey}, {Sarro},
  {de Laverny}, {Lanzafame}, {Lobel}, {Sordo}, {Rybizki}, {Slezak},
  {{\'A}lvarez}, {Drimmel}, {Garabato}, {Delchambre}, {Bailer-Jones},
  {Hatzidimitriou}, {Lorca}, {Le Fustec}, {Pailler}, {Mary}, {Robin},
  {Utrilla}, {Abreu Aramburu}, {Bakker}, {Bellas-Velidis}, {Bijaoui}, {Blomme},
  {Bouret}, {Brouillet}, {Brugaletta}, {Burlacu}, {Carballo}, {Casamiquela},
  {Chaoul}, {Chiavassa}, {Contursi}, {Cooper}, {Dafonte}, {Demouchy},
  {Dharmawardena}, {Garc{\'\i}a-Lario}, {Garc{\'\i}a-Torres}, {Gomez},
  {Gonz{\'a}lez-Santamar{\'\i}a}, {Jean-Antoine Piccolo}, {Kontizas},
  {Lebreton}, {Licata}, {Lindstr{\o}m}, {Livanou}, {Magdaleno Romeo},
  {Manteiga}, {Marocco}, {Martayan}, {Marshall}, {Nicolas}, {Ordenovic},
  {Palicio}, {Pallas-Quintela}, {Pichon}, {Poggio}, {Recio-Blanco}, {Riclet},
  {Santove{\~n}a}, {Schultheis}, {Segol}, {Silvelo}, {Smart}, {S{\"u}veges},
  {Th{\'e}venin}, {Torralba Elipe}, {Ulla}, {van Dillen}, {Zhao}, \&
  {Zorec}}]{2023A&A...674A..28F}
{Fouesneau}, M., {Fr{\'e}mat}, Y., {Andrae}, R., {et~al.} 2023, \aap, 674, A28

\bibitem[{{Fu} {et~al.}(2021){Fu}, {Wu}, {Yang}, {Brown}, {Feng}, {Ma}, \&
  {Li}}]{2021ApJS..254....6F}
{Fu}, Y., {Wu}, X.-B., {Yang}, Q., {et~al.} 2021, \apjs, 254, 6

\bibitem[{{Gaia Collaboration} {et~al.}(2023{\natexlab{a}}){Gaia
  Collaboration}, {Bailer-Jones}, {Teyssier}, {Delchambre}, {Ducourant},
  {Garabato}, {Hatzidimitriou}, {Klioner}, {Rimoldini}, {Bellas-Velidis},
  {Carballo}, {Carnerero}, {Diener}, {Fouesneau}, {Galluccio}, {Gavras},
  {Krone-Martins}, {Raiteri}, {Teixeira}, {Brown}, {Vallenari}, {Prusti}, {de
  Bruijne}, {Arenou}, {Babusiaux}, {Biermann}, {Creevey}, {Evans}, {Eyer},
  {Guerra}, {Hutton}, {Jordi}, {Lammers}, {Lindegren}, {Luri}, {Mignard},
  {Panem}, {Pourbaix}, {Randich}, {Sartoretti}, {Soubiran}, {Tanga}, {Walton},
  {Bastian}, {Drimmel}, {Jansen}, {Katz}, {Lattanzi}, {van Leeuwen}, {Bakker},
  {Cacciari}, {Casta{\~n}eda}, {De Angeli}, {Fabricius}, {Fr{\'e}mat},
  {Guerrier}, {Heiter}, {Masana}, {Messineo}, {Mowlavi}, {Nicolas},
  {Nienartowicz}, {Pailler}, {Panuzzo}, {Riclet}, {Roux}, {Seabroke}, {Sordo},
  {Th{\'e}venin}, {Gracia-Abril}, {Portell}, {Altmann}, {Andrae}, {Audard},
  {Benson}, {Berthier}, {Blomme}, {Burgess}, {Busonero}, {Busso},
  {C{\'a}novas}, {Carry}, {Cellino}, {Cheek}, {Clementini}, {Damerdji},
  {Davidson}, {de Teodoro}, {Nu{\~n}ez Campos}, {Dell'Oro}, {Esquej},
  {Fern{\'a}ndez-Hern{\'a}ndez}, {Fraile}, {Garc{\'\i}a-Lario}, {Gosset},
  {Haigron}, {Halbwachs}, {Hambly}, {Harrison}, {Hern{\'a}ndez}, {Hestroffer},
  {Hodgkin}, {Holl}, {Jan{\ss}en}, {Jevardat de Fombelle}, {Jordan},
  {Lanzafame}, {L{\"o}ffler}, {Marchal}, {Marrese}, {Moitinho}, {Muinonen},
  {Osborne}, {Pancino}, {Pauwels}, {Recio-Blanco}, {Reyl{\'e}}, {Riello},
  {Roegiers}, {Rybizki}, {Sarro}, {Siopis}, {Smith}, {Sozzetti}, {Utrilla},
  {van Leeuwen}, {Abbas}, {{\'A}brah{\'a}m}, {Abreu Aramburu}, {Aerts},
  {Aguado}, {Ajaj}, {Aldea-Montero}, {Altavilla}, {{\'A}lvarez}, {Alves},
  {Anderson}, {Anglada Varela}, {Antoja}, {Baines}, {Baker},
  {Balaguer-N{\'u}{\~n}ez}, {Balbinot}, {Balog}, {Barache}, {Barbato},
  {Barros}, {Barstow}, {Bartolom{\'e}}, {Bassilana}, {Bauchet}, {Becciani},
  {Bellazzini}, {Berihuete}, {Bernet}, {Bertone}, {Bianchi}, {Binnenfeld},
  {Blanco-Cuaresma}, {Boch}, {Bombrun}, {Bossini}, {Bouquillon}, {Bragaglia},
  {Bramante}, {Breedt}, {Bressan}, {Brouillet}, {Brugaletta}, {Bucciarelli},
  {Burlacu}, {Butkevich}, {Buzzi}, {Caffau}, {Cancelliere}, {Cantat-Gaudin},
  {Carlucci}, {Carrasco}, {Casamiquela}, {Castellani}, {Castro-Ginard},
  {Chaoul}, {Charlot}, {Chemin}, {Chiaramida}, {Chiavassa}, {Chornay},
  {Comoretto}, {Contursi}, {Cooper}, {Cornez}, {Cowell}, {Crifo}, {Cropper},
  {Crosta}, {Crowley}, {Dafonte}, {Dapergolas}, {David}, {de Laverny}, {De
  Luise}, {De March}, {De Ridder}, {de Souza}, {de Torres}, {del Peloso}, {del
  Pozo}, {Delbo}, {Delgado}, {Delisle}, {Demouchy}, {Dharmawardena}, {Diakite},
  {Distefano}, {Dolding}, {Enke}, {Fabre}, {Fabrizio}, {Faigler}, {Fedorets},
  {Fernique}, {Figueras}, {Fournier}, {Fouron}, {Fragkoudi}, {Gai},
  {Garcia-Gutierrez}, {Garcia-Reinaldos}, {Garc{\'\i}a-Torres}, {Garofalo},
  {Gavel}, {Gerlach}, {Geyer}, {Giacobbe}, {Gilmore}, {Girona}, {Giuffrida},
  {Gomel}, {Gomez}, {Gonz{\'a}lez-N{\'u}{\~n}ez},
  {Gonz{\'a}lez-Santamar{\'\i}a}, {Gonz{\'a}lez-Vidal}, {Granvik}, {Guillout},
  {Guiraud}, {Guti{\'e}rrez-S{\'a}nchez}, {Guy}, {Hauser}, {Haywood}, {Helmer},
  {Helmi}, {Sarmiento}, {Hidalgo}, {Hilger}, {H{\l}adczuk}, {Hobbs}, {Holland},
  {Huckle}, {Jardine}, {Jasniewicz}, {Jean-Antoine Piccolo},
  {Jim{\'e}nez-Arranz}, {Juaristi Campillo}, {Julbe}, {Karbevska}, {Kervella},
  {Khanna}, {Kontizas}, {Kordopatis}, {Korn}, {K{\'o}sp{\'a}l},
  {Kostrzewa-Rutkowska}, {Kruszy{\'n}ska}, {Kun}, {Laizeau}, {Lambert},
  {Lanza}, {Lasne}, {Le Campion}, {Lebreton}, {Lebzelter}, {Leccia}, {Leclerc},
  {Lecoeur-Taibi}, {Liao}, {Licata}, {Lindstr{\o}m}, {Lister}, {Livanou},
  {Lobel}, {Lorca}, {Loup}, {Madrero Pardo}, {Magdaleno Romeo}, {Managau},
  {Mann}, {Manteiga}, {Marchant}, {Marconi}, {Marcos}, {Marcos Santos},
  {Mar{\'\i}n Pina}, {Marinoni}, {Marocco}, {Marshall}, {Martin Polo},
  {Mart{\'\i}n-Fleitas}, {Marton}, {Mary}, {Masip}, {Massari},
  {Mastrobuono-Battisti}, {Mazeh}, {McMillan}, {Messina}, {Michalik}, {Millar},
  {Mints}, {Molina}, {Molinaro}, {Moln{\'a}r}, {Monari}, {Mongui{\'o}},
  {Montegriffo}, {Montero}, {Mor}, {Mora}, {Morbidelli}, {Morel}, {Morris},
  {Muraveva}, {Murphy}, {Musella}, {Nagy}, {Noval}, {Oca{\~n}a}, {Ogden},
  {Ordenovic}, {Osinde}, {Pagani}, {Pagano}, {Palaversa}, {Palicio},
  {Pallas-Quintela}, {Panahi}, {Payne-Wardenaar}, {Pe{\~n}alosa Esteller},
  {Penttil{\"a}}, {Pichon}, {Piersimoni}, {Pineau}, {Plachy}, {Plum}, {Poggio},
  {Pr{\v{s}}a}, {Pulone}, {Racero}, {Ragaini}, {Rainer}, {Ramos},
  {Ramos-Lerate}, {Re Fiorentin}, {Regibo}, {Richards}, {Rios Diaz}, {Ripepi},
  {Riva}, {Rix}, {Rixon}, {Robichon}, {Robin}, {Robin}, {Roelens}, {Rogues},
  {Rohrbasser}, {Romero-G{\'o}mez}, {Rowell}, {Royer}, {Ruz Mieres}, {Rybicki},
  {Sadowski}, {S{\'a}ez N{\'u}{\~n}ez}, {Sagrist{\`a} Sell{\'e}s}, {Sahlmann},
  {Salguero}, {Samaras}, {Sanchez Gimenez}, {Sanna}, {Santove{\~n}a},
  {Sarasso}, {Schultheis}, {Sciacca}, {Segol}, {Segovia}, {S{\'e}gransan},
  {Semeux}, {Shahaf}, {Siddiqui}, {Siebert}, {Siltala}, {Silvelo}, {Slezak},
  {Slezak}, {Smart}, {Snaith}, {Solano}, {Solitro}, {Souami}, {Souchay},
  {Spagna}, {Spina}, {Spoto}, {Steele}, {Steidelm{\"u}ller}, {Stephenson},
  {S{\"u}veges}, {Surdej}, {Szabados}, {Szegedi-Elek}, {Taris}, {Taylor},
  {Tolomei}, {Tonello}, {Torra}, {Torra}, {Torralba Elipe}, {Trabucchi},
  {Tsounis}, {Turon}, {Ulla}, {Unger}, {Vaillant}, {van Dillen}, {van Reeven},
  {Vanel}, {Vecchiato}, {Viala}, {Vicente}, {Voutsinas}, {Weiler}, {Wevers},
  {Wyrzykowski}, {Yoldas}, {Yvard}, {Zhao}, {Zorec}, {Zucker}, \&
  {Zwitter}}]{2023A&A...674A..41G}
{Gaia Collaboration}, {Bailer-Jones}, C.~A.~L., {Teyssier}, D., {et~al.}
  2023{\natexlab{a}}, \aap, 674, A41

\bibitem[{{Gaia Collaboration} {et~al.}(2016){Gaia Collaboration}, {Prusti},
  {de Bruijne}, {Brown}, {Vallenari}, {Babusiaux}, {Bailer-Jones}, {Bastian},
  {Biermann}, {Evans}, {Eyer}, {Jansen}, {Jordi}, {Klioner}, {Lammers},
  {Lindegren}, {Luri}, {Mignard}, {Milligan}, {Panem}, {Poinsignon},
  {Pourbaix}, {Randich}, {Sarri}, {Sartoretti}, {Siddiqui}, {Soubiran},
  {Valette}, {van Leeuwen}, {Walton}, {Aerts}, {Arenou}, {Cropper}, {Drimmel},
  {H{\o}g}, {Katz}, {Lattanzi}, {O'Mullane}, {Grebel}, {Holland}, {Huc},
  {Passot}, {Bramante}, {Cacciari}, {Casta{\~n}eda}, {Chaoul}, {Cheek}, {De
  Angeli}, {Fabricius}, {Guerra}, {Hern{\'a}ndez}, {Jean-Antoine-Piccolo},
  {Masana}, {Messineo}, {Mowlavi}, {Nienartowicz}, {Ord{\'o}{\~n}ez-Blanco},
  {Panuzzo}, {Portell}, {Richards}, {Riello}, {Seabroke}, {Tanga},
  {Th{\'e}venin}, {Torra}, {Els}, {Gracia-Abril}, {Comoretto},
  {Garcia-Reinaldos}, {Lock}, {Mercier}, {Altmann}, {Andrae}, {Astraatmadja},
  {Bellas-Velidis}, {Benson}, {Berthier}, {Blomme}, {Busso}, {Carry},
  {Cellino}, {Clementini}, {Cowell}, {Creevey}, {Cuypers}, {Davidson}, {De
  Ridder}, {de Torres}, {Delchambre}, {Dell'Oro}, {Ducourant}, {Fr{\'e}mat},
  {Garc{\'\i}a-Torres}, {Gosset}, {Halbwachs}, {Hambly}, {Harrison}, {Hauser},
  {Hestroffer}, {Hodgkin}, {Huckle}, {Hutton}, {Jasniewicz}, {Jordan},
  {Kontizas}, {Korn}, {Lanzafame}, {Manteiga}, {Moitinho}, {Muinonen},
  {Osinde}, {Pancino}, {Pauwels}, {Petit}, {Recio-Blanco}, {Robin}, {Sarro},
  {Siopis}, {Smith}, {Smith}, {Sozzetti}, {Thuillot}, {van Reeven}, {Viala},
  {Abbas}, {Abreu Aramburu}, {Accart}, {Aguado}, {Allan}, {Allasia},
  {Altavilla}, {{\'A}lvarez}, {Alves}, {Anderson}, {Andrei}, {Anglada Varela},
  {Antiche}, {Antoja}, {Ant{\'o}n}, {Arcay}, {Atzei}, {Ayache}, {Bach},
  {Baker}, {Balaguer-N{\'u}{\~n}ez}, {Barache}, {Barata}, {Barbier}, {Barblan},
  {Baroni}, {Barrado y Navascu{\'e}s}, {Barros}, {Barstow}, {Becciani},
  {Bellazzini}, {Bellei}, {Bello Garc{\'\i}a}, {Belokurov}, {Bendjoya},
  {Berihuete}, {Bianchi}, {Bienaym{\'e}}, {Billebaud}, {Blagorodnova},
  {Blanco-Cuaresma}, {Boch}, {Bombrun}, {Borrachero}, {Bouquillon}, {Bourda},
  {Bouy}, {Bragaglia}, {Breddels}, {Brouillet}, {Br{\"u}semeister},
  {Bucciarelli}, {Budnik}, {Burgess}, {Burgon}, {Burlacu}, {Busonero}, {Buzzi},
  {Caffau}, {Cambras}, {Campbell}, {Cancelliere}, {Cantat-Gaudin}, {Carlucci},
  {Carrasco}, {Castellani}, {Charlot}, {Charnas}, {Charvet}, {Chassat},
  {Chiavassa}, {Clotet}, {Cocozza}, {Collins}, {Collins}, \&
  {Costigan}}]{2016A&A...595A...1G}
{Gaia Collaboration}, {Prusti}, T., {de Bruijne}, J.~H.~J., {et~al.} 2016,
  \aap, 595, A1

\bibitem[{{Gaia Collaboration} {et~al.}(2023{\natexlab{b}}){Gaia
  Collaboration}, {Vallenari}, {Brown}, {Prusti}, {de Bruijne}, {Arenou},
  {Babusiaux}, {Biermann}, {Creevey}, {Ducourant}, {Evans}, {Eyer}, {Guerra},
  {Hutton}, {Jordi}, {Klioner}, {Lammers}, {Lindegren}, {Luri}, {Mignard},
  {Panem}, {Pourbaix}, {Randich}, {Sartoretti}, {Soubiran}, {Tanga}, {Walton},
  {Bailer-Jones}, {Bastian}, {Drimmel}, {Jansen}, {Katz}, {Lattanzi}, {van
  Leeuwen}, {Bakker}, {Cacciari}, {Casta{\~n}eda}, {De Angeli}, {Fabricius},
  {Fouesneau}, {Fr{\'e}mat}, {Galluccio}, {Guerrier}, {Heiter}, {Masana},
  {Messineo}, {Mowlavi}, {Nicolas}, {Nienartowicz}, {Pailler}, {Panuzzo},
  {Riclet}, {Roux}, {Seabroke}, {Sordo}, {Th{\'e}venin}, {Gracia-Abril},
  {Portell}, {Teyssier}, {Altmann}, {Andrae}, {Audard}, {Bellas-Velidis},
  {Benson}, {Berthier}, {Blomme}, {Burgess}, {Busonero}, {Busso},
  {C{\'a}novas}, {Carry}, {Cellino}, {Cheek}, {Clementini}, {Damerdji},
  {Davidson}, {de Teodoro}, {Nu{\~n}ez Campos}, {Delchambre}, {Dell'Oro},
  {Esquej}, {Fern{\'a}ndez-Hern{\'a}ndez}, {Fraile}, {Garabato},
  {Garc{\'\i}a-Lario}, {Gosset}, {Haigron}, {Halbwachs}, {Hambly}, {Harrison},
  {Hern{\'a}ndez}, {Hestroffer}, {Hodgkin}, {Holl}, {Jan{\ss}en}, {Jevardat de
  Fombelle}, {Jordan}, {Krone-Martins}, {Lanzafame}, {L{\"o}ffler}, {Marchal},
  {Marrese}, {Moitinho}, {Muinonen}, {Osborne}, {Pancino}, {Pauwels},
  {Recio-Blanco}, {Reyl{\'e}}, {Riello}, {Rimoldini}, {Roegiers}, {Rybizki},
  {Sarro}, {Siopis}, {Smith}, {Sozzetti}, {Utrilla}, {van Leeuwen}, {Abbas},
  {{\'A}brah{\'a}m}, {Abreu Aramburu}, {Aerts}, {Aguado}, {Ajaj},
  {Aldea-Montero}, {Altavilla}, {{\'A}lvarez}, {Alves}, {Anders}, {Anderson},
  {Anglada Varela}, {Antoja}, {Baines}, {Baker}, {Balaguer-N{\'u}{\~n}ez},
  {Balbinot}, {Balog}, {Barache}, {Barbato}, {Barros}, {Barstow},
  {Bartolom{\'e}}, {Bassilana}, {Bauchet}, {Becciani}, {Bellazzini},
  {Berihuete}, {Bernet}, {Bertone}, {Bianchi}, {Binnenfeld}, {Blanco-Cuaresma},
  {Blazere}, {Boch}, {Bombrun}, {Bossini}, {Bouquillon}, {Bragaglia},
  {Bramante}, {Breedt}, {Bressan}, {Brouillet}, {Brugaletta}, {Bucciarelli},
  {Burlacu}, {Butkevich}, {Buzzi}, {Caffau}, {Cancelliere}, {Cantat-Gaudin},
  {Carballo}, {Carlucci}, {Carnerero}, {Carrasco}, {Casamiquela}, {Castellani},
  {Castro-Ginard}, {Chaoul}, {Charlot}, {Chemin}, {Chiaramida}, {Chiavassa},
  {Chornay}, {Comoretto}, {Contursi}, {Cooper}, {Cornez}, {Cowell}, {Crifo},
  {Cropper}, {Crosta}, {Crowley}, {Dafonte}, {Dapergolas}, {David}, {David},
  {de Laverny}, {De Luise}, {De March}, {De Ridder}, {de Souza}, {de Torres},
  {del Peloso}, {del Pozo}, {Delbo}, {Delgado}, {Delisle}, {Demouchy},
  {Dharmawardena}, {Di Matteo}, {Diakite}, {Diener}, {Distefano}, {Dolding},
  {Edvardsson}, {Enke}, {Fabre}, {Fabrizio}, {Faigler}, {Fedorets}, {Fernique},
  {Fienga}, {Figueras}, {Fournier}, {Fouron}, {Fragkoudi}, {Gai},
  {Garcia-Gutierrez}, {Garcia-Reinaldos}, {Garc{\'\i}a-Torres}, {Garofalo},
  {Gavel}, {Gavras}, {Gerlach}, {Geyer}, {Giacobbe}, {Gilmore}, {Girona},
  {Giuffrida}, {Gomel}, {Gomez}, {Gonz{\'a}lez-N{\'u}{\~n}ez},
  {Gonz{\'a}lez-Santamar{\'\i}a}, {Gonz{\'a}lez-Vidal}, {Granvik}, {Guillout},
  {Guiraud}, {Guti{\'e}rrez-S{\'a}nchez}, {Guy}, {Hatzidimitriou}, {Hauser},
  {Haywood}, {Helmer}, {Helmi}, {Sarmiento}, {Hidalgo}, {Hilger},
  {H{\l}adczuk}, {Hobbs}, {Holland}, {Huckle}, {Jardine}, {Jasniewicz},
  {Jean-Antoine Piccolo}, {Jim{\'e}nez-Arranz}, {Jorissen}, {Juaristi
  Campillo}, {Julbe}, {Karbevska}, {Kervella}, {Khanna}, {Kontizas},
  {Kordopatis}, {Korn}, {K{\'o}sp{\'a}l}, {Kostrzewa-Rutkowska},
  {Kruszy{\'n}ska}, {Kun}, {Laizeau}, {Lambert}, {Lanza}, {Lasne}, {Le
  Campion}, {Lebreton}, {Lebzelter}, {Leccia}, {Leclerc}, {Lecoeur-Taibi},
  {Liao}, {Licata}, {Lindstr{\o}m}, {Lister}, {Livanou}, {Lobel}, {Lorca},
  {Loup}, {Madrero Pardo}, {Magdaleno Romeo}, {Managau}, {Mann}, {Manteiga},
  {Marchant}, {Marconi}, {Marcos}, {Marcos Santos}, {Mar{\'\i}n Pina},
  {Marinoni}, {Marocco}, {Marshall}, {Martin Polo}, {Mart{\'\i}n-Fleitas},
  {Marton}, {Mary}, {Masip}, {Massari}, {Mastrobuono-Battisti}, {Mazeh},
  {McMillan}, {Messina}, {Michalik}, {Millar}, {Mints}, {Molina}, {Molinaro},
  {Moln{\'a}r}, {Monari}, {Mongui{\'o}}, {Montegriffo}, {Montero}, {Mor},
  {Mora}, {Morbidelli}, {Morel}, {Morris}, {Muraveva}, {Murphy}, {Musella},
  {Nagy}, {Noval}, {Oca{\~n}a}, {Ogden}, {Ordenovic}, {Osinde}, {Pagani},
  {Pagano}, {Palaversa}, {Palicio}, {Pallas-Quintela}, {Panahi},
  {Payne-Wardenaar}, {Pe{\~n}alosa Esteller}, {Penttil{\"a}}, {Pichon},
  {Piersimoni}, {Pineau}, {Plachy}, {Plum}, {Poggio}, {Pr{\v{s}}a}, {Pulone},
  {Racero}, {Ragaini}, {Rainer}, {Raiteri}, {Rambaux}, {Ramos}, {Ramos-Lerate},
  {Re Fiorentin}, {Regibo}, {Richards}, {Rios Diaz}, {Ripepi}, {Riva}, {Rix},
  {Rixon}, {Robichon}, {Robin}, {Robin}, {Roelens}, {Rogues}, {Rohrbasser},
  {Romero-G{\'o}mez}, {Rowell}, {Royer}, {Ruz Mieres}, {Rybicki}, {Sadowski},
  {S{\'a}ez N{\'u}{\~n}ez}, {Sagrist{\`a} Sell{\'e}s}, {Sahlmann}, {Salguero},
  {Samaras}, {Sanchez Gimenez}, {Sanna}, {Santove{\~n}a}, {Sarasso},
  {Schultheis}, {Sciacca}, {Segol}, {Segovia}, {S{\'e}gransan}, {Semeux},
  {Shahaf}, {Siddiqui}, {Siebert}, {Siltala}, {Silvelo}, {Slezak}, {Slezak},
  {Smart}, {Snaith}, {Solano}, {Solitro}, {Souami}, {Souchay}, {Spagna},
  {Spina}, {Spoto}, {Steele}, {Steidelm{\"u}ller}, {Stephenson}, {S{\"u}veges},
  {Surdej}, {Szabados}, {Szegedi-Elek}, {Taris}, {Taylor}, {Teixeira},
  {Tolomei}, {Tonello}, {Torra}, {Torra}, {Torralba Elipe}, {Trabucchi},
  {Tsounis}, {Turon}, {Ulla}, {Unger}, {Vaillant}, {van Dillen}, {van Reeven},
  {Vanel}, {Vecchiato}, {Viala}, {Vicente}, {Voutsinas}, {Weiler}, {Wevers},
  {Wyrzykowski}, {Yoldas}, {Yvard}, {Zhao}, {Zorec}, {Zucker}, \&
  {Zwitter}}]{2023A&A...674A...1G}
{Gaia Collaboration}, {Vallenari}, A., {Brown}, A.~G.~A., {et~al.}
  2023{\natexlab{b}}, \aap, 674, A1

\bibitem[{{Gama} {et~al.}(2016){Gama}, {Lepine}, {Mendoza}, {Wu}, \&
  {Yuan}}]{2016ApJ...830...57G}
{Gama}, D.~R.~G., {Lepine}, J.~R.~D., {Mendoza}, E., {Wu}, Y., \& {Yuan}, J.
  2016, \apj, 830, 57

\bibitem[{{Gezer} {et~al.}(2025){Gezer}, {Marton}, {Roquette}, {Audard},
  {Hernandez}, {Madar{\'a}sz}, \& {Dionatos}}]{2025A&A...696A.196G}
{Gezer}, I., {Marton}, G., {Roquette}, J., {et~al.} 2025, \aap, 696, A196

\bibitem[{{Girard} {et~al.}(2011){Girard}, {van Altena}, {Zacharias}, {Vieira},
  {Casetti-Dinescu}, {Castillo}, {Herrera}, {Lee}, {Beers}, {Monet}, \&
  {L{\'o}pez}}]{2011AJ....142...15G}
{Girard}, T.~M., {van Altena}, W.~F., {Zacharias}, N., {et~al.} 2011, \aj, 142,
  15

\bibitem[{Glorot {et~al.}(2011)Glorot, Bordes, \& Bengio}]{glorot2011relu}
Glorot, X., Bordes, A., \& Bengio, Y. 2011, in Proceedings of the 14th
  International Conference on Artificial Intelligence and Statistics (AISTATS),
  315--323

\bibitem[{G{\'o}rski {et~al.}(2005)G{\'o}rski, Hivon, Banday, Wandelt, Hansen,
  Reinecke, \& Bartelmann}]{Gorski2005}
G{\'o}rski, K.~M., Hivon, E., Banday, A.~J., {et~al.} 2005, \apj, 622, 759

\bibitem[{{Green}(2018)}]{2018JOSS....3..695M}
{Green}, G. 2018, The Journal of Open Source Software, 3, 695

\bibitem[{{Gro{\ss}schedl} {et~al.}(2019){Gro{\ss}schedl}, {Alves}, {Teixeira},
  {Bouy}, {Forbrich}, {Lada}, {Meingast}, {Hacar}, {Ascenso}, {Ackerl},
  {Hasenberger}, {K{\"o}hler}, {Kubiak}, {Larreina}, {Linhardt}, {Lombardi}, \&
  {M{\"o}ller}}]{Grossschedl2019}
{Gro{\ss}schedl}, J.~E., {Alves}, J., {Teixeira}, P.~S., {et~al.} 2019, \aap,
  622, A149

\bibitem[{{G{\"u}nther} {et~al.}(2014){G{\"u}nther}, {Cody}, {Covey},
  {Hillenbrand}, {Plavchan}, {Poppenhaeger}, {Rebull}, {Stauffer}, {Wolk},
  {Allen}, {Bayo}, {Gutermuth}, {Hora}, {Meng}, {Morales-Calder{\'o}n},
  {Parks}, \& {Song}}]{2014AJ....148..122G}
{G{\"u}nther}, H.~M., {Cody}, A.~M., {Covey}, K.~R., {et~al.} 2014, \aj, 148,
  122

\bibitem[{{Gutermuth} {et~al.}(2009){Gutermuth}, {Megeath}, {Myers}, {Allen},
  {Pipher}, \& {Fazio}}]{2009ApJS..184...18G}
{Gutermuth}, R.~A., {Megeath}, S.~T., {Myers}, P.~C., {et~al.} 2009, \apjs,
  184, 18

\bibitem[{{Gutermuth} {et~al.}(2008){Gutermuth}, {Myers}, {Megeath}, {Allen},
  {Pipher}, {Muzerolle}, {Porras}, {Winston}, \& {Fazio}}]{2008ApJ...674..336G}
{Gutermuth}, R.~A., {Myers}, P.~C., {Megeath}, S.~T., {et~al.} 2008, \apj, 674,
  336

\bibitem[{{Hartmann} {et~al.}(1998){Hartmann}, {Calvet}, {Gullbring}, \&
  {D'Alessio}}]{1998ApJ...495..385H}
{Hartmann}, L., {Calvet}, N., {Gullbring}, E., \& {D'Alessio}, P. 1998, \apj,
  495, 385

\bibitem[{He {et~al.}(2015)He, Zhang, Ren, \& Sun}]{he2015init}
He, K., Zhang, X., Ren, S., \& Sun, J. 2015, in Proceedings of the IEEE/CVF
  International Conference on Computer Vision (ICCV), 1026--1034

\bibitem[{He {et~al.}(2016)He, Zhang, Ren, \& Sun}]{he2016resnet}
He, K., Zhang, X., Ren, S., \& Sun, J. 2016, in Proceedings of the IEEE/CVF
  Conference on Computer Vision and Pattern Recognition (CVPR), 770--778

\bibitem[{Hendrycks \& Gimpel(2016)}]{hendrycks2016silu}
Hendrycks, D. \& Gimpel, K. 2016, arXiv preprint arXiv:1606.08415

\bibitem[{{Hernandez} {et~al.}(2026){Hernandez}, {Dionatos}, {Audard},
  {Marton}, {Roquette}, {Gezer}, {Madar{\'a}sz}, \&
  {Polsterer}}]{2026A&A...707A..23H}
{Hernandez}, D., {Dionatos}, O., {Audard}, M., {et~al.} 2026, \aap, 707, A23

\bibitem[{{Hezaveh} {et~al.}(2017){Hezaveh}, {Perreault Levasseur}, \&
  {Marshall}}]{2017Natur.548..555H}
{Hezaveh}, Y.~D., {Perreault Levasseur}, L., \& {Marshall}, P.~J. 2017, \nat,
  548, 555

\bibitem[{{Hillenbrand} {et~al.}(2022){Hillenbrand}, {Kiker}, {Gee}, {Lester},
  {Braunfeld}, {Rebull}, \& {Kuhn}}]{2022AJ....163..263H}
{Hillenbrand}, L.~A., {Kiker}, T.~J., {Gee}, M., {et~al.} 2022, \aj, 163, 263

\bibitem[{{Hodgkin} {et~al.}(2021){Hodgkin}, {Harrison}, {Breedt}, {Wevers},
  {Rixon}, {Delgado}, {Yoldas}, {Kostrzewa-Rutkowska}, {Wyrzykowski}, {van
  Leeuwen}, {Blagorodnova}, {Campbell}, {Eappachen}, {Fraser}, {Ihanec},
  {Koposov}, {Kruszy{\'n}ska}, {Marton}, {Rybicki}, {Brown}, {Burgess},
  {Busso}, {Cowell}, {De Angeli}, {Diener}, {Evans}, {Gilmore}, {Holland},
  {Jonker}, {van Leeuwen}, {Mignard}, {Osborne}, {Portell}, {Prusti},
  {Richards}, {Riello}, {Seabroke}, {Walton}, {{\'A}brah{\'a}m}, {Altavilla},
  {Baker}, {Bastian}, {O'Brien}, {de Bruijne}, {Butterley}, {Carrasco},
  {Casta{\~n}eda}, {Clark}, {Clementini}, {Copperwheat}, {Cropper},
  {Damljanovic}, {Davidson}, {Davis}, {Dennefeld}, {Dhillon}, {Dolding},
  {Dominik}, {Esquej}, {Eyer}, {Fabricius}, {Fridman}, {Froebrich}, {Garralda},
  {Gomboc}, {Gonz{\'a}lez-Vidal}, {Guerra}, {Hambly}, {Hardy}, {Holl},
  {Hourihane}, {Japelj}, {Kann}, {Kiss}, {Knigge}, {Kolb}, {Komossa},
  {K{\'o}sp{\'a}l}, {Kov{\'a}cs}, {Kun}, {Leto}, {Lewis}, {Littlefair},
  {Mahabal}, {Mundell}, {Nagy}, {Padeletti}, {Palaversa}, {Pigulski},
  {Pretorius}, {van Reeven}, {Ribeiro}, {Roelens}, {Rowell}, {Schartel},
  {Scholz}, {Schwope}, {Sip{\H{o}}cz}, {Smartt}, {Smith}, {Serraller},
  {Steeghs}, {Sullivan}, {Szabados}, {Szegedi-Elek}, {Tisserand}, {Tomasella},
  {van Velzen}, {Whitelock}, {Wilson}, \& {Young}}]{2021A&A...652A..76H}
{Hodgkin}, S.~T., {Harrison}, D.~L., {Breedt}, E., {et~al.} 2021, \aap, 652,
  A76

\bibitem[{Howard {et~al.}(2019)Howard, Sandler, Chu, Chen, Chen, Tan, Wang,
  Zhu, Pang, Vasudevan, Le, \& Adam}]{howard2019searching}
Howard, A., Sandler, M., Chu, G., {et~al.} 2019, in Proceedings of the IEEE/CVF
  International Conference on Computer Vision (ICCV), 1314--1324

\bibitem[{{Hunt} \& {Reffert}(2024)}]{2024A&A...686A..42H}
{Hunt}, E.~L. \& {Reffert}, S. 2024, \aap, 686, A42

\bibitem[{Iandola {et~al.}(2016)Iandola, Han, Moskewicz, Ashraf, Dally, \&
  Keutzer}]{iandola2016squeezenet}
Iandola, F.~N., Han, S., Moskewicz, M.~W., {et~al.} 2016, arXiv preprint
  arXiv:1602.07360

\bibitem[{Ioffe \& Szegedy(2015)}]{ioffe2015batchnorm}
Ioffe, S. \& Szegedy, C. 2015, in Proceedings of the 32nd International
  Conference on Machine Learning (ICML), 448--456

\bibitem[{{Iwanek} {et~al.}(2022){Iwanek}, {Soszy{\'n}ski}, {Koz{\l}owski},
  {Poleski}, {Pietrukowicz}, {Skowron}, {Wrona}, {Mr{\'o}z}, {Udalski},
  {Szyma{\'n}ski}, {Skowron}, {Ulaczyk}, {Gromadzki}, {Rybicki}, \&
  {Ratajczak}}]{2022ApJS..260...46I}
{Iwanek}, P., {Soszy{\'n}ski}, I., {Koz{\l}owski}, S., {et~al.} 2022, \apjs,
  260, 46

\bibitem[{{Janson} {et~al.}(2012){Janson}, {Hormuth}, {Bergfors}, {Brandner},
  {Hippler}, {Daemgen}, {Kudryavtseva}, {Schmalzl}, {Schnupp}, \&
  {Henning}}]{2012ApJ...754...44J}
{Janson}, M., {Hormuth}, F., {Bergfors}, C., {et~al.} 2012, \apj, 754, 44

\bibitem[{{Jayasinghe} {et~al.}(2018){Jayasinghe}, {Kochanek}, {Stanek},
  {Shappee}, {Holoien}, {Thompson}, {Prieto}, {Dong}, {Pawlak}, {Shields},
  {Pojmanski}, {Otero}, {Britt}, \& {Will}}]{2018MNRAS.477.3145J}
{Jayasinghe}, T., {Kochanek}, C.~S., {Stanek}, K.~Z., {et~al.} 2018, \mnras,
  477, 3145

\bibitem[{{Kang} {et~al.}(2009){Kang}, {Bieging}, {Povich}, \&
  {Lee}}]{2009ApJ...706...83K}
{Kang}, M., {Bieging}, J.~H., {Povich}, M.~S., \& {Lee}, Y. 2009, \apj, 706, 83

\bibitem[{{Karska} {et~al.}(2022){Karska}, {Koprowski}, {Solarz}, {Szczerba},
  {Sewi{\l}o}, {Si{\'o}dmiak}, {Elia}, {Gawro{\'n}ski}, {Grzesiak}, {Yung},
  {Fischer}, \& {Kristensen}}]{2022A&A...663A.133K}
{Karska}, A., {Koprowski}, M., {Solarz}, A., {et~al.} 2022, \aap, 663, A133

\bibitem[{Kennicutt \& Evans(2012)}]{KennicuttEvans2012}
Kennicutt, Jr., R.~C. \& Evans, II, N.~J. 2012, \araa, 50, 531

\bibitem[{{Kerr} {et~al.}(2021){Kerr}, {Rizzuto}, {Kraus}, \&
  {Offner}}]{2021ApJ...917...23K}
{Kerr}, R. M.~P., {Rizzuto}, A.~C., {Kraus}, A.~L., \& {Offner}, S. S.~R. 2021,
  \apj, 917, 23

\bibitem[{{Kharchenko} \& {Kilpio}(2000)}]{2000BaltA...9..646K}
{Kharchenko}, N. \& {Kilpio}, E. 2000, Baltic Astronomy, 9, 646

\bibitem[{{Kim} \& {L{\'e}pine}(2022)}]{2022MNRAS.510.4308K}
{Kim}, B. \& {L{\'e}pine}, S. 2022, \mnras, 510, 4308

\bibitem[{{Kim} \& {Brunner}(2017)}]{2017MNRAS.464.4463K}
{Kim}, E.~J. \& {Brunner}, R.~J. 2017, \mnras, 464, 4463

\bibitem[{{Koenig} \& {Allen}(2011)}]{2011ApJ...726...18K}
{Koenig}, X.~P. \& {Allen}, L.~E. 2011, \apj, 726, 18

\bibitem[{{Koenig} \& {Leisawitz}(2014)}]{2014ApJ...791..131K}
{Koenig}, X.~P. \& {Leisawitz}, D.~T. 2014, \apj, 791, 131

\bibitem[{Kohonen(1989)}]{Kohonen1989}
Kohonen, T. 1989, Self-Organization and Associative Memory, 3rd edn. (Berlin,
  Heidelberg: Springer-Verlag)

\bibitem[{Kohonen(2001)}]{Kohonen2001}
Kohonen, T. 2001, Springer Series in Information Sciences, Vol.~30,
  Self-Organizing Maps, 3rd edn. (Berlin, Heidelberg: Springer)

\bibitem[{{Konietzka} {et~al.}(2024){Konietzka}, {Goodman}, {Zucker},
  {Burkert}, {Alves}, {Foley}, {Swiggum}, {Koller}, \&
  {Miret-Roig}}]{2024Natur.628...62K}
{Konietzka}, R., {Goodman}, A.~A., {Zucker}, C., {et~al.} 2024, \nat, 628, 62

\bibitem[{{Kounkel} {et~al.}(2022){Kounkel}, {Deng}, \&
  {Stassun}}]{2022AJ....164...57K}
{Kounkel}, M., {Deng}, T., \& {Stassun}, K.~G. 2022, \aj, 164, 57

\bibitem[{{Kounkel} {et~al.}(2017){Kounkel}, {Hartmann}, {Calvet}, \&
  {Megeath}}]{2017AJ....154...29K}
{Kounkel}, M., {Hartmann}, L., {Calvet}, N., \& {Megeath}, T. 2017, \aj, 154,
  29

\bibitem[{{Kourkchi} \& {Tully}(2017)}]{2017ApJ...843...16K}
{Kourkchi}, E. \& {Tully}, R.~B. 2017, \apj, 843, 16

\bibitem[{Krumholz(2014)}]{Krumholz2014}
Krumholz, M.~R. 2014, Physics Reports, 539, 49

\bibitem[{{Kryukova} {et~al.}(2014){Kryukova}, {Megeath}, {Hora}, {Gutermuth},
  {Bontemps}, {Kraemer}, {Hennemann}, {Schneider}, {Smith}, \&
  {Motte}}]{2014AJ....148...11K}
{Kryukova}, E., {Megeath}, S.~T., {Hora}, J.~L., {et~al.} 2014, \aj, 148, 11

\bibitem[{{Kuhn} {et~al.}(2021){Kuhn}, {de Souza}, {Krone-Martins},
  {Castro-Ginard}, {Ishida}, {Povich}, {Hillenbrand}, \& {COIN
  Collaboration}}]{2021ApJS..254...33K}
{Kuhn}, M.~A., {de Souza}, R.~S., {Krone-Martins}, A., {et~al.} 2021, \apjs,
  254, 33

\bibitem[{{Kumar} {et~al.}(2014){Kumar}, {Sharma}, {Manfroid}, {Gosset},
  {Rauw}, {Naz{\'e}}, \& {Kesh Yadav}}]{2014A&A...567A.109K}
{Kumar}, B., {Sharma}, S., {Manfroid}, J., {et~al.} 2014, \aap, 567, A109

\bibitem[{{Lada}(1987)}]{1987IAUS..115....1L}
{Lada}, C.~J. 1987, in IAU Symposium, Vol. 115, Star Forming Regions, ed.
  M.~{Peimbert} \& J.~{Jugaku}, 1

\bibitem[{{Lada} {et~al.}(2017){Lada}, {Lewis}, {Lombardi}, \&
  {Alves}}]{2017A&A...606A.100L}
{Lada}, C.~J., {Lewis}, J.~A., {Lombardi}, M., \& {Alves}, J. 2017, \aap, 606,
  A100

\bibitem[{{Lasker} {et~al.}(2008){Lasker}, {Lattanzi}, {McLean}, {Bucciarelli},
  {Drimmel}, {Garcia}, {Greene}, {Guglielmetti}, {Hanley}, {Hawkins},
  {Laidler}, {Loomis}, {Meakes}, {Mignani}, {Morbidelli}, {Morrison},
  {Pannunzio}, {Rosenberg}, {Sarasso}, {Smart}, {Spagna}, {Sturch},
  {Volpicelli}, {White}, {Wolfe}, \& {Zacchei}}]{2008AJ....136..735L}
{Lasker}, B.~M., {Lattanzi}, M.~G., {McLean}, B.~J., {et~al.} 2008, \aj, 136,
  735

\bibitem[{{Lawrence} {et~al.}(2007){Lawrence}, {Warren}, {Almaini}, {Edge},
  {Hambly}, {Jameson}, {Lucas}, {Casali}, {Adamson}, {Dye}, {Emerson},
  {Foucaud}, {Hewett}, {Hirst}, {Hodgkin}, {Irwin}, {Lodieu}, {McMahon},
  {Simpson}, {Smail}, {Mortlock}, \& {Folger}}]{2007MNRAS.379.1599L}
{Lawrence}, A., {Warren}, S.~J., {Almaini}, O., {et~al.} 2007, \mnras, 379,
  1599

\bibitem[{Lecun {et~al.}(1998)Lecun, Bottou, Bengio, \& Haffner}]{CNN726791}
Lecun, Y., Bottou, L., Bengio, Y., \& Haffner, P. 1998, Proceedings of the
  IEEE, 86, 2278

\bibitem[{{L{\'e}pine} \& {Gaidos}(2011)}]{2011AJ....142..138L}
{L{\'e}pine}, S. \& {Gaidos}, E. 2011, \aj, 142, 138

\bibitem[{{Lim} {et~al.}(2022){Lim}, {Naz{\'e}}, {Hong}, {Yoon}, {Lee},
  {Hwang}, {Park}, \& {Lee}}]{2022AJ....163..266L}
{Lim}, B., {Naz{\'e}}, Y., {Hong}, J., {et~al.} 2022, \aj, 163, 266

\bibitem[{Lin {et~al.}(2014)Lin, Chen, \& Yan}]{lin2014nin}
Lin, M., Chen, Q., \& Yan, S. 2014, in International Conference on Learning
  Representations (ICLR)

\bibitem[{Loshchilov \& Hutter(2019)}]{loshchilov2019adamw}
Loshchilov, I. \& Hutter, F. 2019, in International Conference on Learning
  Representations (ICLR)

\bibitem[{{Lucas} {et~al.}(2008){Lucas}, {Hoare}, {Longmore}, {Schr{\"o}der},
  {Davis}, {Adamson}, {Bandyopadhyay}, {de Grijs}, {Smith}, {Gosling},
  {Mitchison}, {G{\'a}sp{\'a}r}, {Coe}, {Tamura}, {Parker}, {Irwin}, {Hambly},
  {Bryant}, {Collins}, {Cross}, {Evans}, {Gonzalez-Solares}, {Hodgkin},
  {Lewis}, {Read}, {Riello}, {Sutorius}, {Lawrence}, {Drew}, {Dye}, \&
  {Thompson}}]{2008MNRAS.391..136L}
{Lucas}, P.~W., {Hoare}, M.~G., {Longmore}, A., {et~al.} 2008, \mnras, 391, 136

\bibitem[{{Luhman}(2022)}]{2022AJ....163...25L}
{Luhman}, K.~L. 2022, \aj, 163, 25

\bibitem[{{Luhman} \& {Esplin}(2020)}]{2020AJ....160...44L}
{Luhman}, K.~L. \& {Esplin}, T.~L. 2020, \aj, 160, 44

\bibitem[{Ma {et~al.}(2018)Ma, Zhang, Zheng, \& Sun}]{ma2018shufflenet}
Ma, N., Zhang, X., Zheng, H.-T., \& Sun, J. 2018, in Proceedings of the
  European Conference on Computer Vision (ECCV), 116

\bibitem[{{Madar{\'a}sz} {et~al.}(2025){Madar{\'a}sz}, {Marton}, {Gezer},
  {Lehner}, {Roquette}, {Audard}, {Hernandez}, \&
  {Dionatos}}]{2025A&A...696A..37M}
{Madar{\'a}sz}, M., {Marton}, G., {Gezer}, I., {et~al.} 2025, \aap, 696, A37

\bibitem[{Mahabal {et~al.}(2017)Mahabal, Sheth, Gieseke, Pai, Djorgovski,
  Drake, \& Graham}]{2017SSCI....1M}
Mahabal, A., Sheth, K., Gieseke, F., {et~al.} 2017, in 2017 IEEE Symposium
  Series on Computational Intelligence (SSCI), 1--8

\bibitem[{{Marton} {et~al.}(2023){Marton}, {{\'A}brah{\'a}m}, {Rimoldini},
  {Audard}, {Kun}, {Nagy}, {K{\'o}sp{\'a}l}, {Szabados}, {Holl}, {Gavras},
  {Mowlavi}, {Nienartowicz}, {de Fombelle}, {Lecoeur-Ta{\"\i}bi}, {Karbevska},
  {Lario}, \& {Eyer}}]{2023A&A...674A..21M}
{Marton}, G., {{\'A}brah{\'a}m}, P., {Rimoldini}, L., {et~al.} 2023, \aap, 674,
  A21

\bibitem[{{Marton} {et~al.}(2019){Marton}, {{\'A}brah{\'a}m}, {Szegedi-Elek},
  {Varga}, {Kun}, {K{\'o}sp{\'a}l}, {Varga-Vereb{\'e}lyi}, {Hodgkin},
  {Szabados}, {Beck}, \& {Kiss}}]{2019MNRAS.487.2522M}
{Marton}, G., {{\'A}brah{\'a}m}, P., {Szegedi-Elek}, E., {et~al.} 2019, \mnras,
  487, 2522

\bibitem[{{Marton} {et~al.}(2024){Marton}, {Gezer}, {Madar{\'a}sz}, {Dionatos},
  {Audard}, {Roquette}, {Hernandez}, {Paladini}, \&
  {Altieri}}]{2024A&A...688A.203M}
{Marton}, G., {Gezer}, I., {Madar{\'a}sz}, M., {et~al.} 2024, \aap, 688, A203

\bibitem[{{Marton} {et~al.}(2016){Marton}, {T{\'o}th}, {Paladini}, {Kun},
  {Zahorecz}, {McGehee}, \& {Kiss}}]{2016MNRAS.458.3479M}
{Marton}, G., {T{\'o}th}, L.~V., {Paladini}, R., {et~al.} 2016, \mnras, 458,
  3479

\bibitem[{{Mas} {et~al.}(2026){Mas}, {Roquette}, {Audard}, {Madar{\'a}sz},
  {Marton}, {Hernandez}, {Gezer}, \& {Dionatos}}]{2026A&A...705A.172M}
{Mas}, C., {Roquette}, J., {Audard}, M., {et~al.} 2026, \aap, 705, A172

\bibitem[{{Matsunaga} {et~al.}(2009){Matsunaga}, {Kawadu}, {Nishiyama},
  {Nagayama}, {Hatano}, {Tamura}, {Glass}, \& {Nagata}}]{2009MNRAS.399.1709M}
{Matsunaga}, N., {Kawadu}, T., {Nishiyama}, S., {et~al.} 2009, \mnras, 399,
  1709

\bibitem[{{Mayne} {et~al.}(2007){Mayne}, {Naylor}, {Littlefair}, {Saunders}, \&
  {Jeffries}}]{2007MNRAS.375.1220M}
{Mayne}, N.~J., {Naylor}, T., {Littlefair}, S.~P., {Saunders}, E.~S., \&
  {Jeffries}, R.~D. 2007, \mnras, 375, 1220

\bibitem[{{McBride} {et~al.}(2021){McBride}, {Lingg}, {Kounkel}, {Covey}, \&
  {Hutchinson}}]{2021AJ....162..282M}
{McBride}, A., {Lingg}, R., {Kounkel}, M., {Covey}, K., \& {Hutchinson}, B.
  2021, \aj, 162, 282

\bibitem[{{McKee} \& {Ostriker}(2007)}]{2007ARA&A..45..565M}
{McKee}, C.~F. \& {Ostriker}, E.~C. 2007, \araa, 45, 565

\bibitem[{{McMahon} {et~al.}(2013){McMahon}, {Banerji}, {Gonzalez}, {Koposov},
  {Bejar}, {Lodieu}, {Rebolo}, \& {VHS Collaboration}}]{2013Msngr.154...35M}
{McMahon}, R.~G., {Banerji}, M., {Gonzalez}, E., {et~al.} 2013, The Messenger,
  154, 35

\bibitem[{{Medan} {et~al.}(2021){Medan}, {L{\'e}pine}, \&
  {Hartman}}]{2021AJ....161..234M}
{Medan}, I., {L{\'e}pine}, S., \& {Hartman}, Z. 2021, \aj, 161, 234

\bibitem[{{Medina} {et~al.}(2022){Medina}, {Winters}, {Irwin}, \&
  {Charbonneau}}]{2022ApJ...935..104M}
{Medina}, A.~A., {Winters}, J.~G., {Irwin}, J.~M., \& {Charbonneau}, D. 2022,
  \apj, 935, 104

\bibitem[{{Megeath} {et~al.}(2012){Megeath}, {Gutermuth}, {Muzerolle},
  {Kryukova}, {Flaherty}, {Hora}, {Allen}, {Hartmann}, {Myers}, {Pipher},
  {Stauffer}, {Young}, \& {Fazio}}]{2012AJ....144..192M}
{Megeath}, S.~T., {Gutermuth}, R., {Muzerolle}, J., {et~al.} 2012, \aj, 144,
  192

\bibitem[{{Melton}(2020)}]{2020AJ....159..200M}
{Melton}, E. 2020, \aj, 159, 200

\bibitem[{{Meng} {et~al.}(2016){Meng}, {Plavchan}, {Rieke}, {Cody}, {G{\"u}th},
  {Stauffer}, {Covey}, {Carey}, {Ciardi}, {Duran-Rojas}, {Gutermuth},
  {Morales-Calder{\'o}n}, {Rebull}, \& {Watson}}]{2016ApJ...823...58M}
{Meng}, H. Y.~A., {Plavchan}, P., {Rieke}, G.~H., {et~al.} 2016, \apj, 823, 58

\bibitem[{{Meng} {et~al.}(2017){Meng}, {Rieke}, {Su}, \&
  {G{\'a}sp{\'a}r}}]{2017ApJ...836...34M}
{Meng}, H. Y.~A., {Rieke}, G.~H., {Su}, K. Y.~L., \& {G{\'a}sp{\'a}r}, A. 2017,
  \apj, 836, 34

\bibitem[{{Mintz} {et~al.}(2021){Mintz}, {Hora}, \&
  {Winston}}]{2021AJ....162..236M}
{Mintz}, A., {Hora}, J.~L., \& {Winston}, E. 2021, \aj, 162, 236

\bibitem[{{Morrell} \& {Naylor}(2019)}]{2019MNRAS.489.2615M}
{Morrell}, S. \& {Naylor}, T. 2019, \mnras, 489, 2615

\bibitem[{{Newton} {et~al.}(2017){Newton}, {Irwin}, {Charbonneau}, {Berlind},
  {Calkins}, \& {Mink}}]{2017ApJ...834...85N}
{Newton}, E.~R., {Irwin}, J., {Charbonneau}, D., {et~al.} 2017, \apj, 834, 85

\bibitem[{{Nguyen} {et~al.}(2025){Nguyen}, {Costa}, {Bressan}, {Girardi},
  {Cescutti}, {Korn}, {Volpato}, {Chen}, {Pastorelli}, {Trabucchi}, {Shepherd},
  {Ettorre}, \& {Zaggia}}]{2025A&A...701A.258N}
{Nguyen}, C.~T., {Costa}, G., {Bressan}, A., {et~al.} 2025, \aap, 701, A258

\bibitem[{{Nguyen} {et~al.}(2022){Nguyen}, {Costa}, {Girardi}, {Volpato},
  {Bressan}, {Chen}, {Marigo}, {Fu}, \& {Goudfrooij}}]{2022A&A...665A.126N}
{Nguyen}, C.~T., {Costa}, G., {Girardi}, L., {et~al.} 2022, \aap, 665, A126

\bibitem[{{Nu{\~n}ez} {et~al.}(2021){Nu{\~n}ez}, {Povich}, {Binder},
  {Townsley}, \& {Broos}}]{2021AJ....162..153N}
{Nu{\~n}ez}, E.~H., {Povich}, M.~S., {Binder}, B.~A., {Townsley}, L.~K., \&
  {Broos}, P.~S. 2021, \aj, 162, 153

\bibitem[{{O'Neill} {et~al.}(2024){O'Neill}, {Zucker}, {Goodman}, \&
  {Edenhofer}}]{2024ApJ...973..136O}
{O'Neill}, T.~J., {Zucker}, C., {Goodman}, A.~A., \& {Edenhofer}, G. 2024,
  \apj, 973, 136

\bibitem[{Padoan {et~al.}(2014)Padoan, Federrath, Chabrier, Evans, Johnstone,
  J{\o}rgensen, McKee, \& Nordlund}]{Padoan2014}
Padoan, P., Federrath, C., Chabrier, G., {et~al.} 2014, in Protostars and
  Planets VI, ed. H.~Beuther, R.~S. Klessen, C.~P. Dullemond, \& T.~Henning, 77

\bibitem[{{Paegert} {et~al.}(2021){Paegert}, {Stassun}, {Collins}, {Pepper},
  {Torres}, {Jenkins}, {Twicken}, \& {Latham}}]{2021arXiv210804778P}
{Paegert}, M., {Stassun}, K.~G., {Collins}, K.~A., {et~al.} 2021, arXiv
  e-prints, arXiv:2108.04778

\bibitem[{{Paturel} {et~al.}(2003){Paturel}, {Petit}, {Prugniel}, {Theureau},
  {Rousseau}, {Brouty}, {Dubois}, \& {Cambr{\'e}sy}}]{2003A&A...412...45P}
{Paturel}, G., {Petit}, C., {Prugniel}, P., {et~al.} 2003, \aap, 412, 45

\bibitem[{{P{\'e}rez-D{\'\i}az} {et~al.}(2024){P{\'e}rez-D{\'\i}az},
  {P{\'e}rez-Montero}, {Fern{\'a}ndez-Ontiveros}, {V{\'\i}lchez},
  {Hern{\'a}n-Caballero}, \& {Amor{\'\i}n}}]{2024A&A...685A.168P}
{P{\'e}rez-D{\'\i}az}, B., {P{\'e}rez-Montero}, E., {Fern{\'a}ndez-Ontiveros},
  J.~A., {et~al.} 2024, \aap, 685, A168

\bibitem[{{Perraudin} {et~al.}(2019){Perraudin}, {Defferrard}, {Kacprzak}, \&
  {Sgier}}]{2019A&C....27..130P}
{Perraudin}, N., {Defferrard}, M., {Kacprzak}, T., \& {Sgier}, R. 2019,
  Astronomy and Computing, 27, 130

\bibitem[{{Prisinzano} {et~al.}(2018){Prisinzano}, {Damiani}, {Guarcello},
  {Micela}, {Sciortino}, {Tognelli}, \& {Venuti}}]{2018A&A...617A..63P}
{Prisinzano}, L., {Damiani}, F., {Guarcello}, M.~G., {et~al.} 2018, \aap, 617,
  A63

\bibitem[{{Prisinzano} {et~al.}(2022){Prisinzano}, {Damiani}, {Sciortino},
  {Flaccomio}, {Guarcello}, {Micela}, {Tognelli}, {Jeffries}, \&
  {Alcal{\'a}}}]{2022A&A...664A.175P}
{Prisinzano}, L., {Damiani}, F., {Sciortino}, S., {et~al.} 2022, \aap, 664,
  A175

\bibitem[{{Qiu} {et~al.}(2008){Qiu}, {Zhang}, {Megeath}, {Gutermuth},
  {Beuther}, {Shepherd}, {Sridharan}, {Testi}, \& {De
  Pree}}]{2008ApJ...685.1005Q}
{Qiu}, K., {Zhang}, Q., {Megeath}, S.~T., {et~al.} 2008, \apj, 685, 1005

\bibitem[{Radosavovic {et~al.}(2020)Radosavovic, Kosaraju, Girshick, He, \&
  Doll{\'a}r}]{radosavovic2020regnet}
Radosavovic, I., Kosaraju, R.~P., Girshick, R., He, K., \& Doll{\'a}r, P. 2020,
  in Proceedings of the IEEE/CVF Conference on Computer Vision and Pattern
  Recognition (CVPR), 10428--10436

\bibitem[{Ramachandran {et~al.}(2018)Ramachandran, Zoph, \&
  Le}]{ramachandran2017swish}
Ramachandran, P., Zoph, B., \& Le, Q.~V. 2018, in International Conference on
  Learning Representations (ICLR)

\bibitem[{{Rapson} {et~al.}(2014){Rapson}, {Pipher}, {Gutermuth}, {Megeath},
  {Allen}, {Myers}, \& {Allen}}]{2014ApJ...794..124R}
{Rapson}, V.~A., {Pipher}, J.~L., {Gutermuth}, R.~A., {et~al.} 2014, \apj, 794,
  124

\bibitem[{{Rebull} {et~al.}(2011){Rebull}, {Guieu}, {Stauffer}, {Hillenbrand},
  {Noriega-Crespo}, {Stapelfeldt}, {Carey}, {Carpenter}, {Cole}, {Padgett},
  {Strom}, \& {Wolff}}]{2011ApJS..193...25R}
{Rebull}, L.~M., {Guieu}, S., {Stauffer}, J.~R., {et~al.} 2011, \apjs, 193, 25

\bibitem[{{Rebull} {et~al.}(2015){Rebull}, {Stauffer}, {Cody}, {G{\"u}nther},
  {Hillenbrand}, {Poppenhaeger}, {Wolk}, {Hora}, {Hernandez}, {Bayo}, {Covey},
  {Forbrich}, {Gutermuth}, {Morales-Calder{\'o}n}, {Plavchan}, {Song}, {Bouy},
  {Terebey}, {Cuillandre}, \& {Allen}}]{2015AJ....150..175R}
{Rebull}, L.~M., {Stauffer}, J.~R., {Cody}, A.~M., {et~al.} 2015, \aj, 150, 175

\bibitem[{{Rebull} {et~al.}(2018){Rebull}, {Stauffer}, {Cody}, {Hillenbrand},
  {David}, \& {Pinsonneault}}]{2018AJ....155..196R}
{Rebull}, L.~M., {Stauffer}, J.~R., {Cody}, A.~M., {et~al.} 2018, \aj, 155, 196

\bibitem[{{Rebull} {et~al.}(2022){Rebull}, {Stauffer}, {Hillenbrand}, {Cody},
  {Kruse}, \& {Powell}}]{2022AJ....164...80R}
{Rebull}, L.~M., {Stauffer}, J.~R., {Hillenbrand}, L.~A., {et~al.} 2022, \aj,
  164, 80

\bibitem[{{Riedel} {et~al.}(2017){Riedel}, {Blunt}, {Lambrides}, {Rice},
  {Cruz}, \& {Faherty}}]{2017AJ....153...95R}
{Riedel}, A.~R., {Blunt}, S.~C., {Lambrides}, E.~L., {et~al.} 2017, \aj, 153,
  95

\bibitem[{{Rimoldini} {et~al.}(2023){Rimoldini}, {Holl}, {Gavras}, {Audard},
  {De Ridder}, {Mowlavi}, {Nienartowicz}, {Jevardat de Fombelle},
  {Lecoeur-Ta{\"\i}bi}, {Karbevska}, {Evans}, {{\'A}brah{\'a}m}, {Carnerero},
  {Clementini}, {Distefano}, {Garofalo}, {Garc{\'\i}a-Lario}, {Gomel},
  {Klioner}, {Kruszy{\'n}ska}, {Lanzafame}, {Lebzelter}, {Marton}, {Mazeh},
  {Molinaro}, {Panahi}, {Raiteri}, {Ripepi}, {Szabados}, {Teyssier},
  {Trabucchi}, {Wyrzykowski}, {Zucker}, \& {Eyer}}]{2023A&A...674A..14R}
{Rimoldini}, L., {Holl}, B., {Gavras}, P., {et~al.} 2023, \aap, 674, A14

\bibitem[{{Ripepi} {et~al.}(2023){Ripepi}, {Clementini}, {Molinaro}, {Leccia},
  {Plachy}, {Moln{\'a}r}, {Rimoldini}, {Musella}, {Marconi}, {Garofalo},
  {Audard}, {Holl}, {Evans}, {Jevardat de Fombelle}, {Lecoeur-Taibi},
  {Marchal}, {Mowlavi}, {Muraveva}, {Nienartowicz}, {Sartoretti}, {Szabados},
  \& {Eyer}}]{2023A&A...674A..17R}
{Ripepi}, V., {Clementini}, G., {Molinaro}, R., {et~al.} 2023, \aap, 674, A17

\bibitem[{{Robberto} {et~al.}(2013){Robberto}, {Soderblom}, {Bergeron},
  {Kozhurina-Platais}, {Makidon}, {McCullough}, {McMaster}, {Panagia}, {Reid},
  {Levay}, {Frattare}, {Da Rio}, {Andersen}, {O'Dell}, {Stassun}, {Simon},
  {Feigelson}, {Stauffer}, {Meyer}, {Reggiani}, {Krist}, {Manara},
  {Romaniello}, {Hillenbrand}, {Ricci}, {Palla}, {Najita}, {Ananna},
  {Scandariato}, \& {Smith}}]{2013ApJS..207...10R}
{Robberto}, M., {Soderblom}, D.~R., {Bergeron}, E., {et~al.} 2013, \apjs, 207,
  10

\bibitem[{{Rodrigues} {et~al.}(2024){Rodrigues}, {Paliya}, {Garrappa},
  {Omeliukh}, {Franckowiak}, \& {Winter}}]{2024A&A...681A.119R}
{Rodrigues}, X., {Paliya}, V.~S., {Garrappa}, S., {et~al.} 2024, \aap, 681,
  A119

\bibitem[{{Rodriguez} {et~al.}(2017){Rodriguez}, {Ansdell}, {Oelkers},
  {Cargile}, {Gaidos}, {Cody}, {Stevens}, {Somers}, {James}, {Beatty},
  {Siverd}, {Lund}, {Kuhn}, {Gaudi}, {Pepper}, \&
  {Stassun}}]{2017ApJ...848...97R}
{Rodriguez}, J.~E., {Ansdell}, M., {Oelkers}, R.~J., {et~al.} 2017, \apj, 848,
  97

\bibitem[{{Romero} {et~al.}(2012){Romero}, {Schreiber}, {Cieza},
  {Rebassa-Mansergas}, {Mer{\'\i}n}, {Smith Castelli}, {Allen}, \&
  {Morrell}}]{2012ApJ...749...79R}
{Romero}, G.~A., {Schreiber}, M.~R., {Cieza}, L.~A., {et~al.} 2012, \apj, 749,
  79

\bibitem[{{Roquette} {et~al.}(2025){Roquette}, {Audard}, {Hernandez}, {Gezer},
  {Marton}, {Mas}, {Madar{\'a}sz}, \& {Dionatos}}]{2025A&A...702A..63R}
{Roquette}, J., {Audard}, M., {Hernandez}, D., {et~al.} 2025, \aap, 702, A63

\bibitem[{{Roslund} {et~al.}(2005){Roslund}, {Larsson}, {Schalander}, \&
  {Radbo}}]{2005IBVS.5633....1R}
{Roslund}, C., {Larsson}, A., {Schalander}, P., \& {Radbo}, M. 2005,
  Information Bulletin on Variable Stars, 5633, 1

\bibitem[{{S{\'a}nchez-S{\'a}ez} {et~al.}(2023){S{\'a}nchez-S{\'a}ez},
  {Arredondo}, {Bayo}, {Ar{\'e}valo}, {Bauer}, {Cabrera-Vives}, {Catelan},
  {Coppi}, {Est{\'e}vez}, {F{\"o}rster}, {Hern{\'a}ndez-Garc{\'\i}a}, {Huijse},
  {Kurtev}, {Lira}, {Mu{\~n}oz Arancibia}, \& {Pignata}}]{2023A&A...675A.195S}
{S{\'a}nchez-S{\'a}ez}, P., {Arredondo}, J., {Bayo}, A., {et~al.} 2023, \aap,
  675, A195

\bibitem[{{S{\'a}nchez-S{\'a}ez} {et~al.}(2021){S{\'a}nchez-S{\'a}ez}, {Reyes},
  {Valenzuela}, {F{\"o}rster}, {Eyheramendy}, {Elorrieta}, {Bauer},
  {Cabrera-Vives}, {Est{\'e}vez}, {Catelan}, {Pignata}, {Huijse}, {De Cicco},
  {Ar{\'e}valo}, {Carrasco-Davis}, {Abril}, {Kurtev}, {Borissova}, {Arredondo},
  {Castillo-Navarrete}, {Rodriguez}, {Ruz-Mieres}, {Moya},
  {Sabatini-Gacit{\'u}a}, {Sep{\'u}lveda-Cobo}, \&
  {Camacho-I{\~n}iguez}}]{2021AJ....161..141S}
{S{\'a}nchez-S{\'a}ez}, P., {Reyes}, I., {Valenzuela}, C., {et~al.} 2021, \aj,
  161, 141

\bibitem[{{Saral} {et~al.}(2017){Saral}, {Hora}, {Audard}, {Koenig},
  {Mart{\'\i}nez-Galarza}, {Motte}, {Nguyen-Luong}, {Saygac}, \&
  {Smith}}]{2017ApJ...839..108S}
{Saral}, G., {Hora}, J.~L., {Audard}, M., {et~al.} 2017, \apj, 839, 108

\bibitem[{{Sevenster}(2002)}]{2002AJ....123.2772S}
{Sevenster}, M.~N. 2002, \aj, 123, 2772

\bibitem[{{Sewi{\l}o} {et~al.}(2019){Sewi{\l}o}, {Whitney}, {Yung},
  {Robitaille}, {Elia}, {Indebetouw}, {Schisano}, {Szczerba}, {Karska},
  {Wiseman}, {Babler}, {Boyer}, {Fischer}, {Meade}, {Olmi}, {Padgett}, \&
  {Si{\'o}dmiak}}]{2019ApJS..240...26S}
{Sewi{\l}o}, M., {Whitney}, B.~A., {Yung}, B. H.~K., {et~al.} 2019, \apjs, 240,
  26

\bibitem[{{Shappee} {et~al.}(2014){Shappee}, {Prieto}, {Grupe}, {Kochanek},
  {Stanek}, {De Rosa}, {Mathur}, {Zu}, {Peterson}, {Pogge}, {Komossa}, {Im},
  {Jencson}, {Holoien}, {Basu}, {Beacom}, {Szczygie{\l}}, {Brimacombe},
  {Adams}, {Campillay}, {Choi}, {Contreras}, {Dietrich}, {Dubberley},
  {Elphick}, {Foale}, {Giustini}, {Gonzalez}, {Hawkins}, {Howell}, {Hsiao},
  {Koss}, {Leighly}, {Morrell}, {Mudd}, {Mullins}, {Nugent}, {Parrent},
  {Phillips}, {Pojmanski}, {Rosing}, {Ross}, {Sand}, {Terndrup}, {Valenti},
  {Walker}, \& {Yoon}}]{2014ApJ...788...48S}
{Shappee}, B.~J., {Prieto}, J.~L., {Grupe}, D., {et~al.} 2014, \apj, 788, 48

\bibitem[{{Shu} {et~al.}(1987){Shu}, {Adams}, \&
  {Lizano}}]{1987ARA&A..25...23S}
{Shu}, F.~H., {Adams}, F.~C., \& {Lizano}, S. 1987, \araa, 25, 23

\bibitem[{{Silverberg} {et~al.}(2018){Silverberg}, {Kuchner}, {Wisniewski},
  {Bans}, {Debes}, {Kenyon}, {Baranec}, {Riddle}, {Law}, {Teske},
  {Burns-Kaurin}, {Bosch}, {Cernohous}, {Doll}, {Durantini Luca}, {Hyogo},
  {Hamilton}, {Finnemann}, {Lau}, \& {Disk Detective
  Collaboration}}]{2018ApJ...868...43S}
{Silverberg}, S.~M., {Kuchner}, M.~J., {Wisniewski}, J.~P., {et~al.} 2018,
  \apj, 868, 43

\bibitem[{{Soszy{\'n}ski} {et~al.}(2017){Soszy{\'n}ski}, {Udalski},
  {Szyma{\'n}ski}, {Wyrzykowski}, {Ulaczyk}, {Poleski}, {Pietrukowicz},
  {Koz{\l}owski}, {Skowron}, {Skowron}, {Mr{\'o}z}, {Pawlak}, {Rybicki}, \&
  {Jacyszyn-Dobrzeniecka}}]{2017AcA....67..297S}
{Soszy{\'n}ski}, I., {Udalski}, A., {Szyma{\'n}ski}, M.~K., {et~al.} 2017,
  \actaa, 67, 297

\bibitem[{{Soszy{\'n}ski} {et~al.}(2019){Soszy{\'n}ski}, {Udalski}, {Wrona},
  {Szyma{\'n}ski}, {Pietrukowicz}, {Skowron}, {Skowron}, {Poleski},
  {Koz{\l}owski}, {Mr{\'o}z}, {Ulaczyk}, {Rybicki}, {Iwanek}, \&
  {Gromadzki}}]{2019AcA....69..321S}
{Soszy{\'n}ski}, I., {Udalski}, A., {Wrona}, M., {et~al.} 2019, \actaa, 69, 321

\bibitem[{{Souchay} {et~al.}(2024){Souchay}, {Secrest}, {Sexton}, \&
  {Barache}}]{2024A&A...683A.112S}
{Souchay}, J., {Secrest}, N., {Sexton}, R., \& {Barache}, C. 2024, \aap, 683,
  A112

\bibitem[{{Spitzer Science Center (SSC)} \& {Infrared Science Archive
  (IRSA)}(2021)}]{2021yCat.2368....0S}
{Spitzer Science Center (SSC)} \& {Infrared Science Archive (IRSA)}. 2021,
  {VizieR Online Data Catalog: The Spitzer (SEIP) source list (SSTSL2) (Spitzer
  Science Center, 2021)}, VizieR On-line Data Catalog: II/368. Originally
  published in: Spitzer Science Center (SSC), IRSA (2021)

\bibitem[{{Stassun} {et~al.}(2019){Stassun}, {Oelkers}, {Paegert}, {Torres},
  {Pepper}, {De Lee}, {Collins}, {Latham}, {Muirhead}, {Chittidi},
  {Rojas-Ayala}, {Fleming}, {Rose}, {Tenenbaum}, {Ting}, {Kane}, {Barclay},
  {Bean}, {Brassuer}, {Charbonneau}, {Ge}, {Lissauer}, {Mann}, {McLean},
  {Mullally}, {Narita}, {Plavchan}, {Ricker}, {Sasselov}, {Seager}, {Sharma},
  {Shiao}, {Sozzetti}, {Stello}, {Vanderspek}, {Wallace}, \&
  {Winn}}]{2019AJ....158..138S}
{Stassun}, K.~G., {Oelkers}, R.~J., {Paegert}, M., {et~al.} 2019, \aj, 158, 138

\bibitem[{{Stauffer} {et~al.}(2014){Stauffer}, {Cody}, {Baglin}, {Alencar},
  {Rebull}, {Hillenbrand}, {Venuti}, {Turner}, {Carpenter}, {Plavchan},
  {Findeisen}, {Carey}, {Terebey}, {Morales-Calder{\'o}n}, {Bouvier}, {Micela},
  {Flaccomio}, {Song}, {Gutermuth}, {Hartmann}, {Calvet}, {Whitney}, {Barrado},
  {Vrba}, {Covey}, {Herbst}, {Furesz}, {Aigrain}, \&
  {Favata}}]{2014AJ....147...83S}
{Stauffer}, J., {Cody}, A.~M., {Baglin}, A., {et~al.} 2014, \aj, 147, 83

\bibitem[{{Sullivan} \& {Kraus}(2022)}]{2022ApJ...928..134S}
{Sullivan}, K. \& {Kraus}, A.~L. 2022, \apj, 928, 134

\bibitem[{{Szil{\'a}gyi} {et~al.}(2023){Szil{\'a}gyi}, {Kun},
  {{\'A}brah{\'a}m}, \& {Marton}}]{2023MNRAS.520.1390S}
{Szil{\'a}gyi}, M., {Kun}, M., {{\'A}brah{\'a}m}, P., \& {Marton}, G. 2023,
  \mnras, 520, 1390

\bibitem[{Tan {et~al.}(2019)Tan, Chen, Pang, Vasudevan, Sandler, Howard, \&
  Le}]{tan2019mnasnet}
Tan, M., Chen, B., Pang, R., {et~al.} 2019, in Proceedings of the IEEE/CVF
  Conference on Computer Vision and Pattern Recognition (CVPR), 2820--2828

\bibitem[{Tan \& Le(2019)}]{tan2019efficientnet}
Tan, M. \& Le, Q.~V. 2019, in Proceedings of the 36th International Conference
  on Machine Learning (ICML), 6105--6114

\bibitem[{Tan \& Le(2021)}]{tan2021efficientnetv2}
Tan, M. \& Le, Q.~V. 2021, in Proceedings of the 38th International Conference
  on Machine Learning (ICML), 10096--10106

\bibitem[{{Taylor}(2005)}]{2005ASPC..347...29T}
{Taylor}, M.~B. 2005, in Astronomical Society of the Pacific Conference Series,
  Vol. 347, Astronomical Data Analysis Software and Systems XIV, ed.
  P.~{Shopbell}, M.~{Britton}, \& R.~{Ebert}, 29

\bibitem[{{Teixeira} {et~al.}(2021){Teixeira}, {Alves}, {Sicilia-Aguilar},
  {Hacar}, \& {Scholz}}]{2021MNRAS.504L..17T}
{Teixeira}, P.~S., {Alves}, J., {Sicilia-Aguilar}, A., {Hacar}, A., \&
  {Scholz}, A. 2021, \mnras, 504, L17

\bibitem[{{Teixeira} {et~al.}(2020){Teixeira}, {Scholz}, \&
  {Alves}}]{2020A&A...642A..86T}
{Teixeira}, P.~S., {Scholz}, A., \& {Alves}, J. 2020, \aap, 642, A86

\bibitem[{{Tobin} {et~al.}(2016){Tobin}, {Looney}, {Li}, {Chandler}, {Dunham},
  {Segura-Cox}, {Sadavoy}, {Melis}, {Harris}, {Kratter}, \&
  {Perez}}]{2016ApJ...818...73T}
{Tobin}, J.~J., {Looney}, L.~W., {Li}, Z.-Y., {et~al.} 2016, \apj, 818, 73

\bibitem[{{T{\'o}th} {et~al.}(2014){T{\'o}th}, {Marton}, {Zahorecz},
  {Bal{\'a}zs}, {Ueno}, {Tamura}, {Kawamura}, {Kiss}, \&
  {Kitamura}}]{2014PASJ...66...17T}
{T{\'o}th}, L.~V., {Marton}, G., {Zahorecz}, S., {et~al.} 2014, \pasj, 66, 17

\bibitem[{Touvron {et~al.}(2021)Touvron, Cord, Sablayrolles, Synnaeve, \&
  J{\'e}gou}]{touvron2021layerscale}
Touvron, H., Cord, M., Sablayrolles, A., Synnaeve, G., \& J{\'e}gou, H. 2021,
  in Proceedings of the IEEE/CVF International Conference on Computer Vision
  (ICCV), 32--42

\bibitem[{{Tranin} {et~al.}(2024){Tranin}, {Webb}, {Godet}, \&
  {Quintin}}]{2024A&A...681A..16T}
{Tranin}, H., {Webb}, N., {Godet}, O., \& {Quintin}, E. 2024, \aap, 681, A16

\bibitem[{{Usatov} \& {Nosulchik}(2008)}]{2008OEJV...87....1U}
{Usatov}, M. \& {Nosulchik}, A. 2008, Open European Journal on Variable Stars,
  0087, 1

\bibitem[{Vaswani {et~al.}(2017)Vaswani, Shazeer, Parmar, Uszkoreit, Jones,
  Gomez, Kaiser, \& Polosukhin}]{vaswani2017attention}
Vaswani, A., Shazeer, N., Parmar, N., {et~al.} 2017, in Advances in Neural
  Information Processing Systems (NeurIPS), Vol.~30, 5998

\bibitem[{{Vejlgaard} {et~al.}(2024){Vejlgaard}, {Fynbo}, {Heintz}, {Krogager},
  {M{\o}ller}, {Geier}, {Christensen}, \& {Ma}}]{2024A&A...683A.157V}
{Vejlgaard}, S., {Fynbo}, J.~P.~U., {Heintz}, K.~E., {et~al.} 2024, \aap, 683,
  A157

\bibitem[{{Venuti} {et~al.}(2017){Venuti}, {Bouvier}, {Cody}, {Stauffer},
  {Micela}, {Rebull}, {Alencar}, {Sousa}, {Hillenbrand}, \&
  {Flaccomio}}]{2017A&A...599A..23V}
{Venuti}, L., {Bouvier}, J., {Cody}, A.~M., {et~al.} 2017, \aap, 599, A23

\bibitem[{{Venuti} {et~al.}(2015){Venuti}, {Bouvier}, {Irwin}, {Stauffer},
  {Hillenbrand}, {Rebull}, {Cody}, {Alencar}, {Micela}, {Flaccomio}, \&
  {Peres}}]{2015A&A...581A..66V}
{Venuti}, L., {Bouvier}, J., {Irwin}, J., {et~al.} 2015, \aap, 581, A66

\bibitem[{{Venuti} {et~al.}(2021){Venuti}, {Cody}, {Rebull}, {Beccari},
  {Irwin}, {Thanvantri}, {Howell}, \& {Barentsen}}]{2021AJ....162..101V}
{Venuti}, L., {Cody}, A.~M., {Rebull}, L.~M., {et~al.} 2021, \aj, 162, 101

\bibitem[{{Venuti} {et~al.}(2018){Venuti}, {Prisinzano}, {Sacco}, {Flaccomio},
  {Bonito}, {Damiani}, {Micela}, {Guarcello}, {Randich}, {Stauffer}, {Cody},
  {Jeffries}, {Alencar}, {Alfaro}, {Lanzafame}, {Pancino}, {Bayo}, {Carraro},
  {Costado}, {Frasca}, {Jofr{\'e}}, {Morbidelli}, {Sousa}, \&
  {Zaggia}}]{2018A&A...609A..10V}
{Venuti}, L., {Prisinzano}, L., {Sacco}, G.~G., {et~al.} 2018, \aap, 609, A10

\bibitem[{{Vioque} {et~al.}(2023){Vioque}, {Cavieres}, {Pantaleoni
  Gonz{\'a}lez}, {Ribas}, {Oudmaijer}, {Mendigut{\'\i}a}, {Kilian},
  {C{\'a}novas}, \& {Kuhn}}]{2023AJ....166..183V}
{Vioque}, M., {Cavieres}, M., {Pantaleoni Gonz{\'a}lez}, M., {et~al.} 2023,
  \aj, 166, 183

\bibitem[{{Vogt} {et~al.}(2016){Vogt}, {Contreras-Quijada}, {Fuentes-Morales},
  {Vogt-Geisse}, {Arcos}, {Abarca}, {Agurto-Gangas}, {Caviedes}, {DaSilva},
  {Flores}, {Gotta}, {Pe{\~n}aloza}, {Rojas}, \&
  {Villase{\~n}or}}]{2016ApJS..227....6V}
{Vogt}, N., {Contreras-Quijada}, A., {Fuentes-Morales}, I., {et~al.} 2016,
  \apjs, 227, 6

\bibitem[{{Wahhaj} {et~al.}(2010){Wahhaj}, {Cieza}, {Koerner}, {Stapelfeldt},
  {Padgett}, {Case}, {Keller}, {Mer{\'\i}n}, {Evans}, {Harvey}, {Sargent}, {van
  Dishoeck}, {Allen}, {Blake}, {Brooke}, {Chapman}, {Mundy}, \&
  {Myers}}]{2010ApJ...724..835W}
{Wahhaj}, Z., {Cieza}, L., {Koerner}, D.~W., {et~al.} 2010, \apj, 724, 835

\bibitem[{{Wang} {et~al.}(2023){Wang}, {Fang}, {Herczeg}, {Gao}, {Tian},
  {Zhou}, {Zhang}, \& {Chen}}]{2023RAA....23g5015W}
{Wang}, X.-L., {Fang}, M., {Herczeg}, G.~J., {et~al.} 2023, Research in
  Astronomy and Astrophysics, 23, 075015

\bibitem[{{Wenger} {et~al.}(2000){Wenger}, {Ochsenbein}, {Egret}, {Dubois},
  {Bonnarel}, {Borde}, {Genova}, {Jasniewicz}, {Lalo{\"e}}, {Lesteven},
  {Monier}, {Rochat}, {Schneider}, {Chemin}, {Fernique}, {Buras}, {Micol},
  {Moreau}, {Bouret}, {Bouy}, {Duche{\^e}ne}, {H{\o}g}, \&
  {Pau}}]{2000A&AS..143....9W}
{Wenger}, M., {Ochsenbein}, F., {Egret}, D., {et~al.} 2000, \aaps, 143, 9

\bibitem[{{Wilson} {et~al.}(2023){Wilson}, {Lakeland}, {Wilson}, \&
  {Naylor}}]{2023MNRAS.521..354W}
{Wilson}, A.~J., {Lakeland}, B.~S., {Wilson}, T.~J., \& {Naylor}, T. 2023,
  \mnras, 521, 354

\bibitem[{{Winston} {et~al.}(2019){Winston}, {Hora}, {Gutermuth}, \&
  {Tolls}}]{2019ApJ...880....9W}
{Winston}, E., {Hora}, J., {Gutermuth}, R., \& {Tolls}, V. 2019, \apj, 880, 9

\bibitem[{{Winston} {et~al.}(2007){Winston}, {Megeath}, {Wolk}, {Muzerolle},
  {Gutermuth}, {Hora}, {Allen}, {Spitzbart}, {Myers}, \&
  {Fazio}}]{2007ApJ...669..493W}
{Winston}, E., {Megeath}, S.~T., {Wolk}, S.~J., {et~al.} 2007, \apj, 669, 493

\bibitem[{{Wolk} {et~al.}(2018){Wolk}, {G{\"u}nther}, {Poppenhaeger},
  {Winston}, {Rebull}, {Stauffer}, {Gutermuth}, {Cody}, {Hillenbrand},
  {Plavchan}, {Covey}, \& {Song}}]{2018AJ....155...99W}
{Wolk}, S.~J., {G{\"u}nther}, H.~M., {Poppenhaeger}, K., {et~al.} 2018, \aj,
  155, 99

\bibitem[{Wu \& He(2018)}]{wu2018groupnorm}
Wu, Y. \& He, K. 2018, in Proceedings of the European Conference on Computer
  Vision (ECCV), 3--19

\bibitem[{Xie {et~al.}(2017)Xie, Girshick, Doll{\'a}r, Tu, \&
  He}]{xie2017resnext}
Xie, S., Girshick, R., Doll{\'a}r, P., Tu, Z., \& He, K. 2017, in Proceedings
  of the IEEE/CVF Conference on Computer Vision and Pattern Recognition (CVPR),
  1492--1500

\bibitem[{{Yao} {et~al.}(2017){Yao}, {Liu}, {Deng}, {de Grijs}, \&
  {Matsunaga}}]{2017ApJS..232...16Y}
{Yao}, Y., {Liu}, C., {Deng}, L., {de Grijs}, R., \& {Matsunaga}, N. 2017,
  \apjs, 232, 16

\bibitem[{{Yoon} {et~al.}(2014){Yoon}, {Cho}, {Kim}, {Yun}, \&
  {Park}}]{2014ApJS..211...15Y}
{Yoon}, D.-H., {Cho}, S.-H., {Kim}, J., {Yun}, Y.~j., \& {Park}, Y.-S. 2014,
  \apjs, 211, 15

\bibitem[{{Zacharias} {et~al.}(2017){Zacharias}, {Finch}, \&
  {Frouard}}]{2017AJ....153..166Z}
{Zacharias}, N., {Finch}, C., \& {Frouard}, J. 2017, \aj, 153, 166

\bibitem[{{Zari} {et~al.}(2018){Zari}, {Hashemi}, {Brown}, {Jardine}, \& {de
  Zeeuw}}]{2018A&A...620A.172Z}
{Zari}, E., {Hashemi}, H., {Brown}, A.~G.~A., {Jardine}, K., \& {de Zeeuw},
  P.~T. 2018, \aap, 620, A172

\bibitem[{{Zhang} {et~al.}(2023){Zhang}, {Zhang}, {Li}, \&
  {Yuan}}]{2023ApJS..264...24Z}
{Zhang}, C., {Zhang}, G.-Y., {Li}, J.-Z., \& {Yuan}, J.-H. 2023, \apjs, 264, 24

\bibitem[{{Zucker} {et~al.}(2022){Zucker}, {Goodman}, {Alves}, {Bialy},
  {Foley}, {Speagle}, {Gro{\^I}{\texttwosuperior}schedl}, {Finkbeiner},
  {Burkert}, {Khimey}, \& {Swiggum}}]{2022Natur.601..334Z}
{Zucker}, C., {Goodman}, A.~A., {Alves}, J., {et~al.} 2022, \nat, 601, 334

\end{thebibliography}

\onecolumn
\begin{appendix}
\section{Source types in the training sample}

\begin{table}[!ht]
\centering
\caption{Number of sources selected for the training, validation, and test
sets from the different catalogues used to distinguish YSOs from non-YSOs.}
\label{tvt}
\small\setlength{\tabcolsep}{4pt}
\begin{tabular}{@{}>{\raggedright\arraybackslash}p{\dimexpr\linewidth-40pt-3.0cm-3.3cm-3.9cm\relax}>{\raggedright\arraybackslash}p{3.0cm}>{\raggedright\arraybackslash}p{3.3cm}*{3}{>{\raggedleft\arraybackslash}p{1.3cm}}@{}}
\hline\hline
Reference & Catalogue/subsample & Object type & Train & Validation & Test\\
\hline
{\citealt{2025A&A...702A..63R}} & NEMESIS OSFC & YSO & 8\,000 & 1\,000 & 1\,000\\
{\citealt{2023A&A...674A..21M}} & KYSO & YSO & 8\,000 & 1\,000 & 1\,000\\
\hline
{\citealt{2015A&A...577A.128A}} & - & M dwarf & 350 & 50 & 50\\
{\citealt{2014ApJ...788...48S,2018MNRAS.477.3145J}} & ASAS-SN & $\delta$~Sct & 6\,000 & 500 & 500\\
{\citealt{2014ApJ...788...48S,2018MNRAS.477.3145J}} & ASAS-SN & Mira & 7\,000 & 1\,000 & 1\,000\\
{\citealt{2014ApJ...788...48S,2018MNRAS.477.3145J}} & ASAS-SN & RR~Lyrae (type ab) & 600 & 100 & 100\\
{\citealt{2023A&A...674A..13E,2023A&A...674A..14R}} & \textit{Gaia} DR3 & Cepheid & 7\,000 & 1\,000 & 1\,000\\
{\citealt{2023A&A...674A..41G}} & \textit{Gaia} DR3 & Galaxy & 4\,000 & 500 & 500\\
{\citealt{2023A&A...674A..13E,2023A&A...674A..14R}} & \textit{Gaia} DR3 & Main-sequence oscillator & 7\,000 & 700 & 700\\
{\citealt{2023A&A...674A..41G}} & \textit{Gaia} DR3 & QSO & 1\,500 & 250 & 250\\
{\citealt{2023A&A...674A..13E,2023A&A...674A..14R}} & \textit{Gaia} DR3 & RR~Lyrae & 5\,000 & 500 & 500\\
{\citealt{2003A&A...412...45P}} & HYPERLEDA I & Galaxy & 10\,000 & 1\,000 & 1\,000\\
{\citealt{2022MNRAS.510.4308K}} & - & Main-sequence star & 25\,000 & 2\,500 & 2\,500\\
{\citealt{2017ApJ...834...85N}} & - & M dwarf & 700 & 70 & 70\\
{\citealt{2017AcA....67..297S,2022ApJS..260...46I}} & OGLE & Long-period variable & 350 & 30 & 30\\
{\citealt{2017AcA....67..297S,2022ApJS..260...46I}} & OGLE & Mira & 15\,000 & 1\,000 & 1\,000\\
{\citealt{2015PASA...32...10F}} & Million Quasars & QSO & 3\,000 & 200 & 200\\
{\citealt{2000A&AS..143....9W}} & Simbad & AGB star & 2\,500 & 250 & 250\\
{\citealt{2000A&AS..143....9W}} & Simbad & Galaxy & 15\,000 & 2\,000 & 2\,000\\
{\citealt{2000A&AS..143....9W}} & Simbad & post-AGB star & 100 & 10 & 10\\
\hline
\end{tabular}

\end{table}

\section{Classification performance of individual CNN architectures}
\begingroup
\scriptsize
\setlength{\tabcolsep}{5pt}
\renewcommand{\arraystretch}{0.9}
\begin{longtable}{@{}lrrrrrr@{}}
\caption{Classification performance of all CNN architectures across the four image-based data products.}\label{sedclassificationcnnresults}\\
\hline\hline
CNN & TP [\%] & TN [\%] & FP [\%] & FN [\%] & TH & $F_1$ \\
\hline
\endfirsthead
\caption{continued.}\\
\hline\hline
CNN & TP [\%] & TN [\%] & FP [\%] & FN [\%] & TH & $F_1$ \\
\hline
\endhead
\hline
\endfoot
\multicolumn{7}{@{}l}{SED}\\
\hline
\texttt{EfficientNet B0} & 95.62 & 99.51 & 0.49 & 4.38 & 0.383 & 96.32 \\
\texttt{EfficientNetV2-S} & 96.27 & 99.56 & 0.44 & 3.73 & 0.508 & 96.80 \\
\texttt{MNASNet} & 95.27 & 99.34 & 0.66 & 4.73 & 0.265 & 95.65 \\
\texttt{MobileNetV3 small} & 94.83 & 99.42 & 0.58 & 5.17 & 0.598 & 95.63 \\
\texttt{RegNetY\_400MF} & 95.02 & 99.51 & 0.49 & 4.98 & 0.655 & 96.00 \\
\texttt{ResNet-18} & 95.62 & 99.52 & 0.48 & 4.38 & 0.598 & 96.37 \\
\texttt{ResNet-50} & 95.57 & 99.49 & 0.51 & 4.43 & 0.484 & 96.24 \\
\texttt{ResNeXt-50} & 95.12 & 99.52 & 0.48 & 4.88 & 0.457 & 96.10 \\
\texttt{ShuffleNetV2x0.5} & 94.08 & 99.47 & 0.53 & 5.92 & 0.402 & 95.38 \\
\texttt{SqueezeNet 1.1} & 95.67 & 99.28 & 0.72 & 4.33 & 0.194 & 95.70 \\
\texttt{SmallResNet} & 95.72 & 99.53 & 0.47 & 4.28 & 0.470 & 96.44 \\
\texttt{RCA} & 95.67 & 99.43 & 0.57 & 4.33 & 0.650 & 96.13 \\
\hline
\multicolumn{7}{@{}l}{SED$_r$}\\
\hline
\texttt{EfficientNet B0} & 90.75 & 99.92 & 0.08 & 9.25 & 0.898 & 94.90 \\
\texttt{EfficientNetV2-S} & 91.19 & 99.65 & 0.34 & 8.81 & 0.493 & 94.39 \\
\texttt{MNASNet} & 92.19 & 99.77 & 0.23 & 7.81 & 0.938 & 95.27 \\
\texttt{MobileNetV3 small} & 90.70 & 99.76 & 0.24 & 9.30 & 0.857 & 94.41 \\
\texttt{RegNetY\_400MF} & 90.75 & 99.80 & 0.20 & 9.25 & 0.558 & 94.56 \\
\texttt{ResNet-18} & 93.73 & 99.82 & 0.18 & 6.27 & 0.653 & 96.25 \\
\texttt{ResNet-50} & 91.54 & 99.89 & 0.11 & 8.46 & 0.801 & 95.26 \\
\texttt{ResNeXt-50} & 89.85 & 99.84 & 0.16 & 10.15 & 0.617 & 94.18 \\
\texttt{ShuffleNetV2x0.5} & 91.29 & 99.72 & 0.28 & 8.71 & 0.757 & 94.64 \\
\texttt{SqueezeNet 1.1} & 91.59 & 99.72 & 0.28 & 8.41 & 0.577 & 94.80 \\
\texttt{SmallResNet} & 93.38 & 99.68 & 0.32 & 6.62 & 0.650 & 95.64 \\
\texttt{RCA} & 91.19 & 99.79 & 0.21 & 8.81 & 0.710 & 94.78 \\
\hline
\multicolumn{7}{@{}l}{AllWISE stamps}\\
\hline
\texttt{EfficientNet B0} & 87.51 & 98.46 & 1.54 & 12.49 & 0.454 & 88.97 \\
\texttt{EfficientNetV2-S} & 86.92 & 98.50 & 1.50 & 13.08 & 0.720 & 88.75 \\
\texttt{MNASNet} & 85.32 & 98.19 & 1.81 & 14.68 & 0.553 & 87.01 \\
\texttt{MobileNetV3 small} & 86.67 & 98.01 & 1.99 & 13.33 & 0.477 & 87.32 \\
\texttt{RegNetY\_400MF} & 89.25 & 98.09 & 1.91 & 10.75 & 0.486 & 88.97 \\
\texttt{ResNet-18} & 87.61 & 98.26 & 1.74 & 12.39 & 0.503 & 88.49 \\
\texttt{ResNet-50} & 88.31 & 97.91 & 2.09 & 11.69 & 0.501 & 87.96 \\
\texttt{ResNeXt-50} & 87.26 & 98.31 & 1.69 & 12.74 & 0.470 & 88.45 \\
\texttt{ShuffleNetV2x0.5} & 87.41 & 98.62 & 1.38 & 12.59 & 0.596 & 89.35 \\
\texttt{SqueezeNet 1.1} & 86.47 & 98.44 & 1.56 & 13.53 & 0.365 & 88.34 \\
\texttt{SmallResNet} & 85.42 & 98.61 & 1.39 & 14.58 & 0.500 & 88.19 \\
\texttt{RCA} & 89.90 & 98.39 & 1.61 & 10.10 & 0.580 & 90.35 \\
\hline
\multicolumn{7}{@{}l}{ZTF DTDM}\\
\hline
\texttt{EfficientNet B0} & 91.77 & 98.00 & 2.00 & 8.23 & 0.585 & 93.13 \ \\
\texttt{EfficientNetV2-S} & 91.77 & 97.28 & 2.72 & 8.23 & 0.334 & 92.22 \ \\
\texttt{MNASNet} & 93.39 & 97.28 & 2.72 & 6.61 & 0.490 & 93.09 \ \\
\texttt{MobileNetV3 small} & 92.66 & 97.61 & 2.39 & 7.34 & 0.477 & 93.11 \ \\
\texttt{RegNetY\_400MF} & 91.77 & 97.49 & 2.51 & 8.23 & 0.489 & 92.48 \ \\
\texttt{ResNet-18} & 92.02 & 98.31 & 1.69 & 7.98 & 0.545 & 93.64 \ \\
\texttt{ResNet-50} & 90.56 & 98.19 & 1.81 & 9.44 & 0.553 & 92.70 \ \\
\texttt{ResNeXt-50} & 92.90 & 98.00 & 2.00 & 7.10 & 0.519 & 93.73 \ \\
\texttt{ShuffleNetV2x0.5} & 91.94 & 96.89 & 3.11 & 8.06 & 0.408 & 91.82 \ \\
\texttt{SqueezeNet 1.1} & 92.82 & 97.04 & 2.96 & 7.18 & 0.482 & 92.49 \ \\
\texttt{SmallResNet} & 91.77 & 97.85 & 2.15 & 8.23 & 0.480 & 92.94 \ \\
\texttt{RCA} & 93.95 & 97.64 & 2.36 & 6.05 & 0.500 & 93.84 \ \\
\hline
\end{longtable}
\noindent\footnotesize TP, TN, FP, and FN are rates in percent. The $F_1$ scores are computed at the optimised probability threshold listed in column TH.
\endgroup

\section{Classification results for literature YSO catalogues}
\small\setlength{\tabcolsep}{4pt}
\begin{longtable}{@{}>{\raggedright\arraybackslash}p{\dimexpr\linewidth-40pt-3.6cm-4.65cm-1.7cm\relax}>{\raggedright\arraybackslash}p{3.6cm}*{3}{>{\raggedleft\arraybackslash}p{1.55cm}}>{\raggedleft\arraybackslash}p{1.7cm}@{}}
\caption{SED-based reclassification statistics for the literature catalogues considered in this work.}
\label{tab:sed_yso}\\
\hline\hline
Reference & Catalogue/subsample & $N_{\rm cat}$ & $N_{\rm SED}$ & $N_{\rm YSO}$ & \shortstack[r]{Confirmed\\{[\%]}} \\
\hline
\endfirsthead

\caption{continued.}\\
\hline\hline
Reference & Catalogue/subsample & $N_{\rm cat}$ & $N_{\rm SED}$ & $N_{\rm YSO}$ & \shortstack[r]{Confirmed\\{[\%]}} \\
\hline
\endhead

\hline
\endfoot

\hline
\multicolumn{6}{@{}p{\linewidth}@{}}{\smallskip \small Notes. $N_{\rm cat}$ is the number of source entries from each literature catalogue included in our input sample. Depending on the structure and content of the original VizieR catalogue, this may correspond to a selected table or subset of the full published catalogue. $N_{\rm SED}$ is the number of sources for which a usable SED could be constructed after applying the photometric cleaning and SED-quality criteria. $N_{\rm YSO}$ is the number of sources classified as YSOs by at least six of the SED-plot CNNs. The reconfirmation fraction is defined as $100\times \,(N_{\rm YSO} / N_{\rm SED})$, and is calculated relative to the subset with usable SEDs, rather than relative to the original catalogue size.}
\endlastfoot

{\citealt{2014A&A...561A...2A}} & - & 36 & 36 & 36 & 100.00 \\
{\citealt{2012ApJ...750..125A}} & - & 2\,575 & 2\,356 & 1\,893 & 80.35 \\
{\citealt{2011ApJ...736..133A}} & - & 107 & 58 & 42 & 72.41 \\
{\citealt{2023MNRAS.519.5271C}} & ASAS YSO candidates & 9\,260 & 6\,805 & 1\,529 & 22.47 \\
{\citealt{2016ApJ...832...87B}} & - & 295 & 286 & 88 & 30.77 \\
{\citealt{2010ApJ...712..797B}} & - & 870 & 663 & 438 & 66.06 \\
{\citealt{2013A&A...557A..29C}} & - & 535 & 317 & 188 & 59.31 \\
{\citealt{2008ApJ...682..445C}} & - & 26\,821 & 3\,221 & 604 & 18.75 \\
{\citealt{2014MNRAS.439.3719C}} & - & 3\,021 & 584 & 446 & 76.37 \\
{\citealt{2014AJ....147...82C}} & - & 162 & 156 & 151 & 96.79 \\
{\citealt{2018AJ....156...71C}} & - & 340 & 338 & 278 & 82.25 \\
{\citealt{2022A&A...665A..76C}} & - & 135 & 124 & 72 & 58.06 \\
{\citealt{2010AJ....140.1214C}} & - & 110 & 92 & 83 & 90.22 \\
{\citealt{2020MNRAS.495.3614C}} & - & 59 & 12 & 12 & 100.00 \\
{\citealt{2025JKAS...58..209C}} & - & 355 & 180 & 170 & 94.44 \\
{\citealt{2021A&A...647A.116C}} & - & 26\,903 & 24\,397 & 15\,192 & 62.27 \\
{\citealt{2025A&A...699A.145D}} & - & 145\,975 & 68\,129 & 4\,264 & 6.26 \\
{\citealt{2023A&A...674A...1G}} & \textit{Gaia} DR3 T~Tauri candidates & 34\,540 & 34\,205 & 3\,037 & 8.88 \\
{\citealt{2023A&A...674A..21M}} & \textit{Gaia} DR3 YSO candidates & 79\,375 & 78\,350 & 58\,080 & 74.13 \\
{\citealt{2015ApJS..220...11D}} & - & 2\,966 & 2\,559 & 1\,404 & 54.87 \\
{\citealt{2009ApJS..181..321E}} & - & 1\,024 & 903 & 738 & 81.73 \\
{\citealt{2020ApJ...904..146F}} & - & 971 & 576 & 433 & 75.17 \\
{\citealt{2016ApJ...827...96F}} & - & 479 & 311 & 52 & 16.72 \\
{\citealt{2013AJ....145...66F}} & - & 377 & 333 & 278 & 83.48 \\
{\citealt{2021A&A...652A..76H}} & \textit{Gaia} Alerts YSO candidates & 11\,158 & 8\,066 & 582 & 7.22 \\
{\citealt{2016ApJ...830...57G}} & - & 234 & 40 & 20 & 50.00 \\
{\citealt{2014AJ....148..122G}} & - & 882 & 140 & 104 & 74.29 \\
{\citealt{2008ApJ...674..336G}} & - & 137 & 116 & 115 & 99.14 \\
{\citealt{2009ApJS..184...18G}} & - & 2\,548 & 1\,878 & 1\,805 & 96.11 \\
{\citealt{2022AJ....163..263H}} & - & 323 & 305 & 253 & 82.95 \\
{\citealt{2024A&A...686A..42H}} & - & 1\,291\,929 & 454\,371 & 21\,802 & 4.80 \\
{\citealt{2009ApJ...706...83K}} & - & 737 & 372 & 324 & 87.10 \\
{\citealt{2022A&A...663A.133K}} & - & 35 & 26 & 23 & 88.46 \\
{\citealt{2021ApJ...917...23K}} & - & 30\,518 & 18\,180 & 2\,598 & 14.29 \\
{\citealt{2011ApJ...726...18K}} & - & 408 & 400 & 318 & 79.50 \\
{\citealt{2017AJ....154...29K}} & - & 148 & 147 & 115 & 78.23 \\
{\citealt{2022AJ....164...57K}} & - & 810 & 669 & 379 & 56.65 \\
{\citealt{2014AJ....148...11K}} & - & 2\,007 & 396 & 346 & 87.37 \\
{\citealt{2014A&A...567A.109K}} & - & 241 & 123 & 76 & 61.79 \\
{\citealt{2017A&A...606A.100L}} & - & 177 & 90 & 74 & 82.22 \\
{\citealt{2022AJ....163..266L}} & - & 728 & 688 & 451 & 65.55 \\
{\citealt{2020AJ....160...44L}} & - & 2\,020 & 1\,818 & 656 & 36.08 \\
{\citealt{2022AJ....163...25L}} & - & 9\,675 & 5\,439 & 1\,291 & 23.74 \\
{\citealt{2016MNRAS.458.3479M}} & - & 742\,586 & 669\,942 & 33\,284 & 4.97 \\
{\citealt{2019MNRAS.487.2522M}} & - & 57\,710 & 50\,330 & 13\,817 & 27.45 \\
{\citealt{2023A&A...674A..21M}} & KYSO & 11\,671 & 11\,336 & 10\,157 & 89.60 \\
{\citealt{2024A&A...688A.203M}} & - & 90\,864 & 27\,117 & 10\,299 & 37.98 \\
{\citealt{2021AJ....162..282M}} & - & 646\,139 & 461\,787 & 119\,722 & 25.93 \\
{\citealt{2020AJ....159..200M}} & - & 404 & 184 & 26 & 14.13 \\
{\citealt{2016ApJ...823...58M}} & - & 27 & 16 & 14 & 87.50 \\
{\citealt{2021AJ....162..236M}} & - & 821 & 362 & 323 & 89.23 \\
{\citealt{2021AJ....162..153N}} & - & 370 & 241 & 53 & 21.99 \\
{\citealt{2018A&A...617A..63P}} & - & 2\,107 & 772 & 495 & 64.12 \\
{\citealt{2022A&A...664A.175P}} & - & 190\,303 & 39\,721 & 8\,081 & 20.34 \\
{\citealt{2008ApJ...685.1005Q}} & - & 417 & 242 & 193 & 79.75 \\
{\citealt{2014ApJ...794..124R}} & - & 10\,454 & 9\,869 & 4\,268 & 43.25 \\
{\citealt{2011ApJS..193...25R}} & - & 1\,286 & 908 & 771 & 84.91 \\
{\citealt{2015AJ....150..175R}} & - & 701 & 257 & 207 & 80.54 \\
{\citealt{2018AJ....155..196R}} & - & 1\,313 & 1\,307 & 761 & 58.22 \\
{\citealt{2022AJ....164...80R}} & - & 3\,714 & 3\,215 & 2\,208 & 68.68 \\
{\citealt{2017AJ....153...95R}} & - & 5\,350 & 4\,011 & 335 & 8.35 \\
{\citealt{2025A&A...702A..63R}} & NEMESIS OSFC & 15\,109 & 13\,685 & 12\,841 & 93.83 \\
{\citealt{2023A&A...675A.195S}} & - & 7\,805 & 7\,123 & 4\,466 & 62.70 \\
{\citealt{2017ApJ...839..108S}} & - & 14\,435 & 7\,076 & 5\,748 & 81.23 \\
{\citealt{2019ApJS..240...26S}} & - & 293 & 161 & 143 & 88.82 \\
{\citealt{2018ApJ...868...43S}} & - & 244 & 233 & 57 & 24.46 \\
{\citealt{2000A&AS..143....9W}} & SIMBAD T~Tauri stars & 5\,171 & 4\,821 & 3\,187 & 66.11 \\
{\citealt{2000A&AS..143....9W}} & SIMBAD YSOs & 66\,723 & 38\,005 & 24\,955 & 65.66 \\
{\citealt{2000A&AS..143....9W}} & SIMBAD YSO candidates & 99\,052 & 67\,752 & 30\,683 & 45.29 \\
{\citealt{2021ApJS..254...33K}} & SPICY catalogue & 117\,446 & 68\,868 & 54\,485 & 79.12 \\
{\citealt{2022ApJ...928..134S}} & - & 163 & 151 & 143 & 94.70 \\
{\citealt{2023MNRAS.520.1390S}} & - & 874 & 374 & 104 & 27.81 \\
{\citealt{2020A&A...642A..86T}} & - & 128 & 65 & 29 & 44.62 \\
{\citealt{2021MNRAS.504L..17T}} & - & 242 & 144 & 0 & 0.00 \\
{\citealt{2016ApJ...818...73T}} & - & 99 & 76 & 74 & 97.37 \\
{\citealt{2014PASJ...66...17T}} & - & 44\,001 & 5\,968 & 3\,951 & 66.20 \\
{\citealt{2015A&A...581A..66V}} & - & 676 & 671 & 507 & 75.56 \\
{\citealt{2017A&A...599A..23V}} & - & 532 & 491 & 270 & 54.99 \\
{\citealt{2018A&A...609A..10V}} & - & 1\,892 & 1\,841 & 957 & 51.98 \\
{\citealt{2021AJ....162..101V}} & - & 278 & 226 & 103 & 45.58 \\
{\citealt{2023AJ....166..183V}} & - & 263 & 243 & 166 & 68.31 \\
{\citealt{2010ApJ...724..835W}} & - & 186 & 183 & 175 & 95.63 \\
{\citealt{2023RAA....23g5015W}} & - & 238 & 179 & 149 & 83.24 \\
{\citealt{2023MNRAS.521..354W}} & - & 3\,734 & 3\,610 & 1\,658 & 45.93 \\
{\citealt{2007ApJ...669..493W}} & - & 137 & 104 & 82 & 78.85 \\
{\citealt{2019ApJ...880....9W}} & - & 4\,648 & 2\,120 & 1\,427 & 67.31 \\
{\citealt{2018AJ....155...99W}} & - & 1\,524 & 68 & 47 & 69.12 \\
{\citealt{2018A&A...620A.172Z}} & - & 130\,851 & 127\,187 & 26\,875 & 21.13 \\
{\citealt{2023ApJS..264...24Z}} & - & 403 & 272 & 247 & 90.81 \\
{\citealt{2021AJ....161..141S}} & ZTF ALeRCE YSO candidates & 2\,007 & 1\,970 & 1\,668 & 84.67 \\
\hline
Total & & 3\,977\,397 & 2\,353\,539 & 501\,051 & 21.28 \\
\end{longtable}

\begin{minipage}{\linewidth}
\section{Distribution of the NGYSO sources in Galactic Cartesian coordinates}
\begin{center}

\includegraphics[width=0.42\hsize]{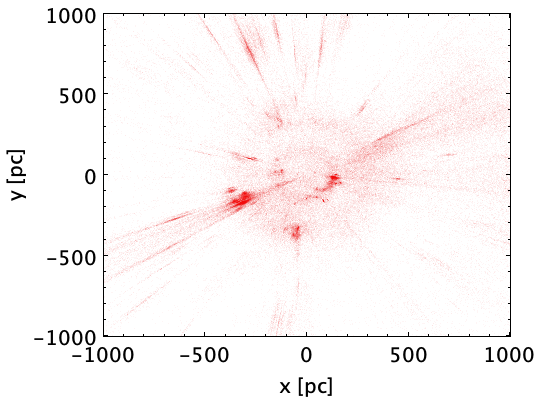}
\includegraphics[width=0.42\hsize]{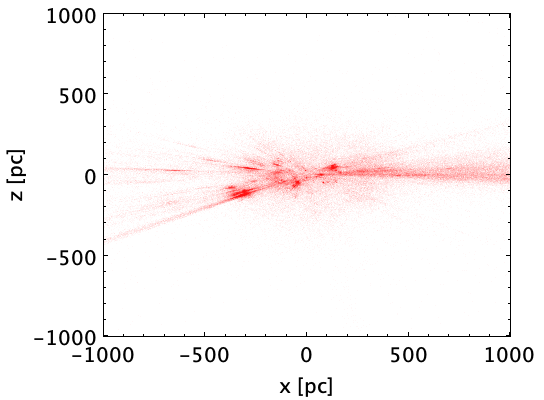}\\
\includegraphics[width=0.42\hsize]{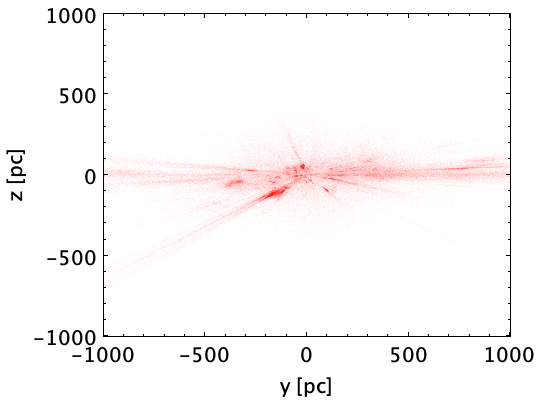}
\includegraphics[width=0.42\hsize]{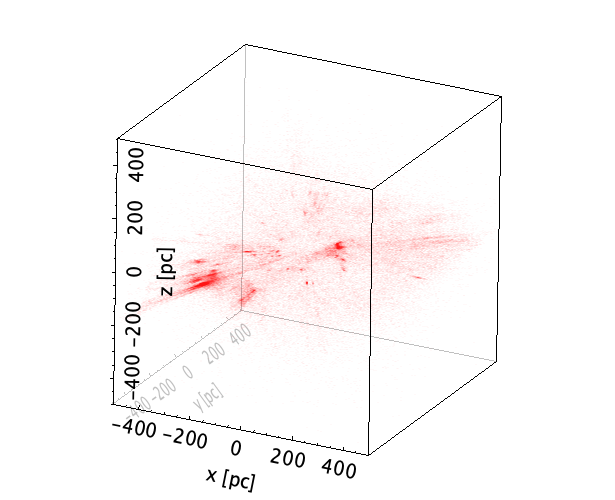}
\captionof{figure}{Distribution of the NGYSO sources in heliocentric Galactic Cartesian coordinates. The upper-left, upper-right, and lower-left panels show the x--y, x--z, and y--z projections, respectively, while the lower-right panel presents the three-dimensional distribution within the central $1\ \mathrm{kpc}$. The strongly flattened distribution in z traces the concentration of young stellar objects towards the Galactic plane. Numerous spatial overdensities associated with nearby star-forming complexes are visible. Several structures appear elongated approximately along heliocentric lines of sight; these radial features are likely enhanced, or in some cases produced, by uncertainties in the inferred rgeo distances \citep[the median of the geometric distance posterior,][]{2021AJ....161..147B}, which propagate primarily along the line-of-sight direction.}
\label{xyz}
\end{center}
\end{minipage}

\section{Misclassification of literature non-YSOs}
\small\setlength{\tabcolsep}{4pt}
\begin{longtable}{@{}>{\raggedright\arraybackslash}p{\dimexpr\linewidth-40pt-3.0cm-2.4cm-3.1cm-1.9cm\relax}>{\raggedright\arraybackslash}p{3.0cm}>{\raggedright\arraybackslash}p{2.4cm}*{2}{>{\raggedleft\arraybackslash}p{1.55cm}}>{\raggedleft\arraybackslash}p{1.9cm}@{}}
\caption{Contamination statistics based on the SED classification method}\label{tab:sedr_nonyso}\\
\hline\hline
Reference & Catalogue/subsample & Object type & $N_{\rm SED}$ & $N_{\rm FP}$ & \shortstack[r]{Misclassified\\{[\%]}} \\
\hline
\endfirsthead
\caption{continued.}\\
\hline\hline
Reference & Catalogue/subsample & Object type & $N_{\rm SED}$ & $N_{\rm FP}$ & \shortstack[r]{Misclassified\\{[\%]}} \\
\hline\endhead

\hline
\endfoot

\hline
\multicolumn{6}{@{}p{\linewidth}@{}}{\smallskip \small Notes. The Catalogue/subsample column gives a separate catalogue or subsample name where available; a dash indicates that no separate name is specified. Object type gives the source class. $N_{\rm SED}$ is the number of sources for which a usable SED could be constructed after SED quality cuts, $N_{\rm FP}$ is the number of false positives misclassified as YSOs by at least six CNN models, and the resulting misclassification fraction (\%) calculated as $100\times \,(N_{\rm FP} / N_{\rm SED})$. AGN denotes active galactic nuclei; CEP, Cepheids; Gal, galaxies; MS, main-sequence stars; QSO, quasars; AGB, asymptotic giant branch stars; and BLA, blazars.}
\endlastfoot

{\citealt{2020MNRAS.494.1784A}} & - & AGN & 161 & 4 & 2.48 \\
{\citealt{2023A&A...674A..17R,2023A&A...674A..13E,2023A&A...674A..14R}} & \textit{Gaia} DR3 & CEP & 5\,404 & 311 & 5.75 \\
{\citealt{2023A&A...674A..41G}} & \textit{Gaia} DR3 & Gal & 1\,555 & 0 & 0.00 \\
{\citealt{2023A&A...674A..17R,2023A&A...674A..13E}} & \textit{Gaia} DR3 & MS & 9\,338 & 69 & 0.74 \\
{\citealt{2023A&A...674A..41G}} & \textit{Gaia} DR3 & QSO & 2\,030 & 0 & 0.00 \\
{\citealt{2023A&A...674A..17R,2023A&A...674A..13E,2023A&A...674A..14R}} & \textit{Gaia} DR3 & RR Lyrae & 7\,006 & 37 & 0.53 \\
{\citealt{2013ApJ...765..154D}} & - & RR Lyrae & 2\,189 & 0 & 0.00 \\
{\citealt{2021ApJS..254....6F}} & - & QSO & 66 & 0 & 0.00 \\
{\citealt{2022ApJS..260...46I}} & - & Mira & 9\,948 & 2 & 0.02 \\
{\citealt{2012ApJ...754...44J}} & - & M dwarf & 678 & 53 & 7.82 \\
{\citealt{2000BaltA...9..646K}} & - & Mira & 932 & 0 & 0.00 \\
{\citealt{2022MNRAS.510.4308K}} & - & MS & 9\,152 & 19 & 0.21 \\
{\citealt{2017ApJ...843...16K}} & - & Gal & 1\,202 & 1 & 0.08 \\
{\citealt{2011AJ....142..138L}} & - & M dwarf & 154 & 0 & 0.00 \\
{\citealt{2009MNRAS.399.1709M}} & - & Mira & 556 & 6 & 1.08 \\
{\citealt{2022ApJ...935..104M}} & - & M dwarf & 129 & 0 & 0.00 \\
{\citealt{2019MNRAS.489.2615M}} & - & M dwarf & 200 & 21 & 10.50 \\
{\citealt{2019AcA....69..321S}} & OGLE & CEP+RR Lyrae & 34\,817 & 333 & 0.96 \\
{\citealt{2024A&A...685A.168P}} & - & Gal & 121 & 1 & 0.83 \\
{\citealt{2024A&A...681A.119R}} & - & BLA & 278 & 1 & 0.36 \\
{\citealt{2012ApJ...749...79R}} & - & AGB & 42 & 0 & 0.00 \\
{\citealt{2005IBVS.5633....1R}} & - & Mira & 66 & 0 & 0.00 \\
{\citealt{2002AJ....123.2772S}} & - & AGB & 497 & 14 & 2.82 \\
{\citealt{2024A&A...683A.112S}} & - & QSO & 422 & 0 & 0.00 \\
{\citealt{2024A&A...681A..16T}} & - & Gal & 32 & 0 & 0.00 \\
{\citealt{2008OEJV...87....1U}} & - & AGB & 659 & 2 & 0.30 \\
{\citealt{2024A&A...683A.157V}} & - & QSO & 507 & 0 & 0.00 \\
{\citealt{2016ApJS..227....6V}} & - & Mira & 787 & 0 & 0.00 \\
{\citealt{2017ApJS..232...16Y}} & - & Mira & 169 & 0 & 0.00 \\
{\citealt{2014ApJS..211...15Y}} & - & AGB & 245 & 9 & 3.67 \\
\hline
Total & & & 89\,342 & 883 & 0.99 \\
\end{longtable}

\begin{minipage}{\linewidth}
\section{NGYSO \textit{Gaia} colour--absolute magnitude distributions}

\begin{center}

\includegraphics[width=0.42\hsize]{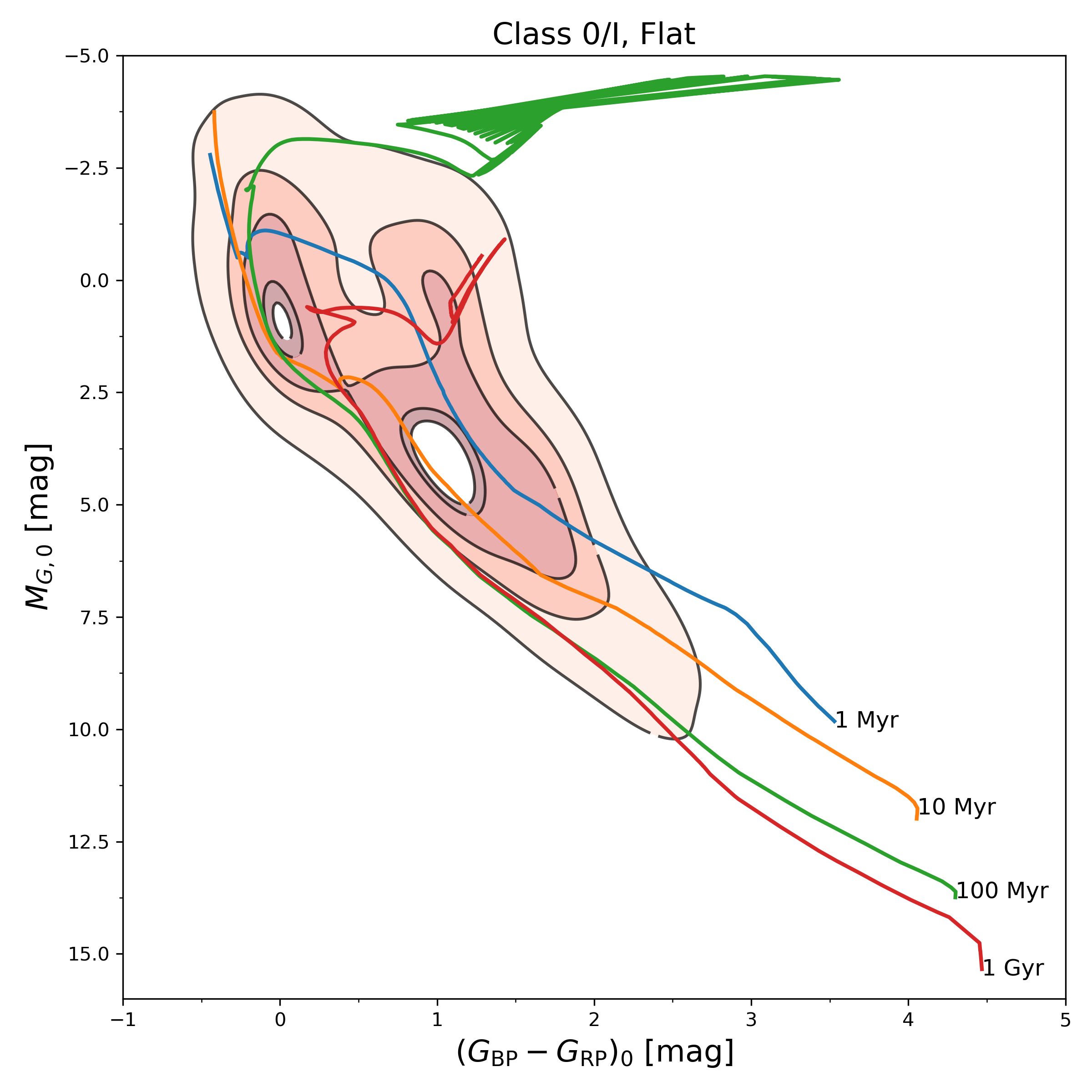}
\includegraphics[width=0.42\hsize]{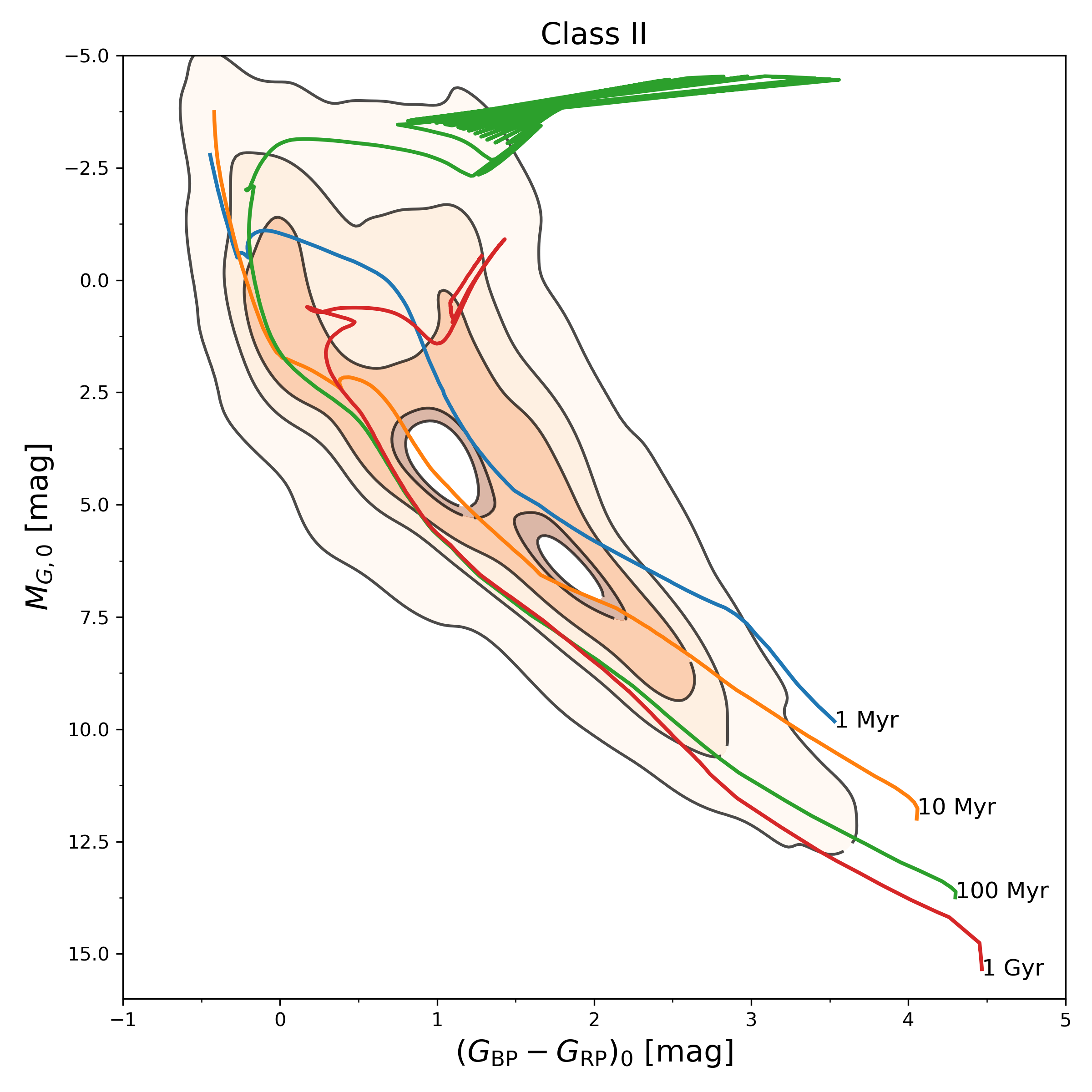}\\
\includegraphics[width=0.42\hsize]{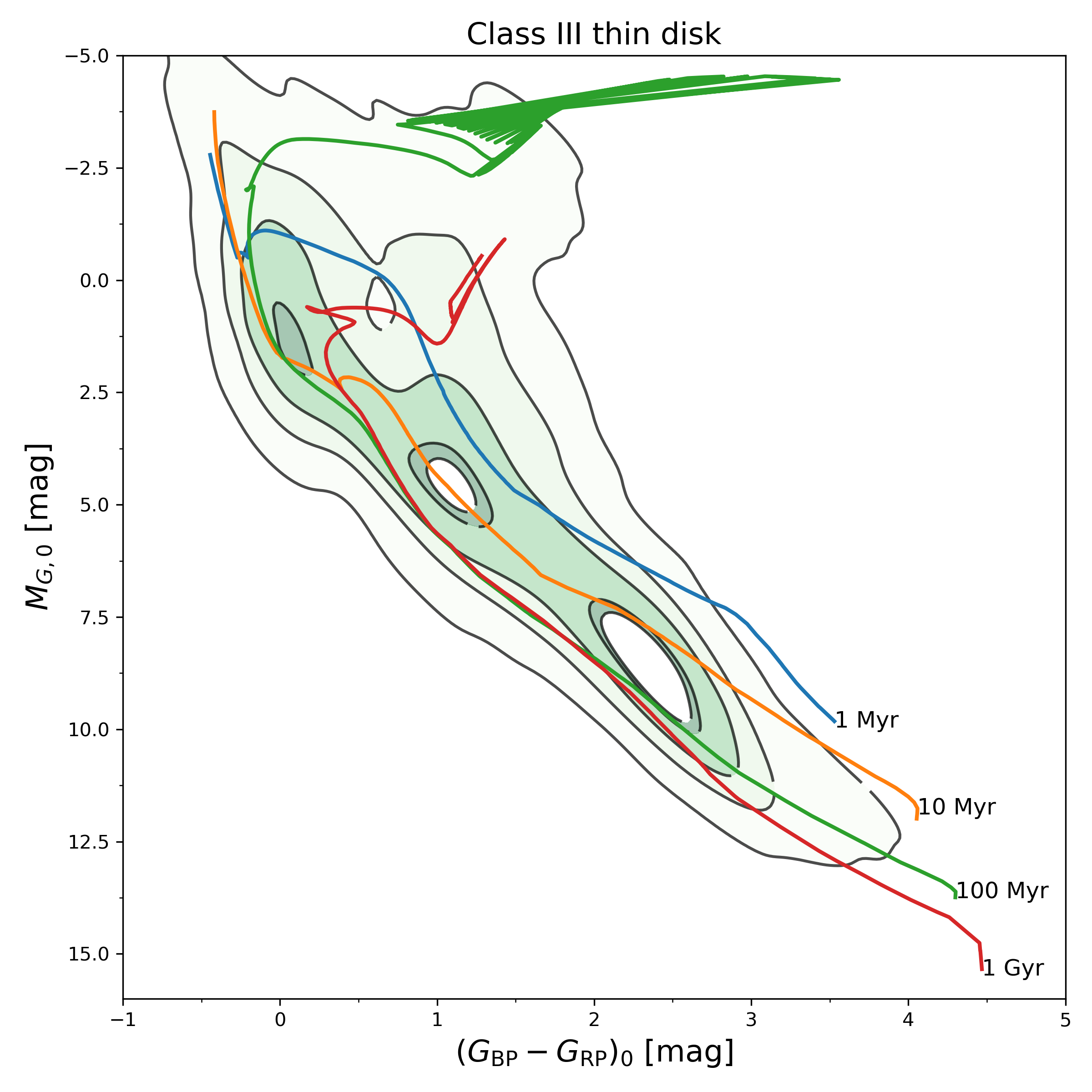}
\includegraphics[width=0.42\hsize]{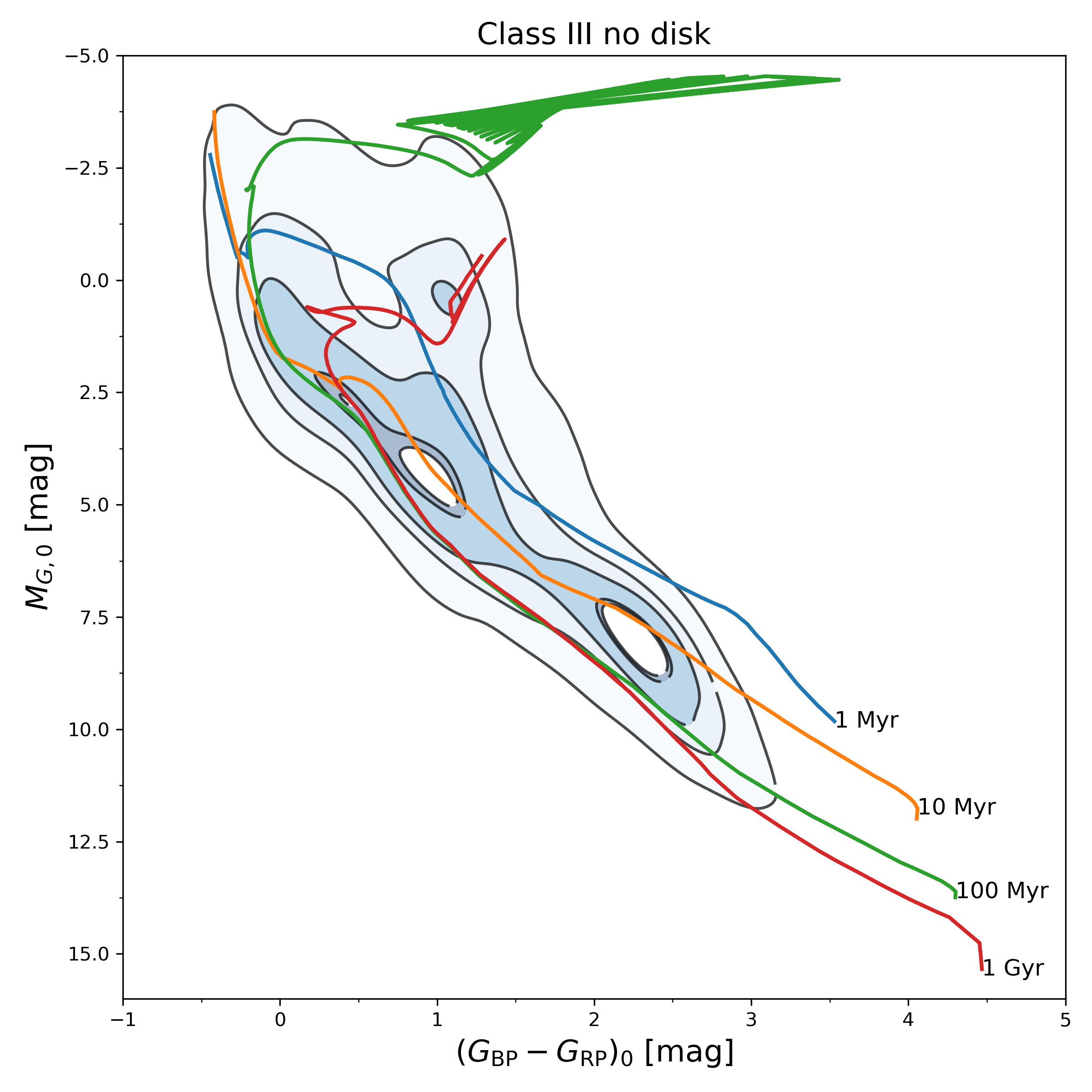}
\captionof{figure}{Density distribution of the YSO candidates in different observational evolutionary classes on the dereddened \textit{Gaia} colour--absolute magnitude diagram, similar to Fig.~\ref{hrd}. Top left: Class 0/I and Flat sources ($\alpha \geq -0.3$). Top right: Class II sources ($-0.3 \geq \alpha \geq -1.6$). Bottom left: Class III thin disk sources ($-1.6 \geq \alpha \geq -2.5$). Bottom right: Class III no disk sources ($\alpha \leq -2.5$).}
\label{hrdclasses}
\end{center}
\end{minipage}

\end{appendix}

\end{document}